\documentclass[twoside,12pt]{article}
\usepackage{graphicx}
\usepackage[utf8]{inputenc}
\pdfoutput=1

\def\Journal#1#2#3#4{{#1} {#2} (#4) #3 }

\def\NPA{{\em Nucl. Phys.} A}

\def\PTP{{\em Prog. Theor. Phys.}}
\def\PTEP{{\em Prog. Theor.Exp. Phys.}}
\def\PPNP{{\em Prog. Part. Nucl. Phys.}}
\def\NPB{{\em Nucl. Phys.} B}

\def\PLB{{\em Phys. Lett.} B}

\def\PRL{{\em Phys. Rev. Lett.}}
\def\PREV{{\em Phys. Rev.}}
\def\PREP{\em Phys. Rep.}

\def\PRD{{\em Phys. Rev.} D}
\def\PRC{{\em Phys. Rev.} C}

\def\ZPA{{\em Z. Phys.} A}
\def\EPJA{{\em Eur. Phys. J.} A}

\newcommand{\be}{\begin{equation}}
\newcommand{\ee}{\end{equation}}
\newcommand{\bea}{\begin{eqnarray}}
\newcommand{\eea}{\end{eqnarray}}

\begin{document}

\title{ \vspace{1cm} On the History of Dibaryons and their Final
  Observation\footnote{To be published in Progress in Particle and Nuclear
    Physics 2017}}
\author{H.\ Clement\\
\\
Physikalisches Institut der Universit\"at T\"ubingen\\
and\\
Kepler Center for Astro and Particle Physics, University of T\"ubingen,\\
Auf der Morgenstelle 14, D-72076 T\"ubingen, Germany\\
\\}
\maketitle
\begin{abstract} A review is given about the long-standing search for dibaryons, {\it
    i.e.} six-quark objects, from the early days until present, when the first
dibaryon resonance has been established, which has the potential of
constituting a compact six-quark object.
\end{abstract}

\tableofcontents

\section{Introduction}
\label{intro}

Strictly speaking a dibaryon denotes just any object with baryon number B =
2. In this sense the first known dibaryon has been the deuteron discovered in
1932 by Urey, Brickwedde and Murphy \cite{Urey}. In terms of
quarks a dibaryon is composed of six valence quarks -- or more generally
speaking the
number of quarks minus the number of antiquarks has to be six. It may be of
molecular type, {\it i.e.} spatially extended with two well separated
interacting quark bags as is the case for the well-known deuteron. Or -- more
exotic and hence more interesting -- a dibaryon could be a spatially compact
hexaquark object, where all six quarks sit in a single quark bag.

The long history of dibaryon searches dating back to the fifties has been a very
changeful one with many ups and downs. Early predictions of a vast number of
dibaryon states initiated endless experimental claims, but finally
none survived careful experimental investigations. For a review about this
dibaryon rush era, which lasted until the eighties, see, {\it e.g.}
Refs. \cite{Yokosawa,Locher,igor,Seth,Seth1,Tatischeffreview,Komarov}. For
more recent reviews from the theoretical point of view see, {\it e.g.}, Ref. \cite{Gal,GalMESON2016}.

Despite their long painful history dibaryon searches have recently
received renewed interest, in particular by the recognition that there are more
complex quark configurations than just the familiar $q\bar q$ and $qqq$
systems -- in favor also of hidden-color aspects \cite{BBC}. Recently two
groups announced that lattice QCD calculations 
\cite{NPLQCD,NPLQCD1,HALQCD,Inoue} provide evidence for a
bound $H$-dibaryon -- as predicted by Jaffe \cite{Jaffe} already in
1977. Nevertheless any experimental evidence for it is still pending despite the
manyfold experimental efforts
\cite{NAGARA,Nakazawa,E224,E522,BELLE,ALICE,STAR,Morita,Ohnishi}.   
 
Recently the WASA-at-COSY collaboration has found that the
double-pionic fusion reaction $pn \to d\pi^0\pi^0$ proceeds dominantly via a
resonance structure observed in the total cross section at $\sqrt s$ = 2.37
GeV with $\Gamma \approx$ 70 MeV and $I(J^P) = 0(3^+)$ \cite{prl2011} -- after
first signs of it had been observed already at Uppsala by the CELSIUS/WASA
collaboration \cite{prl2009}. Meanwhile nearly all possible decay channels have
been investigated \cite{isofus,pp0-,np00}, in particular also the one
into the elastic $np$ channel. And new data on polarized $np$
scattering in the region of interest exhibit a resonance pole in the coupled
$^3D_3-^3G_3$ partial waves in accordance with the resonance hypothesis
\cite{prl2014,npfull}. This gives the first solid evidence 
for the existence of a non-trivial, possibly extraordinary dibaryon.

Since the measurements suggest this resonance to decay dominantly via an
intermediate 
$\Delta\Delta$ system, it constitutes asymptotically a $\Delta\Delta$ system
bound by nearly 100 MeV - as predicted by Dyson and Xuong \cite{Dyson} already
in 1964 and later-on also by Goldman {\it et al.} \cite{Goldman}, who called
it the "inevitable dibaryon" $d^*$ due to its unique symmetry features. Most
recent relativistic three-body calculations based on hadron dynamics
\cite{GG1,GG2} as well as quark model calculations \cite{Nanjing,Yuan,zhang}
succeeded to predict properly a number of characteristics of this
resonance. The latter also postulate a substantial hidden-color component
accounting in particular for the unusually narrow width of this resonance. 

In the following we present a historical review about dibaryon predictions as
well as experimental dibaryon searches. We start with some basics of the
baryon-baryon system, before we enter the early days of dibaryon
initiatives. Then we consider the dibaryon rush era initiated by Jaffe's
prediction of the $H$ dibaryon, a bound $\Lambda\Lambda$ system. The deplorable
experience with unjustified claims of dibaryon discoveries due to statistically
poor, inclusive and/or kinematically incomplete measurements led in the
nineties on the one hand to a big frustration about this subject, but on
the other hand also to a restart of dedicated exclusive and kinematically complete
high-statistics measurements, which will be the topic of the second part of
this review and which finally lead to the first solid evidence for a
non-trivial dibaryon resonance. Finally we give an outlook on future
activities in this field.

In the following we will adopt the convention of the Particle Data Group
\cite{PDG} and quote masses for simplicity in MeV instead of MeV/c$^2$. Also, 
angular momenta between baryons are denoted by capital letters, otherwise by
small letters.

\section{Basics of the Baryon-Baryon System}
\label{sec-basics}

Naturally the quest for dibaryons concentrated first on the nucleon-nucleon
($NN$) system, before later-on also the strangeness sector was explored.

\subsection{\it The Nucleon-Nucleon System}
\subsubsection{\it The deuteron}

As mentioned already in the introduction, formally the oldest known dibaryon is
the deuteron. And until recently it has been the only safely discovered
dibaryon at all.

Since its discovery in 1932 by Urey, Brickwedde and Murphy \cite{Urey}
in the atomic spectrum of hydrogen gas in a discharge tube, the spectroscopic
quantities of the deuteron like mass, charge, spin-parity, dipole and quadrupole
moments as well as form factors and structure functions have been measured with
high precision.  

The deuteron is a very loosely bound object with a binding energy of only 2.2
MeV or 1.1 MeV/A -- which has to be compared to the average binding energy of
about 8 MeV/A in nuclei. Accordingly it possesses a very large charge radius
of 2.1 fm \cite{ADANDT,CODATA}, {\it i.e.} the centers of proton and neutron
are on the average further apart from each other then the range of the pion
exchange $r \approx \frac {\bar hc} m_\pi \approx$ 1.4 fm, which constitutes
the longest range hadronic interaction in the $NN$ system.  

Due to this large internucleon distance the pion exchange is the by far
dominating 
interaction, which completely describes the asymptotic behavior of the
deuteron. Hence the pioneering simple model descriptions of Hulthen
\cite{Hulthen} and Reid \cite{Reid} have been already very successful in
properly accounting for the asymptotic behavior of the deuteron
wavefunction. Only at short distances in the neutron-proton overlap region the
situation is not yet fully understood. In particular it is still an unsolved
problem, how much of a hexaquark configuration within a compact six-quark bag
is present in the deuteron. Whereas early theoretical work \cite{Dijk} on that
claimed up to 1.5$\%$ six-quark content in the deuteron, more recent
evaluations see this number near 0.15 - 0.30 $\%$ \cite{Millerdeuteron,Diaz}.
However, a recent dressed bag model calculation arrives again at 2 - 3 $\%$
\cite{KukulinD}.  

For the $\Delta\Delta$ component in the deuteron there exists an experimental
upper limit of 0.4 $\%$ at the 90 $\%$ confidence level from neutrino-deuteron
interaction studies \cite{Allasia,Dymarz}. Theoretical estimates give
contributions of the same order \cite{Dymarz,Ivanov}.

\subsubsection{\it Dineutron and diproton}

The question, whether a stable diproton, {\it i.e.} a stable $^2He$ nucleus
exists, may be solved easily by mass spectrometer measurements. With 
regard to the dineutron the experimental  situation is not quite as easy,
since it has no charge and hence also no atomic shell, which could reveal its
existence or non-existence.   

However, we know the answer to both questions from NN scattering experiments
and successive partial-wave analyses. Of particular interest in this respect
are the $L = 0$ partial waves $^1S_0$ and $^3S_1$ with isospin 1 and 0,
respectively, most importantly their scattering lengths.  If the scattering
length is negative, then we have an attractive interaction, which however is
not strong enough to create a $NN$ boundstate. Only, if the scattering length is
positive and larger than the effective range of the interaction, then the $NN$
system possesses an boundstate in this particular partial wave. 

For the scattering length in the isoscalar $^3S_1$ partial wave, which occurs
only in the $np$ system, a value of $\approx$ +5 fm \cite{Machleidt} has been
obtained experimentally, {\it i.e.} this system has a boundstate, the well-known
deuteron boundstate. 

For the
scattering length of the isovector $^1S_0$ partial wave the experimental
results are $\approx$ -18 fm in case of $nn$ \cite{Schori,Gabioud} and $pp$
\cite{Noyes} scattering and -24 fm in case of $np$ scattering
\cite{Machleidt,Huhn} -- the difference being due to charge symmetry 
breaking. These results tell us that there is {\bf no} boundstate in
$nn$ and $pp$ systems.

\subsubsection{\it The virtual $^1S_0$ state in the isovector nucleon-nucleon
  system}

Though the $NN$ interaction in the isovector $^1S_0$ channel is too weak to
produce a boundstate, it is strong enough to create at least a virtual state
being unbound by merely 66 keV \cite{Flambaum}. In nuclear reactions
experiments this state 
is sensed as a final state interaction (FSI) between emitted nucleons
\cite{Migdal,Watson}. There it shows up as a low-mass enhancement in $NN$
invariant-mass distributions -- as depicted {\it e.g.} in Fig.~\ref{fig-NNpi}.

\subsection{\it The Hyperon-Nucleon System}

For the Hyperon-Nucleon system there is naturally much less information
available from experimental side, since hyperons are not stable. For the most
long-lived hyperon, the only weakly decaying $\Lambda(1116)$, some kind of
secondary beam may be obtained from $\Lambda$ production processes in nuclear
targets, but this is still of very limited statistics and precision.

\subsubsection{\it The $\Lambda -N$ system}

Most information about the interaction between $N$ and $\Lambda$ may be
obtained by considering the final-state interaction between both particles in
production processes like $pp \to \Lambda p K^+$. There it shows up as an
enhancement at the low-mass threshold in the $\Lambda N$ invariant-mass
distributions -- similar to the case of the $NN$ FSI. For an example see
Fig.~\ref{fig-TOF}, where the $\Lambda N$ FSI is clearly seen as a strong
enhancement at the $\Lambda p$ threshold. From analyses of the FSI enhancement
and of the limited data on elastic $p \Lambda$ scattering a spin-averaged
scattering length of $a_{\Lambda N} \approx$ -2 fm has been deduced.

From precision measurements of this reaction with polarized beam it has been
possible recently to resolve the spin-dependence of the scattering length and to
extract model-independently the spin-triplet scattering length as $a_t =
-2.3_{-1.39~stat}^{+0.72} \pm 0.6_{syst} \pm 0.3_{theor}$~fm \cite{FH}, which
agrees with $a_t = 
-1.6_{-0.8}^{+1.1}$ fm from $p\Lambda$ elastic scattering \cite{pLambda}. For
the spin-singlet scattering length a value of $a_s = -2.43_{-0.25}^{+0.16}$ fm
has been deduced \cite{hires}. These results are in agreement with model
predictions based on meson exchange \cite{Juelich,Nijmegen} and chiral
effective field theory \cite{HaidenbauerLambda}.

The experimental results mean that the $\Lambda N$ interaction is attractive,
but substantially less attractive than the $NN$ interaction, so that no
boundstate can be formed, {\it i.e.}, in particular there is no "strange"
deuteron. 

\subsubsection{\it The $\Sigma -N$ system}

The $\Sigma N$ interaction can be studied by hyperon-production reactions of
the type $pp \to \Sigma^0 p K^+$ and $pp \to \Sigma^+ p K^0$. Both reactions
have been studied in exclusive and kinematically complete measurements at
COSY-TOF \cite{DD,DD1}. No sign of a significant $\Sigma p$ FSI has been
observed. This means that the $\Sigma N$ interaction must be substantially
weaker than the $\Lambda N$ interaction. 

\subsection{\it The $\Lambda - \Lambda$ System}

In heavy-ion collisions $\Lambda$ particles are produced abundantly, so that
also the $\Lambda \Lambda$ FSI can be looked at. In measurements of Au-Au
collisions with the STAR detector at RHIC the $\Lambda\Lambda$ correlation
function has been obtained \cite{STAR}. Whereas in a first analysis a slightly
repulsive interaction has been deduced, an improved analysis obtains a
scattering length of $-1.2 fm  < a_{\Lambda\Lambda} < -0.5 fm$ \cite{Morita},
which still would exclude a boundstate in the $\Lambda\Lambda$
system. However, if in this analysis so-called 
feed-down corrections are included, the limits relax to $a_{\Lambda\Lambda} >$
-1.2 fm, which does not completely rule out a boundstate -- see the discussion
about the $H$ dibaryon in sect.~7.

\subsection{\it Summary}

Summarizing, in the baryon-baryon ($BB$) system we have  so far clear-cut
experimental evidence only for a single boundstate, which is the deuteron
groundstate known since 1932. In particular, there is no boundstate in the
hyperon-nucleon system with strangeness $S$ = -1. In the strangeness $S$ = -2
sector the existence of a possible boundstate, the $H$ dibaryon, has not yet
been ruled out completely at present.

\section{The Early Days of Dibaryon Searches}
\label{sec-early}
The question, whether there are more eigenstates in the system of two baryons
than just the $^3S_1$ deuteron groundstate and the virtual $^1S_0$ state,
has been around in principle since the discovery of the deuteron in
1932. It became clear pretty soon that the loosely bound deuteron would not
have any excited bound states. Hence the search for resonances focused soon on
nucleon-nucleon collision energies close to the pion-production threshold and
above, where new degrees of freedom come into play. Reports on such
measurements \cite{Neganov,Neganov1,Mesh} date back to the
fifties, when suitable particle accelerators became available. In particular
measurements of the reactions $pp \rightleftharpoons d\pi^+$ indicated
a resonant behavior with $I(J^P) = 1(2^+)$ right at the $\Delta(1232) N$
threshold with a width identical to that of the $\Delta(1232)$ resonance. We
will come back to a detailed discussion of this dibaryon candidate in
sect.~4.4.

\subsection{\it The prediction of Oakes }

Already before the publication of the quark model by Gell-Mann \cite{GellMann}
Oakes considered the two-baryon system in a multiplet representation of
$SU(3)$ following the
so-called eightfold way  \cite{Oakes}. This led to the prediction of 10
baryon-baryon states with the following hypercharge (Y) and isospin (I)
combinations:
\begin{itemize}
\item 1 state with (Y,I) = (2,0) containing the baryon-baryon configuration ($np$),
\item 2 states with (Y,I) = (1,1/2) containing the baryon-baryon configurations
  ($\Lambda N, \Sigma N$),
\item 3 states with(Y,I) = (0,1) containing the baryon-baryon configurations
  ($\Sigma\Lambda, \Sigma\Sigma, \Xi N$) and
\item 4 states with (Y,I) = (-1, 3/2) containing the baryon-baryon
  configurations ($\Xi\Sigma$).
\end{itemize}

In the limit of exact unitary symmetry, in which the masses of the eight
baryons $N, \Lambda, \Sigma$ and $\Xi$ are degenerate, also the baryon-baryon
forces are the same and hence the masses of these baryon-baryon states are
degenerate. However, since this symmetry is broken, their masses are
different, but may be calculated by use of a mass formula derived from this
symmetry breaking.

Oakes identified the single state with a $np$ configuration with the
well-known deuteron. Since this has spin-parity $J^P = 1^+$, also the other
nine states should have the same quantum numbers. 

By use of the Gell-Mann-Okubo \cite{GellMann1,Okubo} mass formula and by
identifying the (Y,I) = (1,1/2) doublet with the $\Sigma N$ cusp at 2.13 GeV,
which was already known at that time --– see detailed discussions in sections
4.2.5 and 8.2 --– he predicted the (Y,I) 
= (0,1) triplet and (-1, 3/2) quadruplet states to be at 2357 and 2564 MeV,
respectively. Since these masses are substantially larger than the masses of
their constituents, he concluded that these states are far from being bound
states, {\it i.e.}, could only be resonance states.

\subsection{\it The Prediction of Dyson and Xuong}

The quest for dibaryon states was reinforced, when it got apparent that
baryons and mesons contain substructures, the quarks, and QCD does not
prohibit colorless multiples of three quarks, in particular does not forbid
quarks in a colorless six-pack. In fact, shortly after Gell-Mann's famous
publication \cite{GellMann} of the quark model in 1964 Dyson and Xuong
\cite{Dyson} demonstrated that SU(6) symmetry provides a multiplet of six
non-strange dibaryon states denoted by $D_{IJ}$ with $IJ = 01,10,12,21,03$ and
$30$ as given in Table 1.

The first three states of their sextet Dyson and Xuong identified as the
deuteron groundstate, the virtual $^1S_0$ isovector state -- known at that time
already from the final-state interaction in low-energy nucleon-nucleon
scattering -- and an $I(J^p) = 1(2^+)$ state right at the $\Delta N$ threshold,
for which first experimental indications had been available already at that
time -- see discussion in Ref. \cite{Dyson}.  
This way they  fixed all parameters of
their mass formula allowing the prediction of masses for the remaining three
higher-lying states. 
By identifying the first two states with the deuteron groundstate and
the $^1S_0$ virtual state with both having roughly the same mass the predicted
dibaryon masses become isospin independent, {\it i.e.} $m(D_{21}) = m(D_{12})$
and $m(D_{30}) = m(D_{03})$. 

Therefore $D_{21}$ is expected to be situated similar to $D_{12}$ near the
$\Delta N$ threshold with a mass around 2160 MeV. Due to its isospin $I = 2$
this state is decoupled from the $NN$ system and hence can be produced only
associatedly in $NN$ collisions, {\it e.g.}, in two-pion production via the
process $pp \to D_{21}^{+++}\pi^- \to pp\pi^+\pi^-$. For a more detailed
discussion about this proposed state see sect. 11.2.

According to the mass formula the next higher-lying dibaryon state, $D_{03}$, is expected
to be at 2350 MeV, {\it i.e.} about 110 MeV below the $\Delta\Delta$
threshold. Due to its quantum numbers it should asymptotically -- {\it i.e.} at
large distances as an intermediate step in its decay or formation -- 
conform to an isoscalar $\Delta\Delta$ configuration with $J^P = 3^+$. Most
favorably this means two spin aligned $\Delta$ particles in relative
$S$-wave. Since this 
state is isoscalar, it should couple to the $np$ system and be sensed in its
$^3D_3$ partial wave. And because the $\Delta$ excitation is highly inelastic
due to 
its preferred decay into $N\pi$, the $D_{03}$ excitation will be highly
inelastic, too. Hence it appears to be more favorable to sense this state in the
process $np \to D_{03} \to NN\pi\pi$ rather than searching in $np$
scattering. In fact, such a state was observed experimentally just very
recently. And indeed, it was first found in the 
two-pion production reaction $np \to d\pi^0\pi^0$, before it was confirmed in
$np$ scattering -- see sections 9 and 10.

Finally, the last state, $D_{30}$, predicted by Dyson and Xuong with quantum
numbers just mirrored to $D_{03}$ constitutes a really exotic one. Its
highest charge state $D_{30}^{++++}$ consists of six up quarks, which
asymptotically yield a $\Delta^{++}\Delta^{++}$ configuration. Due to
its large isospin this state needs to be produced associatedly in $NN$
collisions with at least two other pions, {\it e.g} in a four-pion production
process of the form $pp \to D_{30}^{++++}\pi^-\pi^- \to
\Delta^{++}\Delta^{++}\pi^-\pi^- \to pp\pi^+\pi^+\pi^-\pi^-$. For a search for
this state see sect. 11.1.

\begin{table}
\begin{center}
\begin{minipage}[t]{16.5 cm}
\caption{Prediction of Dyson and Xuong \cite{Dyson} about a sextet of
  non-strange dibaryon states based on $SU(6)$ symmetry. The states
  are denoted by $D_{IJ}$, where $I$ denotes the isospin and $J$ the total
  spin of the state. Given are the associated asymptotic baryon-baryon ($BB$)
  configurations and the masses calculated from symmetry breaking and by
  identifying the two lowest-lying states with the deuteron groundstate and
  the virtual $^1S_0$ state.}  
\label{tab:dyson}
\end{minipage}
\begin{tabular}{rrrlll}
\\ 
\hline

notation&I&J&asymptotic baryon-baryon&mass (formula)&mass (value)\\ 
&&&~~~~~configuration&&(MeV)\\

\hline

$D_{01}$&0&1&deuteron&A&1876\\
$D_{10}$&1&0&$^1S_0$ $NN$ virtual state&A&1876\\
$D_{12}$&1&2&$\Delta N$&A + 6B &2160\\
$D_{21}$&2&1&$\Delta N$&A + 6B&2160\\
$D_{03}$&0&3&$\Delta\Delta$&A + 10B&2350\\
$D_{30}$&3&0&$\Delta\Delta$&A + 10B&2350\\

\hline
 \end{tabular}\\
\end{center}
\end{table}

\section{The Dibaryon Rush Era}
\label{sec-3}
   As we have seen, the first three states in Dyson's dibaryon sextet
   constitute quite conventional states, where the two bags containing three
   quarks each do not overlap markedly. Maybe it was this fact that this work
   did not find overwhelming attention 
   -- though as we will see below it turns out now to have quite some
   predictive power in view of the first observation of a non-trivial dibaryon
   resonance. 

A real dibaryon rush started, when Jaffe 1977 \cite{Jaffe} predicted the
so-called $H$ dibaryon, a hadronically bound $\Lambda\Lambda$ system containing
two strange 
quarks. This initiated a flood of further predictions on a multitude of
states in all kind of baryon-baryon systems, which in turn initiated 
world-wide experimental searches for bound and unbound dibaryon states. As a
consequence a huge number of claims have been made. However, finally not a
single one survived a critical inspection. 

A major reason for this striking failure
was certainly the insufficient quality of experimental data obtained by use of
inadequate instrumental equipment. Other reasons may concern the wrong choice
of reaction and/or energy region, where searches have been conducted. 
For reviews on this epoch see
Refs. \cite{Yokosawa,Locher,igor,Tatischeffreview,Komarov}. 
For a
critical, but also very amusing review of this epoch see, {\it e.g.} K. K. Seth
\cite{Seth,Seth1}. For a more recent review from the theoretical point of view
see also A. Gal \cite{Gal}.  
 
\subsection{\it Theoretical Predictions}

Triggered by Jaffe's prediction  \cite{Jaffe} of a bound $\Lambda\Lambda$
system many other theoretical predictions appeared in the following years
based on QCD-inspired models like bag, potential, string or flux-tube models
for the six-quark system 
\cite{Mulders,Aerts,Mulders1,Mulders2,MuldersThomas,Saito,Bickerstaff,AertsDover,Aerts1,Aerts2,Lomon,Lomon1,Lomon2,Schepkin,Schepkin1,Schepkin2,Oka,Oka1,Oka1a,Maltman,Maltman1,GoldmanS=3,Goldman,Goldman1,Barnes},
models of just the hadronic baryon-baryon interaction without any explicit
quark-degrees of freedom
\cite{Kamae,SatoSaito,Garcilazo,Garcilazo1,Gale,Kalbermann,Ueda,Dillig} or just
symmetry 
considerations based on SU(3) \cite{Xie}. Most recently also lattice
QCD calculations \cite{NPLQCD,NPLQCD1,HALQCD,Inoue} joined this subject -- see
discussion in section 7. 

In general, the QCD-based or -inspired model calculations make use of the
color-magnetic interaction between quarks giving rise to hyperfine splitting:

\begin{equation}
v_{color-mag} = - \sum \limits_{i<j} (\lambda_i \cdot \lambda_j) (\sigma_i
\cdot \sigma_j) v(r_{ij}),
\end{equation} 

where $\lambda_i$ and $\sigma_i$ denote color and spin operators,
respectively,  of the quark i and $v(r_{ij})$ is a flavor conserving
short-range interaction between the quarks i and j \cite{Gal}.

As pointed out by Jaffe and others, the color-magnetic interaction is most
attractive in the flavor-singlet state with $I(J^P) = 0(0^+)$ and $S$ = -2,
the $H$ dibaryon, the wavefunction of which corresponds asymptotically to
baryon-baryon 
configurations of $\Lambda\Lambda$, $\Xi N$ and $\Sigma\Sigma$ \cite{Oka1a}.

In this concept the leading dibaryon candidates having the underlying baryons
 in relative $S$ wave are states with \cite{Gal,Oka1a}:
\begin{itemize}
\item $S$ = 0, $I(J^P) = 0(3^+)$ and $BB$ structure $\Delta\Delta$,
\item $S$ = -1, $I(J^P) = 1/2(2^+)$ and $BB$ structure $\Sigma^* N$ and
  $\Sigma\Delta$, 
\item $S$ = -2, $I(J^P) = 0(0^+)$ with $BB$ structure $\Lambda\Lambda$, $\Xi N$
and $\Sigma\Sigma$ (the $H$ dibaryon) and
\item $S$ = -3, $I(J^P) = 1/2(2^+)$ with $BB$ structure $\Omega N$, $\Xi^*
  \Sigma$, $\Xi^*\Lambda$ and $\Xi \Sigma^*$,
\end{itemize} 
where $BB$ means the baryon-baryon configuration, which is asymptotically
closest to the dibaryon state. 

The results based on eq. (1) assume unbroken $SU(3)$ symmetry. However, realistic
considerations, which account for symmetry breaking, lead to significant
changes in the predicted dibaryon masses.

With respect to a possible experimental observation of dibaryon resonances,
predictions of states with low masses are particular interesting because of
their expected narrow width. 

The Nijmegen group \cite{Mulders,Aerts,Mulders1,Mulders2}, well-known for
their partial-wave analyses of $NN$ scattering and 
the therefrom derived Nijmegen potential, predicted a multitude of dibaryon
resonances both in non-strange and strange sectors. They introduced an
elongated bag allowing for finite orbital angular momenta between delocalized
quark clusters in the partitions $q^2 - q^4$ and $q - q^5$. As a consequence
they obtained a multitude of dibaryon states both for non-strange and strange
sectors. In particular, they predicted that the lowest-lying states with
masses around 2.11 GeV would be $NN$-decoupled due to their quantum
numbers. Because of that they were expected to have particular narrow
widths. For a discussion of those states see section 5.

Most calculations agree insofar as they predict no dibaryons with masses below
the $NN\pi$ threshold thus allowing them to decay still hadronically. The
possible existence of an isotensor $NN\pi$ bound state is based on the hope
that the
attractive $\pi N$ interaction in the $P_{33}$ partial wave ($\Delta$ channel)
might provide  enough attraction to bind the barely unbound isovector $NN$
system --
constituting thus a deeply bound $\Delta N$ system. Whereas early calculations 
for this scenario \cite{Mulders,Gale,Kalbermann,Ueda} did not support the existence of
such states,  Garcilazo \cite{Garcilazo} did find theoretical evidence for a
$nn\pi^-$ boundstate in three-body Faddeev calculations, if the
pion-nucleon interaction is sufficiently short-ranged, which, however, could
not be decided given the uncertainty in our knowledge of the short-range
part. By use of a phenomenological $\Delta$ model Dillig \cite{Dillig}
predicted an $I(J^P)=2(1^+)$ state at 2020 MeV, {\it i.e.} just barely unbound
with a width of $\Gamma$ = 1 MeV.   

The Los Alamos theory group predicted dibaryons in the $\Omega N$ system, which
could be so deeply bound that they would be stable with respect to strong
decay \cite{GoldmanS=3}. They also showed that predictions about the binding
energy of the $H$ dibaryon critically depend on the detailed dynamics of the
model under consideration. They rather emphasized the particular importance of
the "inevitable" dibaryon, as they called it \cite{Goldman} and  
argued that certain basic 
features common to all models based on one-gluon exchange and confinement lead
unavoidably to the prediction of a non-strange dibaryon resonance $d^*$ with
$I(J^P)=0(3^+)$ due to its special symmetry. Being asymptotically a bound
$\Delta\Delta$ state it coincides with the $D_{03}$ state predicted by Dyson
and Xuong. However, in the Los Alamos calculations it appears to be very
deeply bound by nearly 400 MeV. In contrast, the MIT and cloudy bag model
calculations \cite{Aerts,MuldersThomas,Saito} obtained for it binding energies
relative to the $\Delta\Delta$ threshold of about 100 MeV, {\it i.e.} close
to the value predicted before by Dyson and Xuong.

Mulders and Thomas \cite{MuldersThomas} as well as Saito \cite{Saito}
demonstrated that pionic corrections ("pion cloud") do not have large
influence on the mass of the "inevitable dibaryon", but large impact on the
mass of the $H$ dibaryon pushing up its mass, possibly even into the
unbound region. Oka, Shimizu and Yazaki showed that also in the non-relativistic
 quark cluster model the $H$ dibaryon gets unbound \cite{Oka1}. Aerts and
 Dover \cite{AertsDover} proposed the double-strangeness-exchange reaction as
 particularly suited for the $H$ search and calculated cross sections expected
 for the $K^-$ $^3$He$\to H n K^+$ reaction. 

\subsection{\it Experimental Searches for Narrow Dibaryons}

The prediction of a copious number of dibaryon states in strange and
non-strange sectors initiated a rush of experimental searches for such
states. Naturally, the search for narrow dibaryons was particularly attractive
for two reasons. 

First, a narrow resonance structure is much easier to
discriminate experimentally against the background from conventional hadronic
processes. The latter will dominate the considered reaction in
general and will show also resonance-like structures due to hadronic excitations of
baryons and mesons during the course of the reaction process. Since the widths of
typical mesonic and baryonic resonances is about 120 MeV and higher,
observation of much narrower structures points to a very interesting exotic
process. 

Second, in order to have a small width, the decay of a dibaryon must be
hindered either 
\begin{itemize}
\item due to its exotic internal quark structure, which provides no large
  overlap with the asymptotic hadronic configuration of its decay products, and / or
\item due to its quantum numbers, which hinder or even forbid the decay into
  lower-lying hadronic channels, and / or
\item due to its mass being below some elsewise favored particle emission
  thresholds  
  or at least nearby.
\end{itemize}

{\it E.g.}, if the mass of the $H$ dibaryon is below the $\Lambda\Lambda$ mass,
it can decay only by weak interaction and hence has a very narrow
width. However, if its mass is above this threshold, 
it may decay by strong interaction in a fall-apart decay into $\Lambda\Lambda$
without any hindrance and hence have a very large width typical for hadronic $S$-wave decays. 

Similarly, the non-strange $NN$-decoupled states $I(J^P)=0(0^-)$ and $0(2^-)$
predicted by Mulders {\it et al.} to lie above the $NN\pi$ threshold, should
still have a fairly narrow width, since their decay into the $NN$ channel is
forbidden due to their quantum numbers. In the $NN$ system such quantum
number combinations are not allowed because of the Pauli principle.  

\subsubsection{\it Search in $NN$ and $\pi d$ scattering}

From the experimental point of view the easiest channels for the dibaryon
search to access with
high precision are elastic proton-proton and pion-deuteron scattering. The
latter got possible with the availability of high-quality and
high-intensity $\pi^+$ and $\pi^-$ beams of kinetic energies up to $T_\pi$ =
500 MeV at the pion factories LAMPF, TRIUMF and PSI (formerly SIN), partly
in combination with polarized deuteron targets. 

These installations also allowed precise measurements of $pp$ scattering up to
$T_p$ = 800 MeV with polarized beam and/or target. Still higher beam energies
have been reached, {\it e.g.}, in Gatchina, Dubna, Argonne National Lab, KEK
and Saclay.  

In order to access the $np$ system experimentally one has to either produce
explicitly a secondary neutron beam or utilize the quasifree reaction process
$p +d \to (p + n) + p_{spectator}$. In both cases deuterons are generally used
as the provider of neutrons. The quasifree process has proven to work very
well at beam energies of several hundred MeV and above. If the spectator
proton is also measured in a kinematically complete experiment, then the
Fermi motion within the deuteron can be utilized to measure the energy
dependence of the observables over quite some range of energies simultaneously.

By the mid-nineties many thousands of experimental data points for differential
and total cross sections as well as for polarization observables were obtained
both for $pp$, $np$ and $\pi d$ scattering. For $pp$ scattering they covered
the beam energy region up to 1.6 GeV (corresponding to a total center-of-mass
energy of $\sqrt s$ = 2.56 GeV), for $np$ scattering up to 1.3 GeV ($\sqrt s$
= 2.44 GeV) and for $\pi d$ scattering up to 500 MeV ($\sqrt s$ = 2.44 GeV). 

These data were collected by centers conducting partial-wave analyses of these
data, see, {\it e.g.} Refs. \cite{Geneva,Nijm93,SM94,SM94pid} - the most prominent
being the SAID data analysis center at Washington, DC. All partial-wave
analyses give comparable results, in particular they give no evidence for
resonances with the exception of the energy region around the $\Delta N$
threshold, which will be discussed in detail in section 4.4.

For a review of odd structures in $NN$ scattering observables and their
possible relation to dibaryons see, {\it e.g.} the review by Yokosawa
\cite{Yokosawa} about the experimental polarized beam program at
Argonne. 

There also two anomalies have been observed in the difference $\Delta\sigma_L$
between $pp$
total cross sections for pure helicity states \cite{ANL} at $\sqrt s$ = 2.7 -
2.9 GeV, {\it i.e.} in the mass range, where there have been predictions for
dibaryon states by LaFrance and Lomon \cite{Lomon,Lomon1,Lomon2}. At the lower
energy also some narrow structure in the spin correlation observable
$A_{00nn}$ was reported from measurements at Saclay \cite{Ball0}, which would
correspond to 
a resonance mass of 2735 MeV and an estimated width of 17 MeV -- in agreement
with the predictions of Refs. \cite{Lomon,Lomon1,Lomon2} for a $I(J^P)=1(0^+)$
state based on cloudy bag and R-matrix calculations.

\subsubsection{\it Search in pion production}

Single-pion production has been the first reaction process to look for
dibaryons, since it was hoped that exotic processes would be sensed
particularly well, if pionic degrees of freedom were explicitely
involved. Indeed, as already mentioned, the $pp \to d\pi^+$ reaction was used
already in the fifties to search for signals from dibaryons
\cite{Neganov,Mesh}. With the availability of pion beams at the pion factories
the reverse reaction $\pi^+ d \to pp$ got easily accessible and hence a wealth
of data on this reaction up to $T_\pi$ = 500 MeV ($\sqrt s$ = 2.44 GeV) stem
from measurements at LAMPF, TRIUMF and PSI (formerly SIN).  For higher
energies measurements of $pp \to d\pi^+$ reaction had to be conducted at
appropriate proton accelerators. 

The data for pion energies up to $T_\pi$ = 500 MeV have been accumulated in 
the SAID database amounting to several thousands of experimental data for
total and differential cross sections as well as polarization observables. The
partial-wave analysis \cite{SAIDpidtopp} of these data does not give any hints
for narrow resonances. However, in the $\Delta N$ threshold region the
analysis exhibits pronounced loopings of particular partial waves in
the Argand diagram -- which usually is a clear signature of resonance
phenomenon. For a detailed discussion on that see section 4.4.

Experimentally more demanding is the study of pion production in reactions,
where three or more particles are emitted. In such cases single-arm
experiments, as usually used in two-body reactions, are no longer sufficient
to record a reaction in an exclusive and kinematically complete way. Hence  
experimental setups with a single detector arm, {\it e.g.}, a magnetic
spectrometer, can measure such reactions only inclusively with the consequence
of insufficient information for a proper analysis of the data, which in
addition contain a much higher amount of background. To the contrary,
bubble-chamber measurements may record all emitted particles, if they are
charged -- but they usually have the problem of low statistics.

Hence it is of no surprise that a number of dibaryon claims have been made
resting upon the observation of some intriguing structures in the observables,
which, however, could not be verified by other dedicated measurements.

\begin{figure} 
\centering
\includegraphics[width=8.2cm,clip]{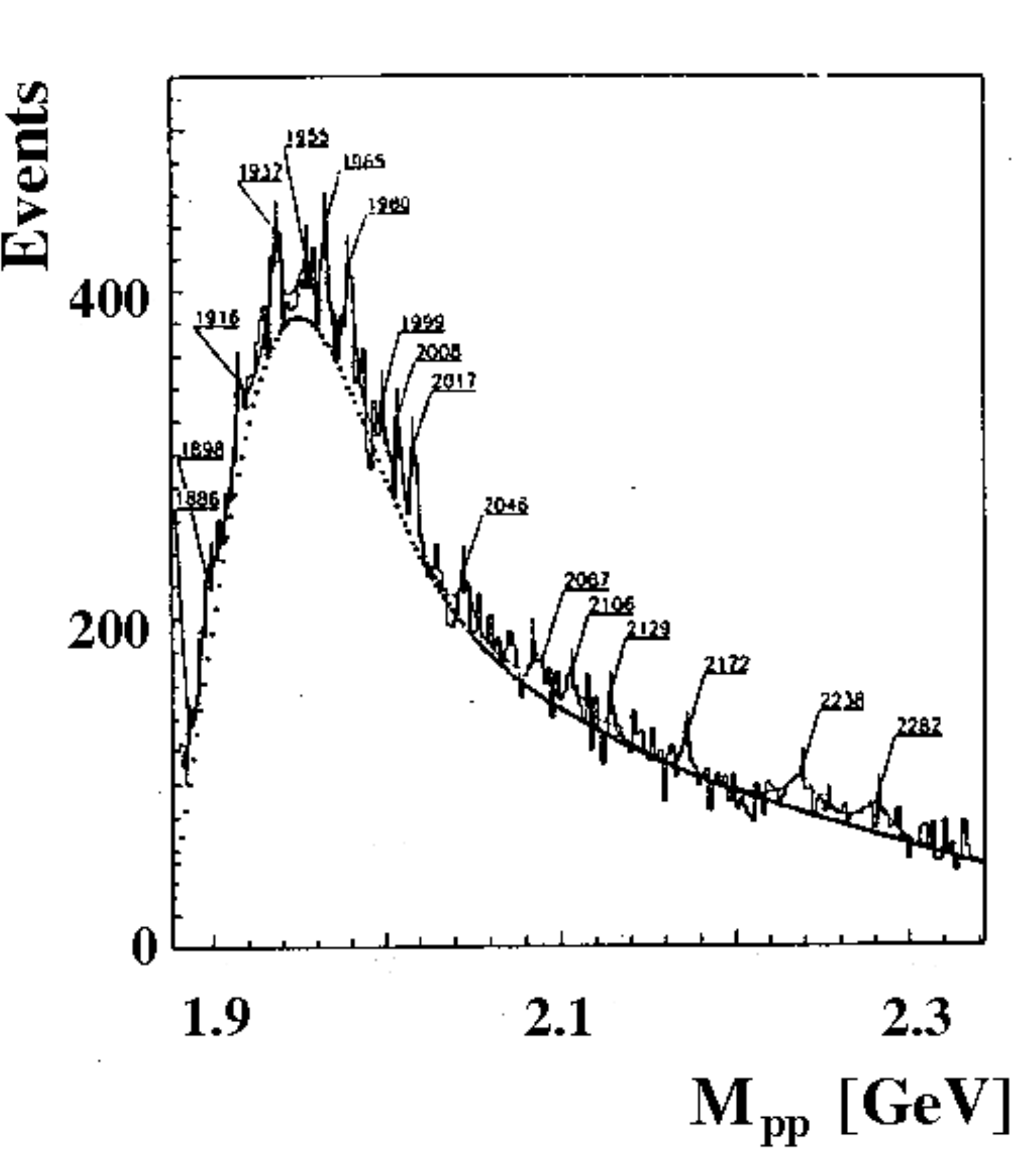}\\
\includegraphics[width=8.2cm,clip]{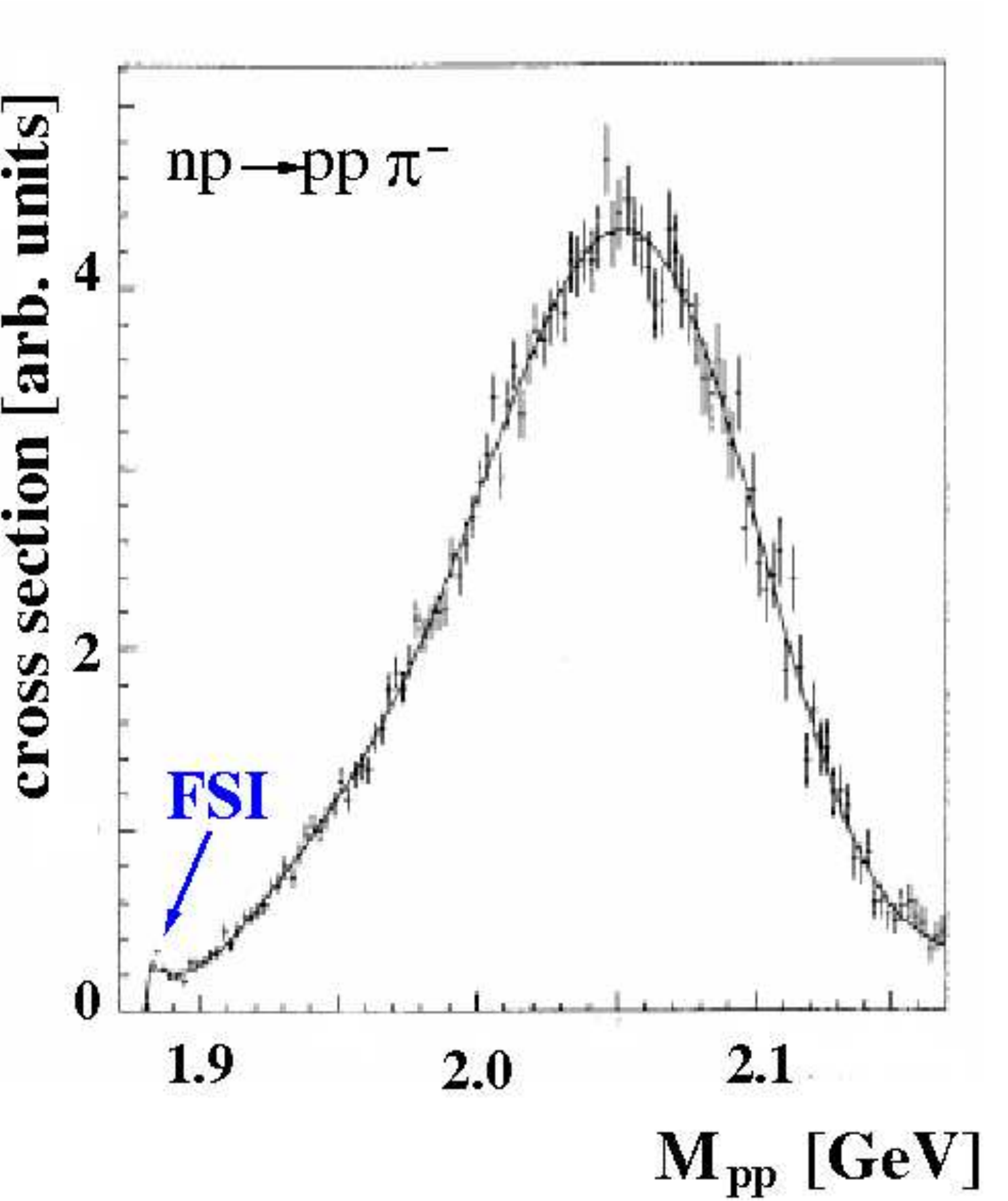}\\
\caption{
Distribution of the $pp$-invariant mass $M_{pp}$ as observed in $np$-initiated
single-pion production by the JINR bubble-chamber at Dubna \cite{Troyan} (top)
and the magnetic spectrometer setup at ITEP \cite{Kulikov} (bottom),
respectively. Whereas in the Dubna spectrum , which includes also events from
multi-pion production, 17 narrow spikes above an assumed smooth background
(dotted line) have been assigned to narrow dibaryon resonances, the ITEP
spectrum shows no statistically significant narrow structures except of the
well-known enhancement at the $pp$ threshold due to the $NN$ FSI. From
Refs. \cite{Troyan,Kulikov}. 
}
\label{fig-NNpi}       
\end{figure}

{\it E.g.}, Troyan and Pechenov found 17 narrow, only few MeV broad structures 
in a $pp$ invariant-mass spectrum, which -- in order to increase statistics --
they constructed out of many separate
bubble-chamber measurements of the reactions $np \to pp\pi^-$, $np \to
pp\pi^-\pi^0$, $np \to pp\pi^0\pi^-$, $np \to pp\pi^+\pi^-\pi^-$ and   $np \to
pp\pi^+\pi^-\pi^-\pi^0$ at neutron beam momenta between 1.26 - 5.24 GeV/c
\cite{Troyan} -- see Fig.~\ref{fig-NNpi}. They associated these structures
with  17 narrow dibaryons with
masses in the range 1886 - 2282 MeV, though their statistical
significance fully depended on the choice of conventional background assumed
in the analysis. In fact, a subsequent exclusive high-resolution and
high-statistics measurement of the $np \to pp\pi$ reaction at ITEP
\cite{Kulikov} revealed that there were no narrow structures in the mass range
up to 2170 MeV except of the threshold structure due to the $pp$ final-state
interaction caused by the well-known virtual $^1S_0$ state, see section 2.3.

The same bubble-chamber setup at Dubna has also been used to search for narrow
$I$ = 2 dibaryon resonances in the reaction $np \to pp\pi^+\pi^-\pi^-$. In the
obtained $pp\pi^+$-invariant mass distribution eight narrow structures have
been identified and claimed to belong to narrow $I$ = 2 dibaryon resonances
\cite{Troyan1}.  

Another claim for dibaryons in this mass range comes from inclusive, though
high-statistics single-arm magnetic spectrometer measurements at
Saclay \cite{Tatischeff}. Investigating the reactions of type $pp \to p\pi^+X$
at beam energies of $T_p$ = 1.5 - 2.1 GeV three small structures were
identified and associated with narrow dibaryons of masses between 
2050 - 2150 MeV. In yet another experiment at Saclay a small oscillatory
pattern has been observed in the tensor analyzing power and associated with a
narrow dibaryon at 1945 MeV \cite{Tatischeff1}. 

Pion-production in nuclei also has been investigated with respect to possible
dibaryon signals. Remarkably, several inclusive experiments conducted at Dubna
\cite{Krasnov}, Saclay \cite{Julien} and Moscow \cite{Aseev} noted a narrow,
only a few MeV broad resonance structure at a mass of about 2220 MeV in
proton initiated subthreshold $\pi^+$ production in nuclei, whereas this
structure was not observed at TRIUMF in $\pi^+$ and $\pi^-$ production
\cite{Yen} as well as in $\pi^0$ production at Saclay \cite{Julien1}. The
non-observation in the latter case could have been due to the insufficient
energy resolution in that experiment. At first this structure was associated
with the $^3F_3$ dibaryon resonance candidate formed by in-medium $pp$
collisions, but it was soon realized that Fermi-motion would have necessarily
smeared out such a narrow resonance as observed. For other hypotheses see,
{\it e.g.} the discussion in Ref. \cite{Julien1}.
Kurepin and Oganessian put the idea forward that a $I(J^P)=1(2^+)$
$\Delta\Delta$ excitation shifted down in energy due to the nuclear medium might
be the reason for this narrow structure \cite{Kurepin}.

A LAMPF experiment looking for very low-lying dibaryons
in the $\pi^+ d \to pp$ reaction found no evidence for such states \cite{Pasyuk}.

\subsubsection{\it Search in photo-induced reactions}

Already as early as in the sixties it was noted that the proton polarization
in deuteron photo-disintegration $\gamma d \to \vec p n$ shows a sharp increase
-- in particular at angles around 90$^\circ$, if the incident photon energy is
increased beyond 350 MeV \cite{Liu,Kose}. A model assuming  $\Delta$
excitation of one of nucleons in the struck deuteron could accommodate the
data taken at lower energies, but not those taken at 400 and 500 MeV. 

In the
seventies measurements of this particular reaction were extended up to 700 MeV
photon energy using Bremsstrahlung photons produced by electrons accelerated
in the synchrotron at Tokyo. As a result it was observed that the polarization
at 90$^\circ$ starts rising at 300 MeV, reaches a maximum around 500 MeV and
starts dropping above 550 MeV describing thus a resonance-like excitation
function \cite{Kamae1}. It was suggested that this behavior could be due to the excitation
of a dibaryon resonance with a mass of 2380 MeV and preferably with quantum
numbers $I(J^P)=0(3^+)$ as predicted in 1964 by Dyson and Xuong \cite{Dyson}
and -- in close connection with the Bremsstrahlung measurements -- by Kamae and
Fujita \cite{Kamae} based on a non-relativistic one-boson exchange potential
model. Subsequent measurements improving the data base allowed for resonance
fitting, which led to the assignment of several broad resonances with widths of
200 MeV and beyond \cite{Kamae2,IkedaKamae,IkedaKamae1}.

Another intriguing structure in inclusive photo-induced pion production on the
deuteron was reported from Saclay measurements, where a bumpy structure was
observed in the energy dependence of the cross section around $E_\gamma$ = 400
MeV, which was associated with a dibaryon of mass 2230 MeV
\cite{Argan}. Subsequent exclusive high-resolution measurements of the $\gamma
d \to pp\pi^-$ reaction carried out at Bonn demonstrated that the $pp\pi^-$
invariant mass distribution is flat showing no evidence for any dibaryon in
the mass range 2160 - 2320 MeV \cite{Ruhm} -- in agreement with corresponding
pion-induced measurements carried out at Saclay \cite{Tamas} and SIN
\cite{Arvieux}. However, the Bonn data surprisingly revealed a small narrow
structure in the $pp$ invariant-mass spectrum at 2014 MeV, which coincides
right with the $pp\pi^-$ threshold \cite{Bock}.

In CLAS measurements of the $\gamma d \to d \pi^0$ reaction at JLAB a
resonance-like structure was observed at backward angles at $E_\gamma$ = 700
MeV, which could be explained by $\eta$ excitation in the intermediate state
connected to the $N^*(1535)$ baryon resonance \cite{Ilieva}. 

Another search by using the $\gamma d \to \pi^0 X$ reaction
has been carried out at MAMI achieving an energy resolution of 0.8 MeV. Again
no statistically significant narrow structure has been found, which could
signal an isoscalar or isovector dibaryon resonance in the mass range below
2100 MeV \cite{Siodlaczek} -- see also section 5.

\subsubsection{\it Search for super-narrow dibaryon resonances below the $NN\pi$
  threshold}

For dibaryons below the pion emission threshold the only hadronic decay channel
is $NN$. However, if the quantum numbers of the dibaryon state are Pauli
forbidden in the $NN$ system, {\it i.e.} if they are of the type $I(J^P) = 0(0^+),
0(0^-), 0(2^-), 0(4^-), ...$ or $I(J^P)=1(1^+),1(3^+), 1(5^+), ...$ or even of
type $I(J^P,s) = 1(1^-,0)$, where $s$ denotes the internal spin, then such
dibaryons can not decay into the $NN$ system -- only electromagnetically by $\gamma$ emission. Therefore the decay widths of such dibaryons should be tiny.
They have been estimated for some examples by Fil'kov
\cite{Filkov,Filkov1}. In an inclusive measurement of the $pd \to
pX$ reaction at the Moscow Meson Facility three narrow peaks have been observed
and associated with supernarrow dibaryons of masses 1904, 1926 and 1942 MeV
\cite{Filkov2,Filkov3}. Support for those lines has been found in an
analysis of LEGS data on the $\overrightarrow{\gamma}d \to nn\pi^+
\gamma$ reaction \cite{cichocki}. However, the mass range 1896 - 1914 MeV was
excluded 
later-on by  inclusive measurements of the $pd \to pdX$ and $pd \to ppX$
reactions at Osaka \cite{Tami,Tami1}.

Similarly Gerasimov and Khrykin proposed Bremsstrahlung connected production
of such dibaryons, in particular favoring the $pp \to pp\gamma\gamma$ reaction
\cite{Gerasimov}. Indeed some bumpy structure corresponding to a dibaryon mass
of about 1920 MeV was recorded in a subsequent inclusive experiment at Dubna
having a 
setup composed of just two photon detectors left and right of the beam-pipe
\cite{Khrykin,Khrykin1}. A follow-up measurement claimed a signal
corresponding to a dibaryon mass shifted to 1956 MeV \cite{Khrykin2,Khrykin3}.
However, an exclusive and kinematically complete high-resolution experiment
of the $pp \to pp\gamma\gamma$ reaction at CELSIUS, Uppsala utilizing the WASA
detector found no evidence for any dibaryon resonance in the mass range 1900 -
1960 MeV \cite{Jozef}. 

Still more exotic would be bound $pp\pi^+$ and $nn\pi^-$ systems. Such
configurations would be members of an isotensor multiplet and could even not
decay electromagnetically into the $NN$ system -- only by weak interaction.
Of these two charge states the $nn\pi^-$ configuration would be bound even
more  likely, since there is no repulsive Coulomb interaction in this
system. The possibility that these systems may be bound is based on the
assumption that -- similar to the case in nuclei, where $^3$H and $^3$He are
bound more strongly than the deuteron --  the attractive $\pi N$ interaction
in the $\Delta$ channel will cause the barely unbound isovector $NN$ system to
get bound. 
Experimentally this scenario can be studied very well at pion factories by
using $\pi^+$ and $\pi^-$ beams and measuring the pionic double charge exchange
reaction on the deuteron, {\it i.e.} $\pi^+ d \to (pp\pi^+) \pi^-$ and $\pi^-
d \to (nn\pi^-) \pi^+$. By measuring the pion emitted associatedly with the
$NN\pi$ system -- preferably by means of a high-resolution magnetic spectrometer
-- its missing mass spectrum may display a peak outside the kinematic range
for an unbound $NN\pi$ system, if a bound $NN\pi$ system exists. In a series 
of such experiments at LAMPF, PSI and TRIUMF no significant signs of $pp\pi^+$
and $nn\pi^-$ bound states have been found
\cite{Piasetzky,Boer,Lichtenstadt,Ashery,Stanislaus,SethT=2}. For the analysis
of such experiments concerning the possibility to observe narrow dibaryons see
Ref. \cite{Azimov}.

\subsubsection{\it Search in hyperon production and hyperon formation}

Early searches for dibaryons in the strange sector have been carried out in
$K^-d$ interactions at rest, later also with low-energetic $K^-$ beams hitting
a deuteron target \cite{Dahl,Cline,Alexander,Tan,Sims,Eastwood,Braun,Pigot}. The
reaction looked at was $K^- d \to \Lambda p \pi^-$. With the exception of
Ref. \cite{Pigot}, which reports on an inclusive measurement by use of a
magnetic spectrometer, bubble chambers were used for the detection of the
reaction ejectiles. The use of bubble chambers
allowed the detection of all charged ejectiles. In the case that also the
$\Lambda \to p\pi^-$ decay was observed as a V signature, a kinematical
fit with seven overconstraints could be performed, otherwise there was only a
single overconstraint. Due to these overconstraints in the kinematic fits very
good resolutions of 1 - 3 MeV were obtained for $\Lambda p$ invariant masses.

\begin{figure} 
\centering
\includegraphics[width=11cm,clip]{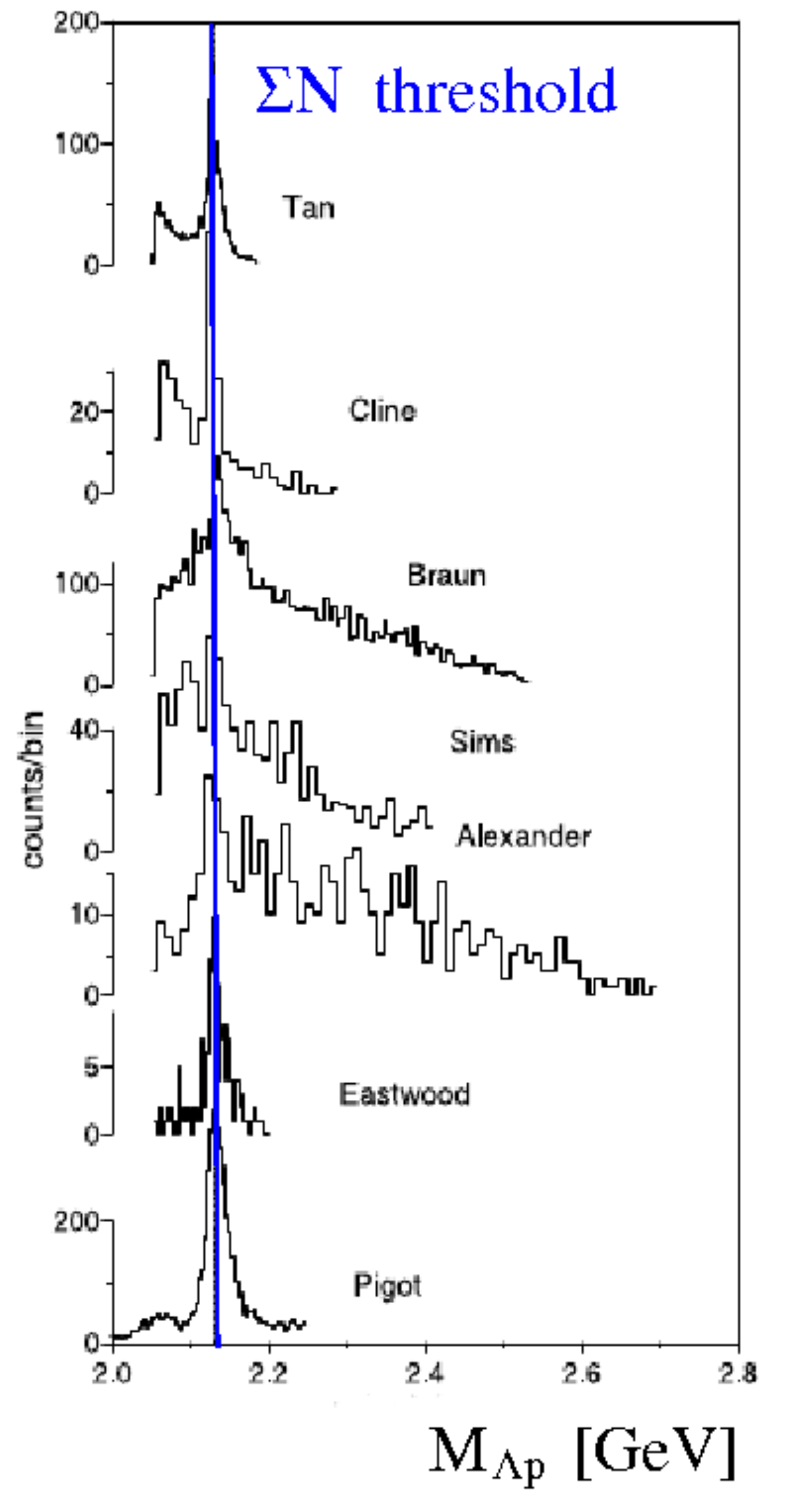}\\
\caption{
Distribution of the $\Lambda p$-invariant mass $M_{\Lambda p}$ as obtained from
the $K^-d \to \Lambda p\pi^-$ reaction studied in
Refs. \cite{Tan,Cline,Braun,Sims,Alexander,Eastwood,Pigot} (from top to
bottom). The vertical line denotes the $\Sigma N$ threshold averaged over the
masses of $\Sigma^+n$ and $\Sigma^0p$.  
From Ref. \cite{Machner}. 
}
\label{fig-SigmaN}       
\end{figure}

All these measurements find a sharp peak in the $\Lambda p$
invariant mass spectrum right at -- or very close to -- the $\Sigma N$
threshold, which in detail consists of two thresholds: the $\Sigma^+ n$
threshold at 2.129 GeV and the $\Sigma^0 p$ threshold at 2.131 GeV. The
observed peak has a width of only few MeV and is skew in shape with a very sharp
rise at its low energy-side and a shallower fall-off towards higher energies
-- see Fig.~\ref{fig-SigmaN}.

In the measurement of Tan \cite{Tan}, which has the best energy resolution of
1 MeV, the high-energy fall-off appears as a kind of shoulder, so that the
full observed structure can be fitted quantitatively by two Breit-Wigner
resonant structures with mass and width $m_1 = 2128.7 \pm 0.2$ MeV, $\Gamma_1
= 7.0 \pm 0.6$ MeV and $m_2 = 2138.8 \pm 0.7$ MeV, $\Gamma_2 = 9.1 \pm 2.4$ MeV.

It is tempting to associate the main structure with a $\Sigma N$ cusp effect
due to the strong coupling between $\Lambda p$ and $\Sigma N$ channels. In
Ref. \cite{Braun} a phenomenological treatment of this scenario has been
sketched by use of asymmetric Flatt$\acute{e}$ \cite{Flatte} distributions,
which assume an 
underlying $S$-wave resonance in the $\Lambda p$ channel. It has been shown
that a satisfying description of the data in the region of the $\Sigma N$
threshold may be obtained, however, without being able to fix mass and width
of the underlying hypothetical resonance in any satisfying manner.

Toker, Gal and Eisenberg \cite{Toker} studied this scenario in the Faddeev
formalism testing various $YN$ interactions. They obtain a good description of
the $\Lambda p$ invariant mass spectrum of Tan \cite{Tan} without invoking a
$\Sigma N$ bound 
state. According to their description the main peak may be regarded as a
genuine three-body cusp, whereas the high-energy shoulder results from the
interference between direct $\Lambda$ production processes and those involving
the intermediate $\Sigma N-\Lambda N$ conversion.
 
Subsequent calculations by Torres, Dalitz and Deloff \cite{Torres} in the
Faddeev formalism, too, come to a similar though in detail somewhat different 
conclusion. According to their result the Tan data imply that there is no
stable bound state -- {\it i.e.} a pole in the second Riemann sheet of the
complex energy plane of $YN$ systems above the $\Sigma N$ threshold -- in the
coupled $\Sigma N - \Lambda N$ system, but the 
data require a virtual state, {\it i.e.} a pole in the fourth sheet near the
$\Sigma N$ threshold. 
Similar conclusions have been reached in Ref. \cite{Badalyan}.
These conclusions support the view that this $^3S_1$
state in the coupled  $\Sigma N - \Lambda N$ system is simply the strange
$S=-1$ analogue of the deuteron as envisaged by Oakes  \cite{Oakes} in 1961,
but with the twist that this deuteron counterpart appears as a virtual state.

With the advance of proton accelerators to deliver high-intensity beams at
energies above the kaon production threshold the $pp \to \Lambda pK^+$ reaction came
into competition to the $K^-d \to \Lambda p \pi^-$ reaction for the
investigation of the $\Sigma N - \Lambda N$ system. After some early
measurements at Brookhaven \cite{Melissinos,Reed} and Princeton \cite{Hogan}
inclusive high-resolution measurements have been carried out at SATURN, Saclay
utilizing the SPES4 magnetic spectrometer for the $K^+$ detection
\cite{Siebert}. The $\Sigma N$ cusp has been again observed in these
measurements,  though it appears only as an
enhancement at the $\Sigma N$ threshold, since in this inclusive experiment
the $\Sigma$ production can not be separated from the $\Lambda$
production. The observed enhancement due to the cusp effect agrees well with
corresponding predictions of Deloff \cite{Deloff} and Laget\cite{Laget}. Aside
from the cusp related enhancement a small, narrow structure with a statistical
significance of 3$\sigma$ has been observed in the Saclay measurements near 2010
MeV. It has
been noted in Ref. \cite{Siebert} that this structure as well as the cusp
effect coincide within 10 MeV to the predictions of Aerts and Dover
\cite{Aerts1,Aerts2} for the mass of $P$-wave the singlet and triplet states
$D_s$ and $D_t$ having a $q^4 - q^2$ quark structure. Also the observed widths
are in accord with the predictions.

In conclusion, whereas it is obviously not unambiguous, whether the cusp
effect at the $\Sigma N$ threshold is correlated with a real or virtual
dibaryon state, the nature of the narrow structure near 2010 MeV can be
clarified by an 
improved measurement with higher resolution and statistics. In fact, such an
improved measurement has been carried out recently at COSY utilizing the BIG
KARL magnetic spectrometer providing a missing mass resolution of 0.84 MeV
\cite{HIRES}. This measurements finds no evidence for a dibaryon state
between 2058 - 2105 MeV yielding very low upper limits in the range of nb/sr.
In the same set of measurements also the cusp related enhancement at the
$\Sigma N$ threshold has been confirmed \cite{HIRES1}.  

In a recent review by Machner {\it et al.} \cite{Machner} all measurements in
the $\Sigma N$ cusp region -- including the more recent exclusive and
kinematical complete COSY-TOF measurements, which will be discussed in section
8.2  -- have been reevaluated in detail. In that review the conclusion is
reached 
that after all a two Breit-Wigner resonance ansatz provides the best overall
description of the data with resonance parameters in agreement with those of
Tan\cite{Tan}.

\subsection{\it Conclusions about this Period}

Animated by the enormous amount of predicted states, a worldwide rush of
experimental dibaryon searches started in the eighties  --
ending finally with a vast number of claims. But unfortunately no single one
survived critical experimental and analytical examinations.

Among the reasons for this striking failure was
certainly the insufficient quality of data, be it low-statistics
bubble-chamber data  or data from inclusive measurements, performed often by
single-arm detectors. For critical, though also very amusing reviews of this
epoch see, {\it e.g.}, those given 1988 by K. K. Seth \cite{Seth,Seth1}, who
pioneered this field by many high-quality measurements un-masking thus many of
the dibaryon claims as statistical fluctuations or detector artifacts in poor
data -- see {\it e.g.} Fig. 22 in Ref. \cite{Seth}, where in a high-statistics
and high-resolution $\vec{p} d \to p X$ measurement many of the dibaryons
claims have been excluded in the mass range 1877 - 2200 MeV . In his conclusions he summarizes:
\begin {itemize}
\item "Nobody, anywhere, has seen a genuine, bona-fide, gold-(silver, nickel-,
  or even un-)plated dibaryon, yet!"
\item "The days of doing quick and dirty (Q $\&$ D) experiments are over."
\item "The days of making Q $\&$ D predictions are over."
\item "The days of inventing dibaryons to explain the difference between poor
  experiment and poorer theory (or vice-versa) are over."
\item "We must do honest hard work, or quit."
\end {itemize}

And he recommends for future research: " We should concentrate
on exclusive experiments ...", a point, which we will pursue in the following.

\subsection{\it A Possible Remnant from this Period -- a Broad Resonance
  structure around the $\Delta N$ Threshold}

Possibly, there exists a survivor from this era -- though not a narrow, but
broad resonance structure around the $\Delta N$ threshold. Already 
in the fifties it was observed \cite{Neganov,Neganov1,Mesh} that the $^1D_2$
partial wave of the $NN$ system indicates a resonant behavior in the $\pi^+ d
\to pp$ reaction corresponding to a resonance with $I(J^P) = 1(2^+)$, m
$\approx$ 2160 MeV and $\Gamma \approx$ 120 MeV. Dyson $\&$ Xuong identified
this candidate with their  asymptotic $N\Delta$ state $D_{12}$ to fix the
remaining parameter B in their ansatz -- see Table 1. The possible existence
of this resonance was confirmed in the sixties 
by Dick Arndt \cite{Arndt} analyzing $pp$ scattering data and calling this
resonance structure a "Spin-2 Regge Recurrence of the $^1S_0~p-p$ Pole". 

\begin{figure} 
\centering
\includegraphics[width=5.9cm,clip]{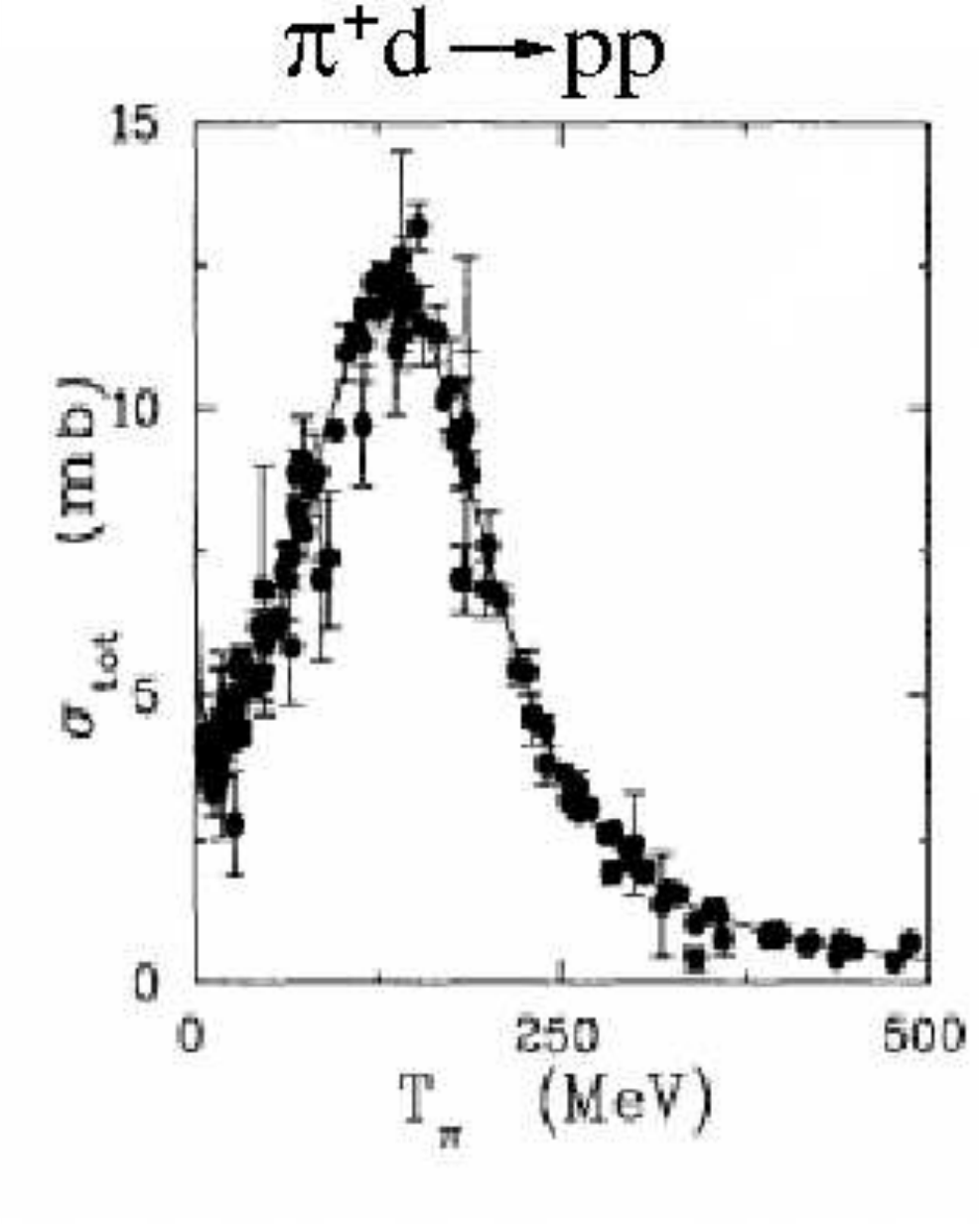}\\
\includegraphics[width=5.9cm,clip]{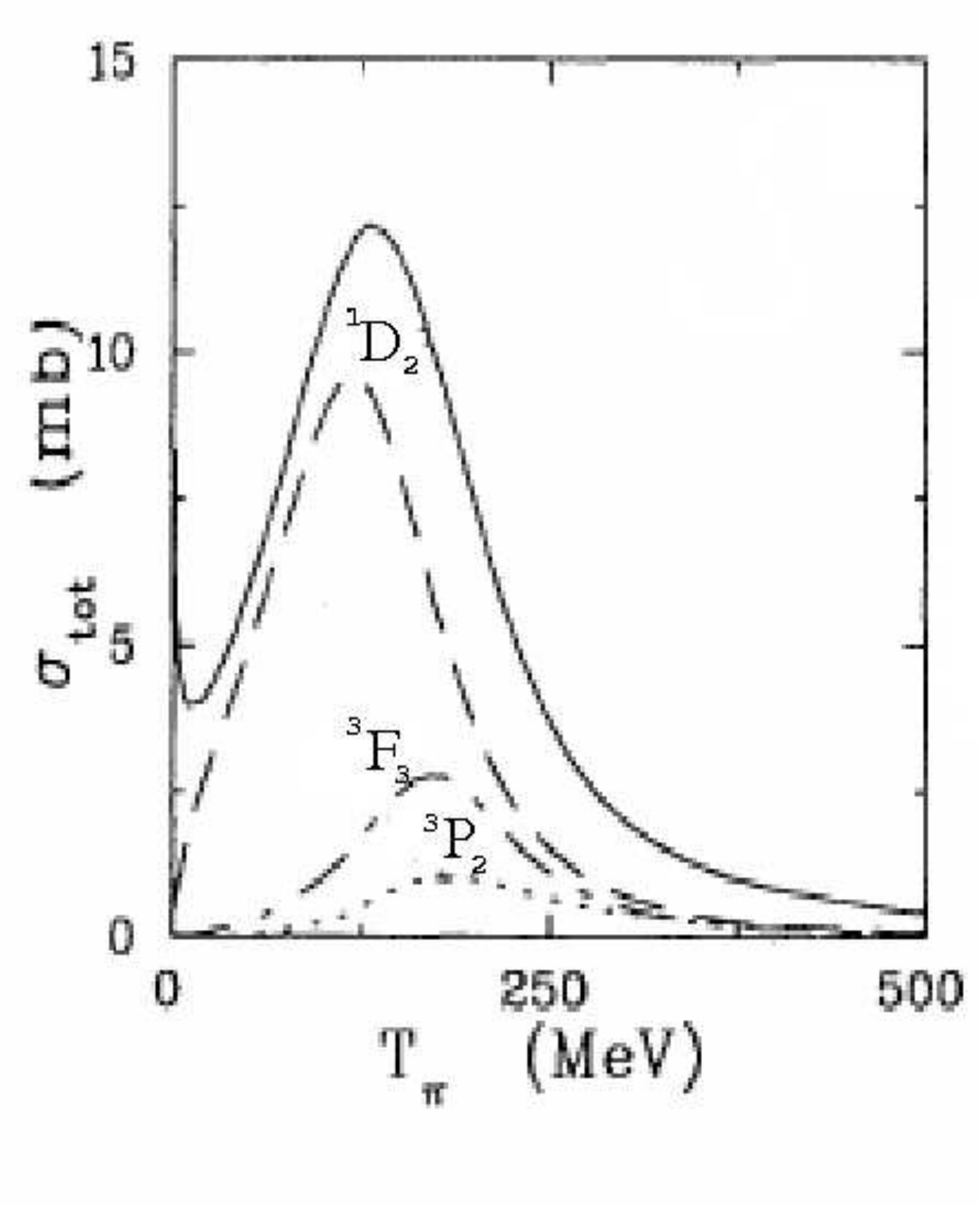}\\
\includegraphics[width=7.8cm,clip]{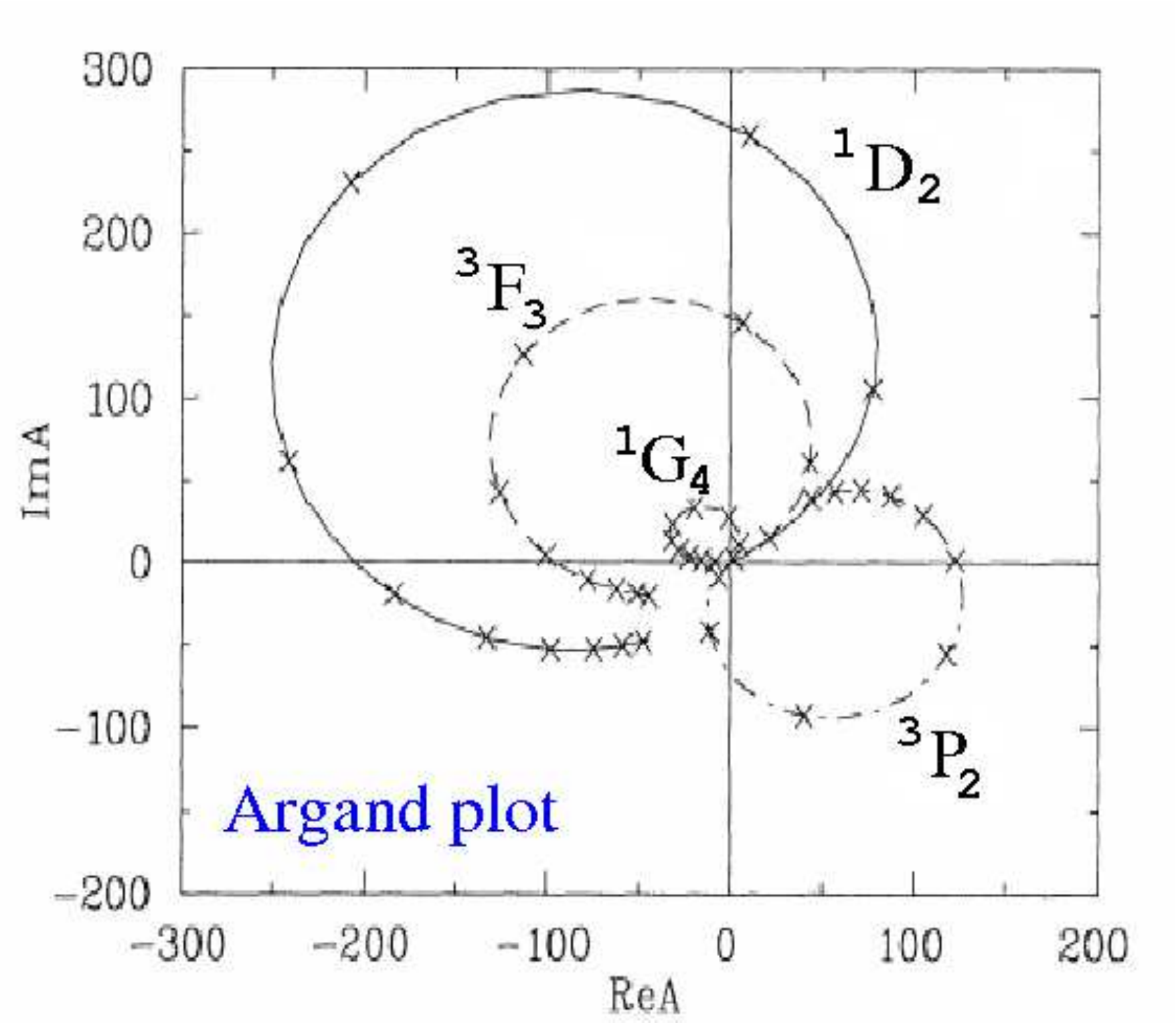}\\
\caption{Energy dependence of the total cross section of the $\pi^+ d \to pp$
  reaction as obtained by 
  measurements (top) and by the SAID partial-wave analysis (middle) together
  with its decomposition into its most prominent contributions from the $pp$
  partial waves $^1D_2$, $^3F_3$ and $^1G_4$. Bottom: Argand plot for selected
  partial waves, which exhibit a pronounced looping. From Ref.~\cite{said1}.
}
\label{fig-DeltaN}       
\end{figure}

Later-on, in the eighties and nineties, the Gatchina group
\cite{Kravtsov,StrakovskyKR,Borkowski} and the SAID data analysis group
\cite{SM94,SM94pid,said0,said1,said3,said4a,said4,FA91} could demonstrate by
partial wave 
analyses based on high-quality cross section and polarization data for $pp \to
pp$, $\pi d \to \pi d$ and $\pi d \to pp$ reactions that the
$^1D_2$ partial wave exhibits a pronounced looping in the Argand diagram --
see Fig.~\ref{fig-DeltaN} -- in
favor of a true $s$-channel resonance. However, since the mass of the
resonance is close to the $N\Delta$ threshold and since in addition  the width
of the resonance is compatible with that of the $\Delta$, it has been argued
that the observed features represent a $N\Delta$ threshold phenomenon rather
than a $s$-channel resonance and the observed looping in the Argand plot is
merely a reflection of the usual $\Delta$ excitation in the
presence of a nucleon with both being forced to be at rest relative to each
other due to the threshold condition \cite{igor,Brayshaw,Bakker,Narodetskii,Simonov,Simonov1,Niskanen}. In this discussion there
have been many pros and cons, see, {\it e.g.}
Refs. \cite{igor,Seth,Narodetskii,Simonov,Simonov1,Niskanen,shypit1,shypit2,sarantsev,strakovsky}. In a series of
papers  \cite{hos1,hos2} Hoshizaki concluded
that interpretations in terms of a $N\Delta$ virtual-state pole or a threshold
cusp are invalid, rather it constitutes a true $S$-matrix pole at (2144 - i 55)
MeV. A similar conclusion has been reached by Ueda {\it et al.} \cite{Ueda1}.

But still we are left with the problem that these values do not deviate
significantly from what is expected for a conventional $\Delta$  excitation by
$t$-channel meson exchange between the two nucleons. 

We note in passing that in the SAID partial wave analyses also the
$^3P_2 - ^3F_2$, $^3F_3$, $^1G_4$ and $^3F_4 - ^3H_4$ $NN$ partial waves exhibit a
looping in the Argand diagram -- see Fig.~\ref{fig-DeltaN}, though their effect in the observables is small
compared to the dominant $^1D_2$ partial wave in the $\Delta N$ region. For the
$^3F_3$ partial wave a pole at (2170 - i 72) MeV has been
found \cite{FA91}, which is right at the $\Delta N$ threshold. Hence it
is also plausible that the deduced width is larger than that of the 
$^1D_2$ resonance. In addition a pole have been located in the threshold
region for the $^3P_2 - ^3F_2$ coupled partial waves at (2167 - i 75) MeV
\cite{FA91}. 

Garcilazo {\it et al.} used local \cite{Mota} as well as non-local
\cite{Mota1} $\Delta N$ and $\Delta\Delta$ potentials derived
from the quark-cluster model to analyze the bound-state problem of the $\Delta
N$ system for the case that the two-body system is in relative $S$-wave. As a
result they find that the $\Delta N$ system has just a single bound state,
which sits right at the $\Delta N$ threshold and which corresponds to the
experimental finding of the resonant $^1D_2$ $NN$ partial wave. Since they
restricted their studies to relative $S$-waves between $N$ and $\Delta$, they
did not investigate the situation with regard to the $^3F_3$ $NN$ partial wave.

More recently the $NN\pi$ system was investigated by Gal and Garcilazo utilizing
Faddeev calculations with hadronic interactions \cite{GG1,GG2}. They also find
a resonance pole at (2147 - i 60) MeV for the $I(J^P) = 1(2^+)$ state.

A completely different  approach for shedding more light onto the longstanding
question, whether there are genuine dibaryon resonances in the $\Delta N$
region or not, has been presented very recently by Platonova and Kukulin
\cite{Platonova} . They
argue that the use of soft meson baryon form factors consistent with $\pi N$
elastic scattering gives too low cross sections in conventional $t$-channel
meson and nucleon exchange calculations for single-pion production, in
particular for the $pp \to d\pi^+$ reaction. Only by inclusion of the formation
and the decay of dibaryon resonances -- in particular the $D_{12}$ resonance
with $I(J^P)=1(2^+)$ in the $^1D_2$ $NN$ 
partial wave -- in the intermediate reaction process a proper
description of the experimental data can be obtained. By additional inclusion of
corresponding resonances in $^3F_3$ and $^3P_2$ partial waves even
polarization observables can be quantitatively described for the first time
\cite{Platonova1}. 

We shall return to the $\Delta N$ threshold phenomena in sections 11.2 and 11.3.

\section{An Intermezzo in Nuclei: The Pionic Double Charge Exchange Reaction}

In the pionic double charge exchange (DCX) reaction on nuclei $\pi^+ A(Z,N) \to
\pi^- B(Z+2,N-2)$  and $\pi^- A(Z,N) \to \pi^+ B(Z-2,N+2)$, respectively -- in
nuclear physics convention denoted as A($\pi^+, \pi^-$)B and A($\pi^-,
\pi^+$)B, respectively --, two
neutrons are converted into two protons or vice versa. This reaction ensures
that the process is a genuine two-nucleon process depending heavily on the
correlations between the two active nucleons. The cross sections have been
found to be largest at low incident pion energies and in addition highly
sensitive to short-range correlations there  -- for a review, see {\it e.g.}
Refs. \cite{Clement,JohnsonMorris}. This is particularly true for so-called
nonanalog 
transitions, for which the isospins of initial and final nuclear states differ
by two units.  
Hence it is not astonishing that this reaction has been considered to be well
suited for the search of dibaryons. 

Since secondary pion beams suffer from background due to pion decay, DCX
measurements have first been carried out at high pion energies, where the
effective pion lifetime in the laboratory system is larger and hence this
background is smaller than at low energies. In the latter case most of the
pions produced at the production target have already decayed before
reaching the final target for the actual measurement.

Hence at the pion factories the first DCX measurements concentrated on pion
energies in the region of the $\Delta$ excitation and also on so-called double
isobaric analog transitions (DIAT) in nuclei. In these reactions the isospin in
initial and final nuclear states stays identical, {\it i.e.} the two active
neutrons in the nucleus are converted into two protons, which stay in the same
nuclear orbit, thus possessing the same wave function in the final nucleus as
the two original neutrons in the initial nucleus had before -- only the 3rd
component of the isospin is changed. Hence the overlap between initial and
final nuclear wave functions is maximal with the consequence that also the
cross section should get comparatively large.

These DIAT measurements in the $\Delta$ region exhibited a diffraction pattern
in the angular distributions, which can be accounted for in strong absorption
model calculations, if an isotensor term is included \cite{Greene}. The
reaction process is assumed to be a two-step process of two sequential single
charge exchanges in the nucleus. In this conception the incoming
positive pion gets absorbed by a valence neutron leading to a (virtual) $\Delta^+$
excitation, which deexcites into a proton in the same orbit and a (virtual)
neutral pion. The latter propagates to another neutron, preferentially in the
same orbit, where it gets absorbed by exciting this neutron to a (virtual)
$\Delta^0$  state, which again deexcites into a proton in the same orbit and a
$\pi^-$ particle, which is then emitted from the final nucleus.

The measured dependence of the DIATs on the mass number A of the investigated
nuclei is in agreement with this two-step process in the strong absorption
limit. In this picture each step of single charge exchange proceeds with
$A^{-1}$ in the amplitude, since the charge of a pion is exchanged on a
specific nucleon out of $A$ nucleons. Due to strong absorption (black
disc) the reaction volume is restricted to the circumference of a disc
providing another factor $A^{1/3}$ in the amplitude. Hence in total we expect a 
$A$ dependence of $A^{-1} A^{-1} A^{1/3} = A^{-5/3}$ in the amplitude and of
$A^{-10/3}$ in the cross section, respectively -- and this is what actually
has been  observed for the DIAT measurements in the $\Delta$ resonance region. 

At incident pion energies above the $\Delta$ resonance region the measured
cross sections get somewhat larger -- as expected from the reduced pion
absorption in nuclei, but basically the two-step concept for DIATs proved to
stay valid there, too. 

Having understood the reaction mechanism for DIATs in the $\Delta$ resonance
region and above it came as a surprise that the first measurements at low
incident pion energies showed the cross sections to be largest in the
region of $T_\pi \approx$ 50 MeV. In contrast to this experimental finding
two-step models had predicted cross sections to be smaller by an order of
magnitude -- among others due to the fact that the forward-angle single-charge
exchange cross section vanishes because of the well-known destructive
interference between $s$- and $p$-waves in the $\pi N$ system.

In this situation G. Miller \cite{Miller} proposed six-quark cluster
components of nuclear wave functions as an explanation for the unexpected
large DIAT cross section. In this scenario a dibaryonic six-cluster would be
formed as an intermediate state. Since the pion-absorption operator is an
axial vector and due to spatial symmetry there are only two possible
intermediate states with $I(J^P) = 0(1^+)$ and $2(1^+)$. Such dibaryon states
had been predicted by Mulders and Thomas \cite{MuldersThomas} with
masses of 2180 and 2460 MeV, respectively. Though Miller's calculations gave a
reasonable description for the forward angle  data for the DIAT in $^{14}$C --
in particular, if pion absorption is taken into account \cite{Leitch} -- see,
{\it e.g.}, Fig~4.18 in Ref. \cite{Clement}, conventional calculations
including isotensor terms subsequently turned out to be at least as successful
or even more successful \cite{Leitch1}
-- see Fig. 4.19 in Ref. \cite{Clement}. Later-on G. Miller \cite{Miller1}
pointed out that the six-quark cluster assumption also necessitates a
resonance-like energy dependence in the total DIAT cross section on $^{14}$C,
which should peak at $T_\pi \approx$ 425 MeV, which again was not observed
\cite{Burleson} -- see, {\it e.g.} Fig. 4.28 in Ref. \cite{Clement}. Among
others a reason for this failure could have been that the two-step process of
a DIAT not only depends on short-range correlations, but also on longer-range
correlations, which possibly blur the six-quark cluster contribution. 

On the contrary, short-range correlations play a central role in non-analog
transitions, which mostly have been measured as a transition between ground
states, {\it e.g.} $^{12}$C($\pi^+, \pi^-$)$^{12}$O. In fact, initially
the cross sections of these non-analog transitions were massively
underestimated. Since no simple 
two-step picture was available for this process, it were thought to have tiny
cross sections. However, first measurements in the $\Delta$ resonance region
revealed these cross sections to be of the same order of magnitude as
the DIATs and partly nearly as large as those. Also unexpectedly, the observed
dependence on the nuclear mass number A was not 
that of a two-step process, namely proportional to $A^{-10/3}$, but that of an
effective one-step process being proportional to $A^{-4/3}$. The latter can be
understood, if the two active neutrons are exchanged into two protons in a
single step, so that in the strong absorption limit we have a DCX amplitude
being proportional to $A^{-1} A^{1/3}$ = $A^{-2/3}$. 

\begin{figure}
\centering

 \includegraphics[width=9cm]{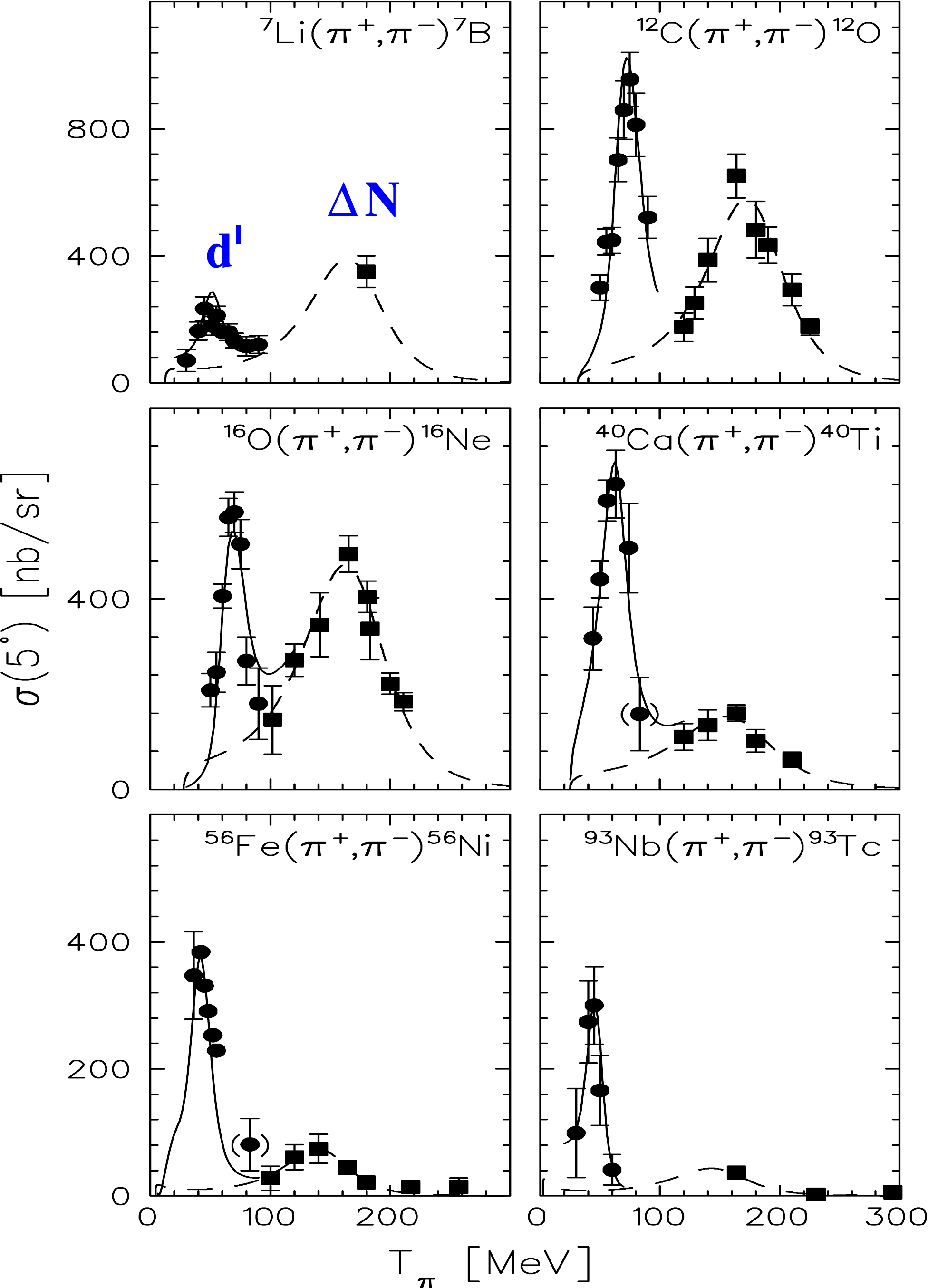}

\caption{Energy dependence of the forward-angle cross sections for nonanalog
  DCX groundstate transitions. The 
  solid lines show calculations assuming the formation of the 
  hypothetical $NN\pi$ resonance $d'$ interfering with the DINT mechanism
  forming a resonating $\Delta N$ system. From
  Refs. \cite{Draeger,ED}. 
\label{fig-DCX}} 
\end{figure}

Another unexpected feature of the non-analog transitions observed in the
$\Delta$ resonance region and above was the energy dependence of the
forward-angle cross sections, which revealed a resonance-like structure reminiscent
of a $\Delta$ excitation in nuclei -- see Fig.~\ref{fig-DCX}. In general, the
peak was found to be just around the $\Delta$ mass, whereas the
width of the observed structure was somewhat smaller in the range of 70 - 90
MeV \cite{Gilman}. 

To accommodate these unexpected features in an appropriate model description,
the so-called $\Delta N$ interaction (DINT) model has been designed
\cite{Johnson,Johnson1}, where -- 
as in the DIAT process -- the incoming $\pi^+$ particle initially excites a
neutron to $\Delta^+$. However, the subsequent $\Delta N$ interaction does not
lead to the exchange of a $\pi^0$ with the neighboring neutron, but to the
exchange of a $\pi^+$ with the consequence that the neighboring neutron is
charge exchanged to a proton, whereas simultaneously the $\Delta^+$ is charge
exchanged to $\Delta^0$, which subsequently decays into proton and emitted
$\pi^-$ particle. Since this effective single-step process certainly is of
very short range, it has been proposed \cite{JohnsonKisslinger,Wirzba} that
the DINT process may involve six-quark clusters in the intermediate state.  
Though this is similar to Miller's proposal, it is now explicitly specified only
for the short-range DINT process. In addition Johnson and Kisslinger
\cite{JohnsonKisslinger} showed that the contribution of the intermediate
$I(J^P) = 0(1^+)$ state is small compared to that of the $I(J^P) =
2(1^+)$ state. 

From the comparison to the experimental results for non-analog
transitions it follows that the dominating intermediate state, which
gives rise to the $\Delta$-like resonance structure in the energy-dependence
of non-analog transitions, does not correspond to a mass of 2460 MeV as
predicted by Mulders and Thomas \cite{MuldersThomas} and adopted by Miller,
but to a mass of 2140 - 2170 MeV,
{\it i.e}, a mass corresponding closely to the $\Delta N$ mass. That way it
would rather represent the $D_{21}$ dibaryonic state predicted to be at the
same mass as $D_{12}$ by Dyson and Xuong \cite{Dyson} -- see sections 3.2 and
11.2.

An even more exciting feature of non-analog transitions has been revealed by
subsequent measurements
\cite{Faucett,Bilger,Bilger1,Foehl,Paetzold,Paetzold1,Draeger} at low
energies. There these transitions exhibit in general not
only larger cross sections than at higher energies, but also a 
very pronounced narrow resonant structure in the energy dependence
(Fig.~\ref{fig-DCX}), which we
will discuss in detail in the following.

\subsection{\it Yet Another Dibaryon Candidate: $d'(2065)$}

On the basis of a QCD string model Schepkin and coworkers at ITEP, Moscow,
predicted the lowest-lying dibaryon states to be a triplet of isoscalar states
with  $J^P = 0^-, 1^-$ and $2^-$ having an l=1 orbitally excited $q^4 - q^2$
structure. This triplet corresponds to the triplet predicted by the
Nijmegen group on the basis of the MIT bag model, however at somewhat smaller
masses, in the range 2050 - 2140 MeV \cite{Schepkin,Schepkin2}. Since the $0^-$
and $2^-$ members of the triplet cannot couple to the $NN$ system due to their
quantum numbers, the only possible hadronic decay channel is $NN\pi$. In
consequence the decay width should be very small in the range of only few MeV
or even below. Due to spin-orbit splitting the $2^-$ state was expected to be
the lowest-lying state. 

Since the predicted mass range corresponded just to the bump structure seen
around $T_\pi \approx$ 50 MeV in the DIATs on $^{14}$C and also $^{18}$O, 
Martemyanov and Schepkin proposed the $2^-$ state to be the reason for this
observed structure \cite{Schepkin1}. Indeed, in subsequent work
\cite{Bilger1,Bilger2,Foehl,Paetzold,Paetzold1,Draeger} it could be
demonstrated that all low-energy data on DIATs and non-analog transitions in
nuclei ranging from $^7$Li to $^{93}$Nb 
could be quantitatively described by the assumption of the excitation of a
narrow dibaryon resonance in the intermediate state: the $d'(2065)$. Only, the
experimental angular distributions requested a spin-parity assignment of $0^-$
instead of the previously assumed $2^-$ assignment. 

In the description of the data the dibaryon resonance effect has been assumed
to sit 
upon a background of non-resonant conventional processes. In case of DIATs these
conventional processes have been described by the model of Auerbach {\it et
  al.} (AGGK), where the two-step process is calculated within the seniority
concept for the nuclear structure aspects \cite{AGGK,AGGK1}. The exceptional role of
the DIAT on $^{48}$Ca, which constitutes the only case, where no bumpy energy
dependence at low pion energies was observed, has been successfully
explained by this model. However, for all other cases this model largely
underestimated the measured cross sections. In particular, it could not explain
the observed steep resonance-like energy dependence there. This failure had
been common to other theoretical investigations at that time.

For the non-analog transitions, which exhibit the resonance-like structure
with a width of only about 20 MeV even much more pronounced, no adequate
conventional description had been available, since the AGGK model predicted
vanishing cross sections in general. The DINT mechanism, which was the only
quasi single-step process to describe the data at higher energies, also
provided only tiny cross sections, since at $T_\pi \approx$ 50 MeV solely the
low-energy tail of the $\Delta$ excitation is left. Hence essentially the full
measured cross section was attributed to the $d'$ resonance excitation.

A consistent successful description of all low-energy DCX data, both DIATs and
non-analog transitions, was achieved with the $d'$ model by adjustment of just 
the resonance parameters for mass and width resulting in a mass of $2065 \pm
5$ MeV, a decay width into the $NN\pi$ system of 0.5 MeV and a spreading width
due to the nuclear surrounding of 10 - 20 MeV \cite{Draeger}. The further
broadening of the resonance due to the Fermi motion of the active nucleon pair
was calculated microscopically. It should be noted that in initial
calculations a spreading width of only 5 MeV was assumed, which gave too large
$d'$ contributions.

The success of the $d'$ hypothesis to describe the low-energy DCX data
initiated a number of theoretical investigations for such a dibaryon
\cite{Glozman,Wagner,Itonaga,Obukhovsky,Buchmann,Obukhovsky1,Ueda1,Ueda2,Garcilazo2,Mathelitsch,Moro,Valcarce,Garcilazo3,Schwesinger}. 
The T\"ubingen theory group
\cite{Glozman,Wagner,Itonaga,Obukhovsky,Buchmann,Obukhovsky1} calculated a
possible dibaryon with the quantum numbers $I(J^P) = 0(0^-)$ of $d'$ in
constituent quark, quark shell-model, colored quark cluster and quark pair
creation models. As a common result the mass of such a dibaryon was found to be
in the range of 2300 - 2400 MeV, if the confinement parameters obtained from
the baryon spectrum were used. Only if the confinement was strongly weakened and
thus the size of such a dibaryon was enlarged, a mass as well as a decay width
as low as needed for the explanation of the DCX data was obtained. 

Ueda calculated such a dibaryon solely by use of $\pi NN$
dynamics. Initially he obtained a mass in the range 2060 - 2090 MeV
\cite{Ueda2}, but in an improved calculation he obtained a mass of only 4 - 5
MeV above the $NN\pi$ threshold, {\it i.e} around 2022 MeV \cite{Ueda3}.
Essentially the same conclusion was reached by Garcilazo solving nonrelativistic
Faddeev equations with hadronic interactions \cite{Garcilazo2}. In further
studies within the framework of $\Delta  N$ and quark cluster models it was
concluded that $d'$ should have isospin $I = 2$ rather than $I = 0$
\cite{Mathelitsch,Valcarce,Garcilazo3}, which also is compatible with the
experimental observations. It also was shown that such a dibaryon resonance
will increase its width in the nuclear medium by an order of magnitude -- in
agreement with experiment -- but change its mass by only a few MeV. 

A $J^P = 0^-$ dibaryon resonance was also looked at in the Skyrme model and it
was demonstrated that this model leads to such a low-lying state only, if the
isospin is $I = 2$ \cite{Schwesinger} -- in agreement with the conclusion
reached by Garcilazo {\it et.al.}.

A crucial disadvantage of using the DCX reaction for the dibaryon issue is that
it can be conducted  only on nuclei, but not on the basic $nn$ system, since the
latter is unbound and hence not available as a target. Therefore any
theoretical comparison with the data suffers from the fact that the nuclear
structure in initial and final states is not fully under control. The
substantial nuclear structure dependence of DCX calculations for DIATs and
also non-analog transitions has been demonstrated impressively in
Refs. \cite{Gibbs,Gibbs1}, where it has been shown that uncertainties in the
nuclear wavefunctions can easily lead to a factor 
of two uncertainty in the predicted cross sections -- in addition to pion wave
distortion, which at least partially is the cause for the rapid fall-off of
the cross section beyond $T_\pi$ = 50 MeV.  Taking all this into account, a
number DIATs and non-analog transitions can be accounted for at least
qualitatively by conventional calculations, though the data remain still
much better reproduced by the $d'$ resonance assumption.

The lightest target nucleus for observing the DCX process in a binary reaction to a discrete
final state is $^7$Li. For lighter nuclei  the DCX process
leads only to the unbound nuclear continuum. With regard to $d'$ this means
that it no longer can be formed off-shell in the presence of $A$ - 2 nucleons,
but only produced on-shell associatedly with the simultaneous release of
spectator nucleons into the continuum -- with the consequence that the $d'$
signature in this now inclusive DCX measurements will be much less
pronounced. Nevertheless, detailed DCX continuum measurements have been
carried out on $^3$He \cite{Graeter} and $^4$He
\cite{Graeter1,Graeter2,Clark}, but the results finally were inconclusive on
the question of $d'$, since it turned out that the initially assumed small
spreading width was strongly underestimated \cite{Nefediev}. With the  more
realistic spreading width the $d'$ effect in these measurements got too small
to be uniquely identified.

Finally, also in this case the existence of a narrow dibaryon resonance could
not be established unambiguously and nature might have bluffed us again. Though
the description of the data by inclusion of the as of yet hypothetical $d'$
resonance works amazingly well, it cannot be excluded that 
by the complexity of nuclear structure and pion-nucleus interaction the
observed resonance structure may finally find its explanation in a capricious
composition of diverse conventional effects.

A possibility to search for $d'$ in a basic reaction is neutral pion
photoproduction from the deuteron, which has been discussed already above in
section 4.2.3. However, the estimated $d'$ production 
cross section in this reaction is in the order of 0.5~-~1~$\mu$b compared to a
background from conventional processes, which is larger by nearly two orders of
magnitude \cite{Bilger3}. A high-resolution measurement \cite{Siodlaczek} of the
$\gamma d \to \pi^0 X$ reaction at MAMI found no statistically significant
signal from a narrow dibaryon resonance at masses below 2100 MeV. But the
deduced upper limits were still an order of magnitude larger than the predicted
$d'$ signal, so that also this measurement could not provide an answer, whether
$d'$ really exists. 

Also electro-production of $d'$ has been looked at by using ARGUS data on
$\gamma^*$$^{16}O \to pp\pi^{\pm}X$. Though the invariant $pp\pi^+$ and
$pp\pi^-$ spectra were in agreement with the $d'$ hypothesis, no firm
conclusion could be drawn due to low statistics \cite{evdargus}.

More promising has been the prediction of a $d'$ signal in another basic
reaction, the $pp \to pp\pi^+\pi^-$ reaction, where the $d'$ cross section
has been estimated to be only one order of magnitude below the cross section
for conventional processes \cite{Schepkin3}. We will discuss the results from
corresponding measurements in section 8.3.

\section{More Recent Searches for Strange ($S=-1$) Dibaryons}

Recently the quest for strange dibaryons experienced a revival, when Akaishi
and Yamazaki \cite{Akaishi,Yamazaki,Yamazaki1} predicted deeply bound
antikaonic nuclear systems. It was argued  that if the attractive $\bar{K} N$
interaction is strong enough to form $\bar K$-nuclear 
sytems with binding energies such large that they reside below the $\bar K N
\to \pi \Sigma$ threshold, then such states will have a very narrow width and
constitute very compact objects. In fact, subsequent measurements at KEK with
$K^-$ ions stopped on $^4$He indicated such deeply bound structures
\cite{Suzuki}. However, an improved follow-up experiment \cite{Sato} did not
confirm the previous findings.

In the context of deeply bound antikaonic nuclear states also the basic system
$ppK^-$ was investigated both theoretically and experimentally. Initial
calculations predicted a deeply nuclear-bound quasistable state with $I(J^P) =
\frac 1 2 (0^-)$, a binding
of about 50 - 80 MeV and a decay width into $\Sigma p \pi$ in the order of 60
- 110 MeV \cite{Yamazaki1,Ikeda,Shevchenko,Shevchenko1}, whereas more
sophisticated recent calculations give very moderate binding energies in the
range of 10 - 20 MeV and widths in the order of 40 - 70 MeV
\cite{Barnea,Dote,Ikeda1}. For a recent review of this state, which
asymptotically can be viewed also as a bound $\Lambda(1405) N$ dibaryon, see
Ref. \cite{GalKaon}. Note that $\Lambda(1405)$ itself is likely to be a
meson-nucleon molecular state as derived by recent lattice QCD \cite{JMMHall}
and chiral $SU(3)$ calculations \cite{Kamiya}.

In initial FINUDA measurements \cite{FINUDA}, where $K^-$ ions were stopped on
$^6$Li, $^7$Li and $^{12}$C targets, a 70 MeV wide structure was found in the
$\Lambda p$ invariant-mass spectrum. From the observed back-to-back angular
correlation it was concluded that this structure corresponded to an
intermediate $ppK^-$ system bound by as much as 115 MeV. However, this
interpretation was later-on heavily criticised \cite{Oset,Magas,Ramos} arguing
that the observed spectrum may be well explained by final-state interaction of
the produced $\Lambda p$ pair with the residual nucleus.

In order to avoid complications due to a nuclear surrounding, the search for a
$ppK^-$ bound state subsequently concentrated on basic reactions, which
contained only two baryons and where the dibaryon system in question was
produced associatedly. Note that no formation reaction is possible in this
case, because there is no $pp$ target.

Since the predicted $ppK^-$ bound
state can decay also into $\Lambda p$, the $pp \to \Lambda p K^+$ raction was
already proposed in Ref. \cite{Yamazaki1} as a reaction well-suited to search
for this predicted state. 
An subsequent analysis \cite{Kienle} of DISTO data on the $pp \to
\Lambda p K^+$ reaction at $T_p$ = 2.85 GeV claimed the observation of a
compact $pp K^-$ system  at 2265 MeV corresponding to a binding energy of 103
MeV and a width of 118 MeV. The DISTO data revealed the corresponding broad
structure in the $\Lambda p$ invariant-mass spectrum only after severe cuts and
assuming pure phase-space distribution for the non-resonant background. The
latter is at variance with the information from many preceding investigations
that this reaction is strongly affected by $N^*$ excitations.
Their reflections may easily produce a bump -like structure in the $\Lambda p$
invariant-mass spectrum. In fact, a recent reanalysis \cite{Epple} of all
published $\Lambda p$ invariant-mass spectra finds no hint for the claimed
resonance with deduced upper limits, which are partly substantially smaller
than the predicted cross section values. In contrast to this finding the
claimed resonance structure corresponded to a cross section four times larger
than the predicted one and exceeded even the $\Lambda(1405)$ production, which
was thought to be part of the intermediate state.

A recent search for the $ppK^-$ bound state via the $\gamma d \to K^+\pi^- X$
reaction at LEPS at Spring-8 also found no evidence for such a state, only
upper limits \cite{LEPS}. 

Surprisingly, a very recent experiment at J-PARC claims evidence for a 
$ppK^-$-like structure in the measurement of the $\pi^+ d \to K^+ X$
reaction at $p_\pi$ = 1.69 GeV/c \cite{KEK}. The measured $K^+$ missing-mass
spectrum shows three prominent peaks corresponding to the quasifree production
of $\Lambda$, $\Sigma$ and excited hyperons $Y^*$ = ($\Sigma(1385)$,
$\Lambda(1405)$). The simulated spectrum deviates in shape from the measured
one only at two positions: at the position of the well-known $\Sigma N$ cusp
-- see sections 4.2.5 and 8.2 -- and in the region of the $Y^*$ excitations, where the
measured spectrum appears to be shifted to lower $K^+$ missing masses. If now
in addition an emitted proton is measured in coincidence with the emitted
$K^+$, then the quasifree production is heavily suppressed, since the Fermi
momentum  of a spectator proton originating from the target deuteron is very
small and hence will not reach the proton detectors. As a result, the $K^+$
missing-mass spectrum with the condition of a coincident fast proton shows no
longer the quasifree peaks, but only the well-known $\Sigma N$ cusp followed
by some broad tail-like strength at higher missing masses -- see Fig. 2b in
Ref. \cite{KEK}. If in a next step
the thus obtained spectrum is divided by the original $K^+$ missing mass
spectrum without the coincidence condition -- the so-called "coincidence
probablity" spectrum --, then a pronounced broad bump structure
arises around 2.3 GeV -- see Fig. 2c in Ref. \cite{KEK}. A Breit-Wigner fit to
this bump results in a resonance 
mass of 2275 MeV and a width of 162 MeV. This structure also persists, if a
coincidence with two fast protons is requested, though the data have now large
statistical uncertainties. Note that this particular structure appears only
after division by the inclusive spectrum, which is dominated by quasifree
processes -- a procedure, which is not easily understandable from  a physics
point of view.

As pointed out by A. Gal \cite{Gal}, a mass for a $ppK^-$ bound state
of 2275 MeV, which corresponds to a binding energy of about 100 MeV with
respect to the $ppK^-$ threshold, is unacceptable theoretically. However, it
might correspond to a meson-assisted dibaryon state with $I(J^P) = \frac 3
2(2^+)$ and strangeness $S=-1$
predicted by Garcilazo and Gal in relativistic three-body Faddeev calculations
for the $\pi \Lambda N - \pi\Sigma N$ coupled system \cite{GarcilazoGal}. This
predicted $\Lambda N \pi$ resonance, which may be viewed as a $^5S_2$
$\Sigma(1385)N - \Delta Y$ quasibound state is according to Gal and Garcilazo
the lowest-lying $S$- wave dibaryon with strangeness $S =-1$.

The
mass of this dibaryon state is predicted to be 10 - 20 MeV below the $\Sigma N
\pi$ threshold, {\it i.e.} in just in the mass region of the structure found
at KEK \cite{KEK}. Whether there is indeed a connection, has to be clarified
by further exclusive and kinematically complete experiments, which -- in case
of a real resonance structure -- must be able to also extract its quantum
numbers. It should be noted that a recent search  by HADES \cite{HADES} at GSI
in the $pp \to \Sigma^+ p K^0$ 
reaction finds no evidence for the predicted $I(J^P) = \frac 3 2(2^+)$ state.  

New light into this matter has been shed by a recent measurement of the
$^3He(K^-,\Lambda p) n$ reaction at J-PARC, where $\Lambda$ and proton have
been detected, whereas the emitted neutron has been identified by its missing
mass \cite{JPARCE15}. These measurements, which currently are being
updated by new data \cite{MIN2016Iwasaki}, exhibit a pronounced bump structure
in the $\Lambda p$ invariant spectrum in the region of the $\Sigma p\pi$,
$\Lambda(1405) p$ and $ppK^-$ thresholds. A Breit-Wigner fit to this structure
results in a mass of about 2355 MeV and a width of about 110 MeV. This mass is
several MeV above the $\Lambda(1405) p$ threshold and 16 MeV below the
$ppK^-$ threshold. In a model description Sekihara, Oset and Ramos
\cite{Sekihara} interpret
the data by a two-bump structure due to a quasi-elastic peak for $K^-$
production in the first collision of the reaction and a peak associated to
the production of a $ppK^-$ quasi-bound state that decays into the $\Lambda
p$ system. The calculated quasi-bound state has a binding energy of about 20
MeV relative to $ppK^-$ threshold and a width of about 80 MeV.

In conclusion, with the possible exception of the currently much discussed
threshold structure near the $ppK^-$ mass a convincing evidence for a dibaryon
with strangeness $S = -1$ has not yet been found experimentally. 

\section{Status on the $H$ Dibaryon $(S = -2)$}

As mentioned already in section 4 the $H$ dibaryon was introduced 1977 by
Jaffe \cite{Jaffe} as a deeply bound $\Lambda\Lambda$ system with quark
structure $uuddss$ and $I(J^P)= 0(0^+)$. Calculating its mass by use of the
attractive short-ranged color-magnetic interaction between quark-pairs in
$SU(3)$-flavor symmetry  he obtained a binding energy of some 80 MeV relative
to the $\Lambda\Lambda$ threshold, {\it i.e.} it would be stable with respect
to a strong decay. Subsequently there have been
numerous theoretical model calculations within various models, {\it e.g.}, the
MIT bag model \cite{Jaffe,Mulders,Rosner}, the constituent quark model
\cite{Lipkin}, potential models \cite{Gignoux}, the Skyrmion model
\cite{Balachandran,Jaffe1,Callen}, the 
hybrid quark-cluster model \cite{Oka1,Straub}, the color-dielectric model
\cite{Nishikawa} and the instanton model \cite{Oka2}. The predicted masses
range from very deeply bound-- much more than in Jaffe's prediction -- to
unbound, {\it i.e.} to masses above the $\Lambda\Lambda$ threshold. In
particular, accounting for $SU(3)$-flavor symmetry breaking lead to a drastic
reduction of the resulting attraction \cite {Rosner,Lipkin,Gignoux}. 

Very recently the question, whether the $H$ dibaryon is bound or not,
received renewed attention, when two state-of-the-art  lattice
QCD calculations \cite{NPLQCD,HALQCD} obtained the $H$ dibaryon to be 
bound by about 8 MeV -- albeit for values of the pion  mass, which are still
large compared to the real pion mass and without paying attention to hadronic
thresholds. Subsequent theoretical investigations
show that such predictions require still quite some fine-tuning 
\cite{NPLQCD1} and that the mass also could well be between $\Lambda\Lambda$
and $\Xi N$ thresholds \cite{Inoue,haidenbauer}.

A study within the chiral effective field theory \cite{Haidenbauer}
arrives at still smaller binding energies, if bound at all, and demonstrates
that $SU(3)$ breaking effects induced by the differences of the pertinent
two-baryon thresholds have a very pronounced impact that need to be properly
incorporated in lattice QCD in order to produce realistic results. It is also
pointed out that, if the $H$ dibaryon is bound, the dominant component should
be $\Xi N$ rather than $\Lambda\Lambda$.

For an experimental search for the $H$ dibaryon, the possible decays of such
an object have to be considered. If the $H$ dibaryon is bound with respect to
the $\Lambda\Lambda$ threshold, it can only decay mediated by the weak
interaction and the dominant decay route will be $H \to \Lambda N\pi$. In this
case a very narrow resonance 
structure is expected in this decay channel. If the $H$ mass is
above the $\Lambda\Lambda$ threshold, then the $H$ will decay predominantly
hadronically via $H \to \Lambda\Lambda$. Such an unhindered fall-apart decay
necessarily will cause the width of such an $H$ dibaryon to be very broad. 
The $H$ dibaryon resonance will get still broader, if
its mass is even above the $\Xi N$ threshold. As pointed out already above, such
broad resonances are very hard to identify uniquely in experiments, since
they can hardly be distinguished from conventional non-resonant background
processes. As demonstrated in section 4.4 regarding the broad resonance
structure at the $\Delta N$ threshold, only detailed partial-wave analyses of
scattering and formation reaction data are able to reveal a real resonance in
such a case -- a situation, which is not feasible in case of the $H$ dibaryon
resonance.  

There have been numerous experiments searching for a $H$ dibaryon. A deeply
bound $H$ dibaryon has been excluded by the so-called 
"NAGARA" event, which exhibits the unambiguous signature of the
double-$\Lambda$ hypernucleus $^6_{\Lambda\Lambda}$He produced via $\Xi^-$
capture in emulsion \cite{NAGARA,Nakazawa}. If the mass $m_H$ of
the $H$  dibaryon would be less than twice the $\Lambda$ mass in the nucleus,
then the  
two $\Lambda$s would be expected to form a $H$ dibaryon. Hence the existence
of a double-$\Lambda$ hypernucleus with a $\Lambda-\Lambda$ binding energy
$B_{\Lambda\Lambda}$ ( in the nucleus) leads to a lower limit for the mass of the $H$ dibaryon
according to
\begin{equation}
m_H > 2m_\Lambda - B_{\Lambda\Lambda}.
\end{equation}
From the double-$\Lambda$ hypernuclei measurements, in particular the NAGARA
event $B_{\Lambda\Lambda} = 6.93 \pm 0.16$~MeV is obtained \cite{Nakazawa},
which gives a lower limit of 7 MeV for the $H$ binding energy and of 2224 MeV
for its mass, respectively, at the 90$\%$ confidence level.

The E224 experiment \cite{E224} at the KEK proton synchrotron searched for $H$
in the  $\Lambda\Lambda$ production in the inclusive reaction $K^-$$^{12}C \to
K^+\Lambda\Lambda X$ and noted an enhancement in the $\Lambda\Lambda$
invariant-mass spectrum right at the $\Lambda\Lambda$ threshold. A follow-up
measurement E522 \cite{E522} with improved statistics confirmed this findings,
but could also demonstrate that the conventional $\Lambda\Lambda$ final-state
interaction can quantitatively describe the observed threshold enhancement
without any need for assuming a $H$ dibaryon resonance.

At BELLE a high-statistics search for $H$ dibaryon production was performed
recently in
inclusive $\Upsilon(1s)$ and $\Upsilon(2s)$ decays \cite{BELLE}. No indication
of an $H$ 
dibaryon with a mass in the range from 35 MeV below 2$m_\Lambda$ to 25 MeV above
2$m_\Lambda$ has been observed in either the $H \to \Lambda\Lambda$ or $H \to
\Lambda p \pi^-$ decay channels. Stringent upper limits for the
branching-fractions have been extracted, which are between one and two orders
below that for inclusive $\Upsilon(1s)$ and $\Upsilon(2s)$ decays to
antideuterons. Since deacys of $\Upsilon(1s)$ and $\Upsilon(2s)$ produce
$SU(3)$ flavor-symmetric decays, these results put stringent constraints on the
dynamical properties of the $H$ dibaryon. {\it I.e.}, it must have dynamic
properties very different from the deuteron or, if $M_H < 2m_\Lambda$, a
strongly suppressed decay mode $H \to \Lambda p \pi^-$.

In heavy-ion collisons hyperons are produced in large numbers, which gives
access to the study of the $\Lambda\Lambda$ interaction and search for the $H$
dibaryon. In measurements of Au-Au collisons at 200 GeV at RHIC by use of the
STAR detector the $\Lambda\Lambda$ correlation function has been determined
\cite{STAR} giving access also to scattering length
$a_{\Lambda\Lambda}$ and effective range  $r_{\Lambda\Lambda}$
of the $\Lambda\Lambda$ interaction. Whereas in the original paper a slightly
repulsive interaction has been derived, an improved analysis \cite{Morita}
arrives at a slightly attractive interaction with -1.2 fm $<
a_{\Lambda\Lambda} <$ -0.5 fm and 3.5 fm $<  r_{\Lambda\Lambda} <$ 7 fm, if
the $\Lambda$ samples do not include the feed-down contributions from
long-lived particles. A negative scattering length excludes the existence of
a bound state in the $\Lambda\Lambda$ system. However, it is found that the
feed-down correction for $\Sigma^0$ decay reduces the sensitivity of the
correlation function to the details of the $\Lambda\Lambda$ interaction such
strongly that only the weaker constraint $a_{\Lambda\Lambda} >$ -1.2 fm is
left -- which unfortunately  does not completely exclude the existence of a
bound state. For a more detailed discussion see also Ref.~\cite{Ohnishi}. 

ALICE has searched for a weakly decaying $H$ dibaryon in central Pb-Pb
collisions at LHC \cite{ALICE} by looking for the decay mode $H \to \Lambda p
\pi^-$. To this end a $\Lambda$ has to be identified by two tracks belonging
to a proton and a pion and originating from a secondary vertex. In addition
another decay pattern reconstructed from a proton and a pion is required to be
found at the decay vertex of the assumed $H$ dibaryon. Since the considered
weak decay will only have a sizeable branching, if the $H$ dibaryon is bound
and hence can not decay hadronically into $\Lambda\Lambda$, this experiment is
sensitive only for the bound case. No evidence has been found for a bound $H$
dibaryon yielding upper limits for the mass range 2200 - 2231 MeV, {\it i.e.}
bindings energies smaller than 31 MeV. These upper limits  are two orders of
magnitude below that estimated in thermal models, which have proven to
quantitatively describe the yields of hadrons and produced nuclei including
hypertriton, which has a binding energy of less than 150 keV.  

In conclusion, none of the numerous dedicated experiments have found any
evidence for the existence of the $H$ dibaryon. From the observation of
double-$\Lambda$ hypernuclei a deeply bound $H$ dibaryon with a binding
energy of more than 7 MeV relative to the $\Lambda\Lambda$ threshold can be
excluded. There have been many dedicated experiments to search for a $H$
dibaryon being either weakly bound or unbound. They provided stringent upper
limits for its production in various processes, which make its existence very 
unlikely. However, as of yet none of them could completely rule out its
existence.

\section{The New Era of Exclusive and Kinematically Complete   High-Statistics
  Measurements at CELSIUS and COSY}
\label{sec-4}

A conclusion drawn from the experiences made in the dibaryon rush era is that
any meaningful dibaryon search should be undertaken with a dedicated
experimental equipment, which is suited to provide exclusive and
kinematically complete measurements of high statistics and high four-momentum
resolution. In the nineties programs were set up at the storage rings CELSIUS
and COSY, 
which allowed the search for dibaryon resonances with dedicated equipment
within the official program for systematic studies of elastic scattering and
meson production. 

The COoler SYnchtron COSY at the research center J\"ulich has been providing
high-brilliance polarized and unpolarized proton and deuteron beams. The beam
cooling has been achieved by electron and stochastic cooling,
respectively. For energy-dependence studies -- like resonance and threshold
issues -- it has been possible to ramp the beam with sub-MeV energy resolution
from lowest to highest energies. That way the energy dependence of observables
could be measured in a single run with a minimum of systematic errors. Whereas
at CELSIUS the highest proton beam energy was 1.45 GeV, it was 2.88 GeV at
COSY.
For a review of the hadron physics program carried out at COSY from its start in
1993 until its  end in 2014, see Ref. \cite{CWmeson2016,COSY}.

\subsection{\it No signal for Narrow Resonances in the $pp$ system}

The EDDA experiment at COSY provided ideal conditions for measuring the
elastic proton-proton scattering over nearly the full angular range both with
polarized beam and polarized target. In energy scan runs the energy
range between $T_p$ = 0.5 - 2.5 GeV, {\it i.e.} between $\pi$ and $\phi$
production thresholds, could be scanned with sub-MeV resolution in with 
high statistics 
\cite{EDDAsig,EDDAsig1,EDDApol,EDDApol1,EDDAcor,EDDAcor1,EDDAdib}. This
provided the instrumental and analytical possibility to reveal even  
very narrow, weakly excited dibaryon resonances \footnote{Occasionally the
  acronym EDDA has been interpreted also as "Elastic Dibaryons Dead or
  Alive" \cite{bisp}.} in the mass range of 2200 - 2800 MeV. 

\begin{figure} 
\centering

 \includegraphics[width=14cm,clip]{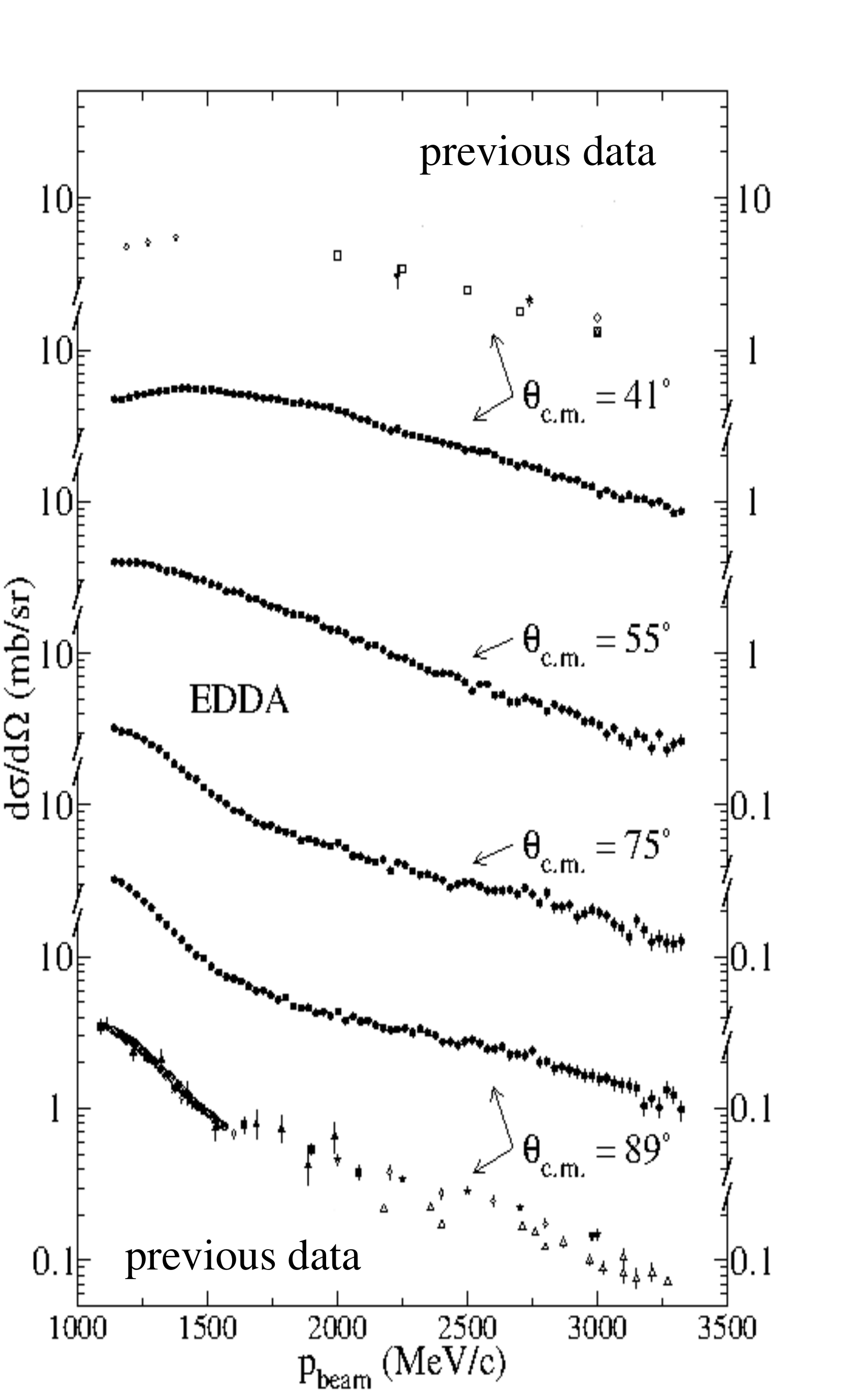}

\caption{Energy dependence of the differential cross section for $pp$ elastic
  scattering in the range of beam momenta $p_{beam}$ = 1.1 - 3.3 GeV/c
  corresponding to beam energies of $T_{beam} \approx$ 0.5 - 2.5 GeV ($\sqrt
  s$ = 2.1 - 2.8 GeV). Shown
  are results for $\Theta_{c.m.}$  = 41$^\circ$, 55$^\circ$, 75$^\circ$ and
  89$^\circ$ The EDDA data are shown by solid squares. For  $\Theta_{c.m.}$
  = 41$^\circ$ and 89$^\circ$ they are compared to previous results. From
  \cite{EDDAsig}. 
}
\label{fig-EDDApp}       
\end{figure}

With EDDA angular distributions over nearly the complete angular range have
been obtained for the differential cross section \cite{EDDAsig,EDDAsig1}, the
analyzing power \cite{EDDApol,EDDApol1} as well as for the spin correlation
observables $A_{NN}$, $A_{SS}$ and $A_{SL}$
\cite{EDDAcor,EDDAcor1,EDDAdib}. As an example Fig.~\ref{fig-EDDApp} shows the
measured energy dependence of the differential 
cross section at center-of-mass polar angles $\Theta_{c.m.}$ = 41$^\circ$,
55$^\circ$, 75$^\circ$ and 89$^\circ$ in comparison with previous
measurements. It gets appearent that the EDDA measurements set a new
standard in the experimental results for $NN$ scattering -- both in quality
and in quantity. 

\begin{figure} 1
\centering

 \includegraphics[width=12cm]{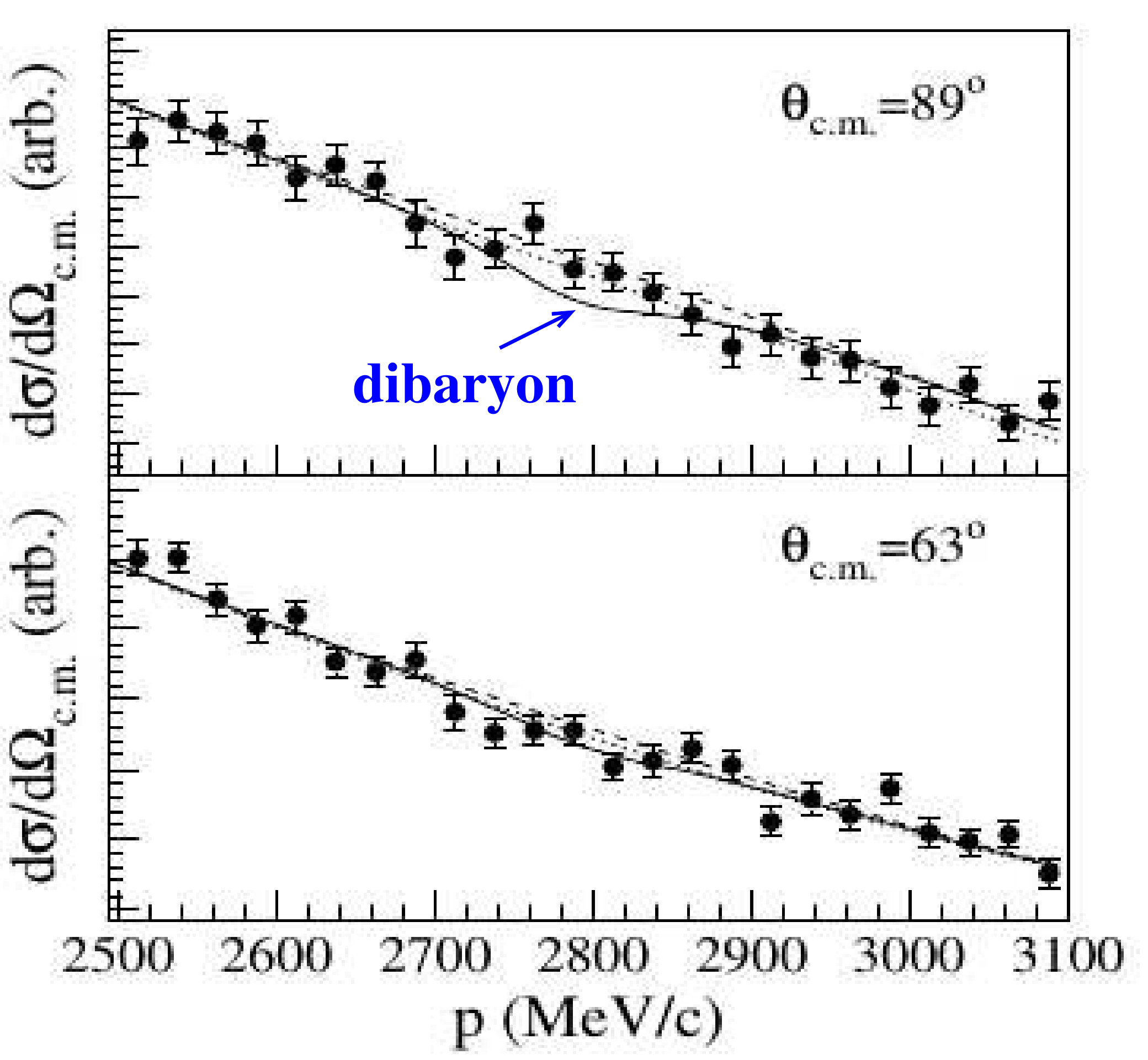}

\caption{EDDA results for the differential cross section for $pp$ elastic
  scattering  in the range of beam momenta $p_{beam}$ = 2.5 - 3.1 GeV/c for
  the scattering angles $\Theta_{c.m.}$ = 89$^\circ$ and 63$^\circ$ in
  comparison with results from phase-shift analyses with (solid lines) and
  without (dashed lines) including
  a putative dibaryon resonance in the $^1S_0$ partial wave  at $\sqrt s$ =
  2.7 GeV with a total width of $\Gamma$ = 50 MeV and an elasticity
  $\eta_{el}$ = 
  0.043. By the $\chi^2$ fit criterion such a resonance is excluded with a
  confidence level of 99$\%$. The dotted line shows the nonresonant background
  described by the phase-shift analysis in presence of such a resonance. From
  \cite{EDDAdib}.  
}
\label{fig-EDDAdib}       
\end{figure}

Similar to the examples shown in Fig.~\ref{fig-EDDApp} all observables measured at EDDA do
not show any statistical significant excursions, which could be a hint for
narrow dibaryons. A quantitative search for dibaryons has been carried
out for resonances in the energy range $\sqrt s$ = 2.2 - 2.8 GeV with total
widths in the range $\Gamma$ = 10 - 100 MeV \cite{EDDAdib}. Such putative
resonances were added to various partial waves in phase-shift analyses of the
data with the only free resonance parameter being the elastic partial width
$\Gamma_{el}$ expressed via the elasticity parameter $\eta_{el} = \Gamma_{el} /
\Gamma$. The upper limits
for $\eta_{el}$ for exclusion with a confidence level of 99$\%$ are given in
Ref. \cite{EDDAdib}. The obtained upper limits for $\eta_{el}$ range between
0.03 - 0.10 allowing thus only isovector resonances, which hardly couple to
the elastic channel. As an example Fig.~\ref{fig-EDDAdib} displays the fit to the data
assuming a resonance in the $^1S_0$ partial wave at $\sqrt s$ = 2.7 GeV with a
width of $\Gamma$ = 50 MeV and an elasticity of  $\eta_{el}$ = 0.043, the
value which is already excluded by the data with a confidence level of
99$\%$. Such a resonance was predicted in Ref. \cite{Lomon1} and indications
for it were claimed by the observation of some odd structure in the spin
correlation parameter $A_{00nn}$ \cite{Ball}. EDDA provides no evidence for
such a resonance.

Though no indications for any narrow resonances in the
proton-proton system have been found, this EDDA data set  is of
indispensable value, since it constitutes the backbone of high-quality
proton-proton scattering data in the SAID data base \cite{SAID} fixing the
empirical partial-wave amplitudes with unprecedented precision. Unfortunately
the EDDA experiment was not 
continued to examine also the proton-neutron scattering with similar precision
-- historically possibly a big mistake, as we will see in section 10.

\subsection{\it No Narrow Resonances in Hyperon Production -- except of
  $\Sigma N$ cusp} 

At COSY the time-of-flight spectrometer TOF was constructed for the dedicated
systematic study of hyperon production in $pp \to \Lambda p K^+$, $pp \to
\Sigma^0 p K^+$ and $pp \to \Sigma^+ p K^0$ reactions. The hermetic detector
has been able 
to cover practically the full reaction phase space. Highly segmented start and
stop scintillation detectors allowed time-of-flight measurements of charged
ejectiles and vertex reconstruction. Fiber and straw-tube detectors in-between
enabled in addition the vertex reconstrcution of decaying hyperons. That way
exclusive and kinematically complete measurements have been possible over the
full phase space. Since most of these measurements were not only
kinematically complete, but even overdetermined, kinematic fits with
overconstraints could be carried out with the consequence of highly improved
energy resolutions in invariant-mass spectra in the order of few MeV and
better.  

The $pp \to \Sigma^+ p K^0$ reaction has been investigated at three beam
energies $T_p$ = 2.16, 2.26 and 2.40 GeV \cite{DD}. It is well described in
all its differential observables by $N^*$ excitation via $t$-channel meson
exchange and its subsequent decay $N^* \to \Sigma^+ K^0$. This is particularly
true for the
$\Sigma^+ p$ invariant-mass spectrum, which spans the mass range from
threshold up to 2290 MeV and which gives no indication for narrow
dibaryon structures of statistical significance.

The $pp \to \Lambda p K^+$ reaction has been measured at numerous beam energies
in the range $T_p$ = 1.92 - 2.49 GeV \cite{ER,ER1,DD1,KE,SJ}, partly even
with polarized beam \cite{MR,FH}. Of particular interest here are the
high-resolution and high-statistics measurements, which cover the $\Lambda p$
invariant-mass range from threshold up to 2.3 GeV \cite{KE,SJ,MR}. 

\begin{figure} 
\centering

 \includegraphics[width=7.4cm,clip]{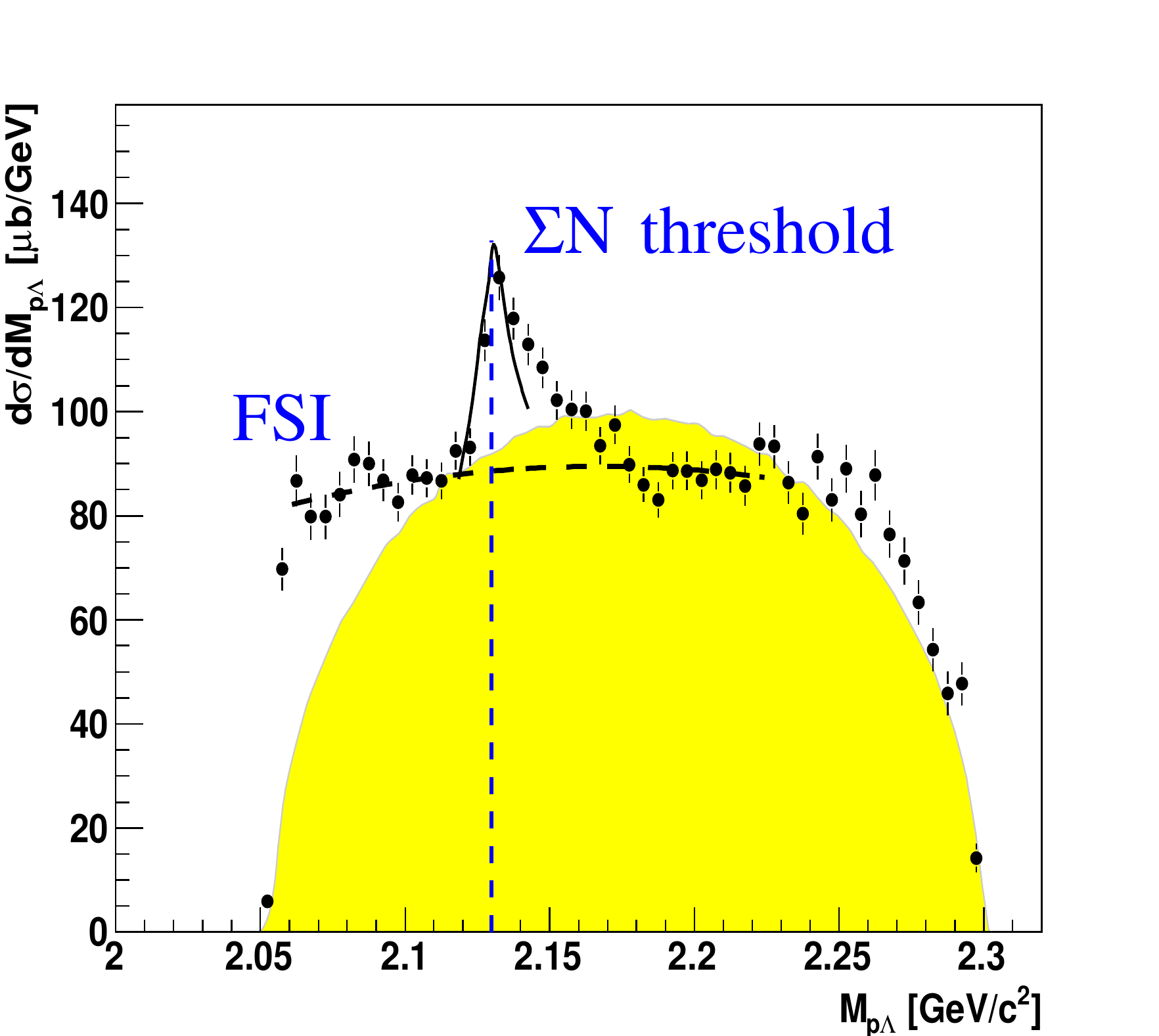}\\
 \includegraphics[width=7.4cm,clip]{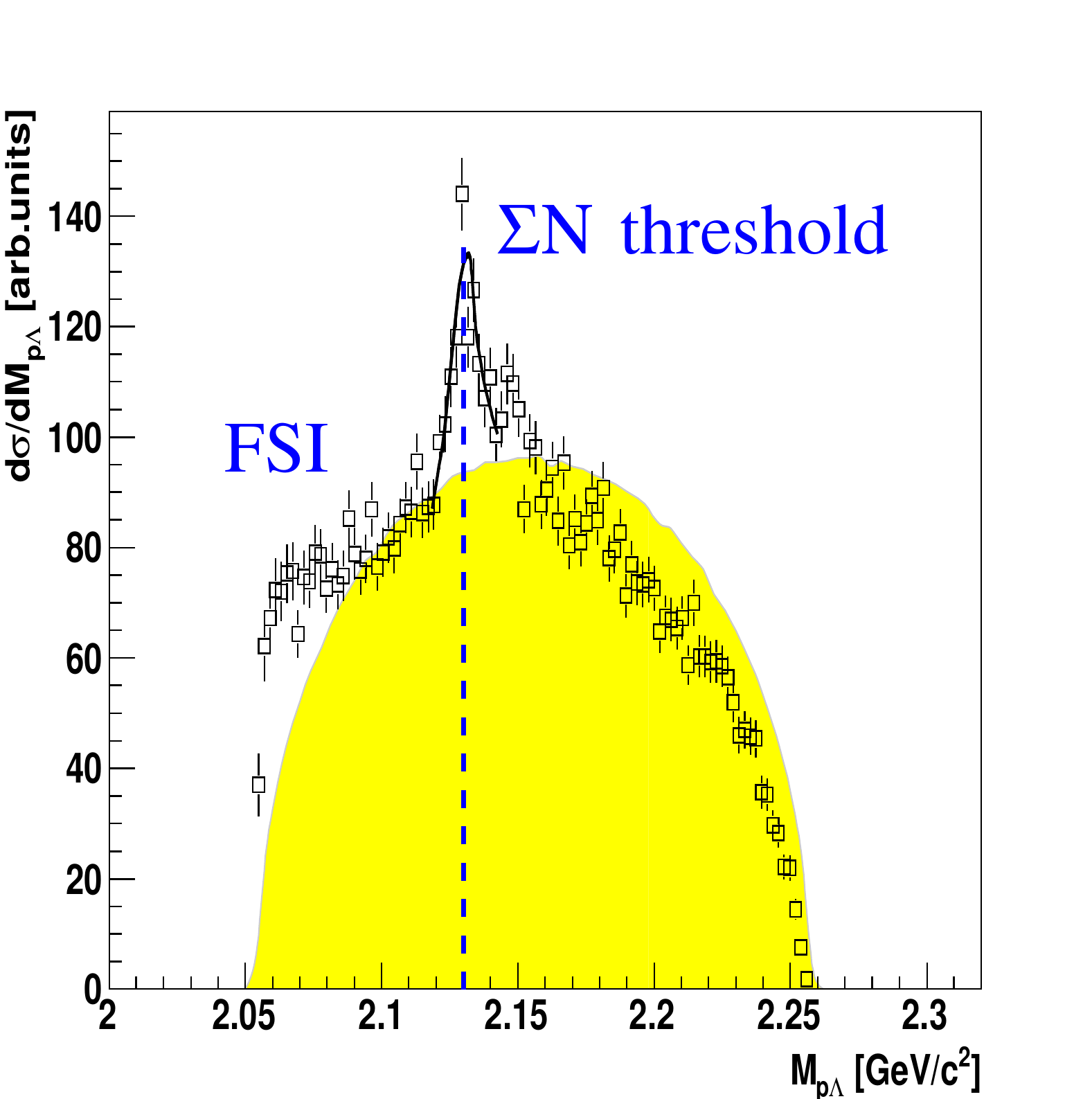}\\
 \includegraphics[width=7.4cm]{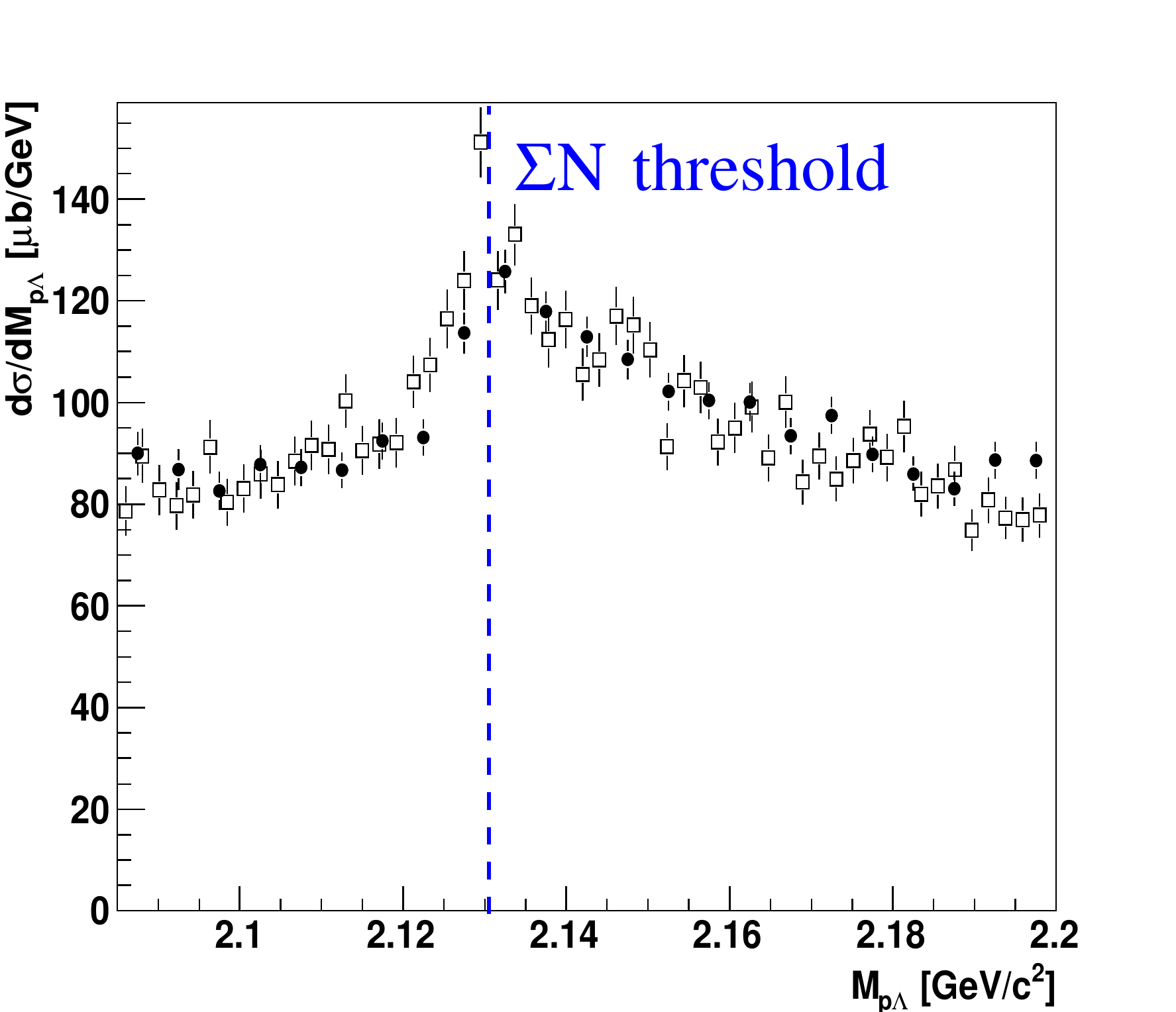}

\caption{Distributions of the $\Lambda p$ invariant mass as obtained from
  COSY-TOF 
  measurements at $T_p$ = 2.28 GeV with an invariant-mass resolution of
  $\sigma_m$ = 2.6 MeV (top) and at $T_p$ = 2.16 GeV with $\sigma_m$ = 1.1~MeV
  (middle). The shaded (yellow) area shows the distribution expected from pure
  phase space. The solid curve gives the shape of the $\Sigma N$ cusp
  distribution as obtained from a coupled-channel treatment of this phenomenon
  -- averaged over the experimental resolution and fitted in height to the
  data at the low-energy side of the cusp. The cusp is assumed to sit upon a
  smooth background represented by the dashed line. At the bottom both data
  sets are compared in the region of the cusp. Aside rom a possible slight
  shift of about 1 Mev, which is within the uncertainty of the energy
  calibrations both data sets coincide, if the different energy resolutions
  are taken into account. From
  \cite{KE}.  
}
\label{fig-TOF}       
\end{figure}

Fig.~\ref{fig-TOF} shows the $\Lambda p$ invariant-mass spectra obtained with
COSY-TOF at $T_p$ = 2.28 GEV with a mass resolution of $\sigma_m$ = 2.6 MeV
and at $T_p$ = 2.16 GeV with $\sigma_m$ = 1.1 MeV. The only obvious structures
are the enhancement at the $\Lambda p$ threshold due to the $\Lambda p$
final-state interaction and the narrow spike at the thresholds of $\Sigma^+ n$
and $\Sigma^0 p$ at 2129 MeV and 2131 MeV, respectively. A coupled-channel
treatment of the cusp due to these thresholds gives a good account for the
low-energy side of the observed structure. However, for the high-energy side
the calculation falls off a bit too fast. 

This situation is very similar to that observed in $K^-$ absorption on the
deuteron -- as discussed already in detail in section 4.2.5. In particular, the
high-resolution measurement of Tan \cite{Tan} displays also such a high-energy
shoulder. The high-resolution data shown in Fig.~\ref{fig-TOF}, middle,
indicate some narrow fluctuation below 3$\sigma$ confidence at 2.146 GeV on
this high-energy slope. A very recent high-statistics and high-resolution
measurement \cite{SJ} shows again such a fluctuation, but now at a slightly
smaller mass. From this we conclude that at present there is no statistical
solid evidence for a narrow structure beyond the $\Sigma N$ cusp. However,
there is solid evidence for a surplus of cross section at the high-energy side
of the cusp, which so far is not understood theoretically.

Also, the peak at the $\Sigma N$ threshold is observed to vanish in the $pp \to
\Lambda pK^+$ reaction towards smaller incident energies. At $T_p$ = 1.92 GeV
it is already invisible in the data \cite {FH}. There the cross section ratio
between $\Lambda$ and $\Sigma^0$ production is five times larger than at $T_p$
= 2.28 GeV due to the $\Lambda p$ FSI. This is in accord with the expection
for a cusp effect, where the channel-coupling depends on this cross section
ratio. 

In Ref. \cite{SJ} it is shown that the shallow fall-off at the high-energy
side of 
the cusp can be accounted for phenomenologically in the framework of a
Flatt$\acute e$ ansatz, if a (broad) virtual state in the $\Lambda p$ system
in the mass range 2200 - 2231 MeV is assumed. Clearly, more profound
theoretical investigations are needed.

The $pp \to \Lambda p K^+$ reaction was also measured exclusively and
kinematically complete with the HADES detector at GSI at a beam energy of 3.5
GeV 
\cite{HADESLambda}. Though the solid angle coverage is substantially less than
$4\pi$, a reliable correction based on partial-wave analysis could be
achieved. As a result \cite{Epple} no statistically significant structures for
narrow $\Lambda p$ dibaryon resonances in the mass range below 2600 MeV have
been revealed -- see Fig. 6 of Ref. \cite{Epple}.  

Summarizing, aside from the $\Lambda N$ FSI and the possibly not yet completely
understood $\Sigma N$ cusp 
effect there are no statistically significant structures giving indications
for resonances in $\Lambda N$ and $\Sigma N$ systems with masses below
2600~MeV.

\subsection{\it No signal from the anticipated $NN\pi$ Resonance $d'(2065)$}

The WASA experiment, until 2005 installed at CELSIUS and thereafter at COSY, provided the
ideal requirements for measuring multi-pion production in unprecedented
quality. Though the $\pi N$ system was known to provide a wealth of baryon
resonances and also first meaningful theoretical calculations predicted the
$\Delta\Delta$ system to play a dominant role in the two-pion production
process, the data base was still scarce at that time. A reason was that afore
no detector was available on a nucleon-nucleon machine, which was able to
cover the full phase space of complex reactions combined with high statistics.

Since $pp$ collisions are experimentally much easier to conduct than $pn$
collisions, the WASA collaboration started to 
systematically measure the $pp \to pp \pi^+\pi^-$ reaction close to
threshold. 

One of the first aims was the search for the anticipated $NN\pi$
resonance $d'(2065)$, which was introduced for the explanation of the
resonance-like forward-angle cross section observed in the pionic double-charge
exchange reaction on nuclei at incident pion energies around 50 MeV -- see
section 5. 

\begin{figure} 
\centering

 \includegraphics[width=15cm,clip]{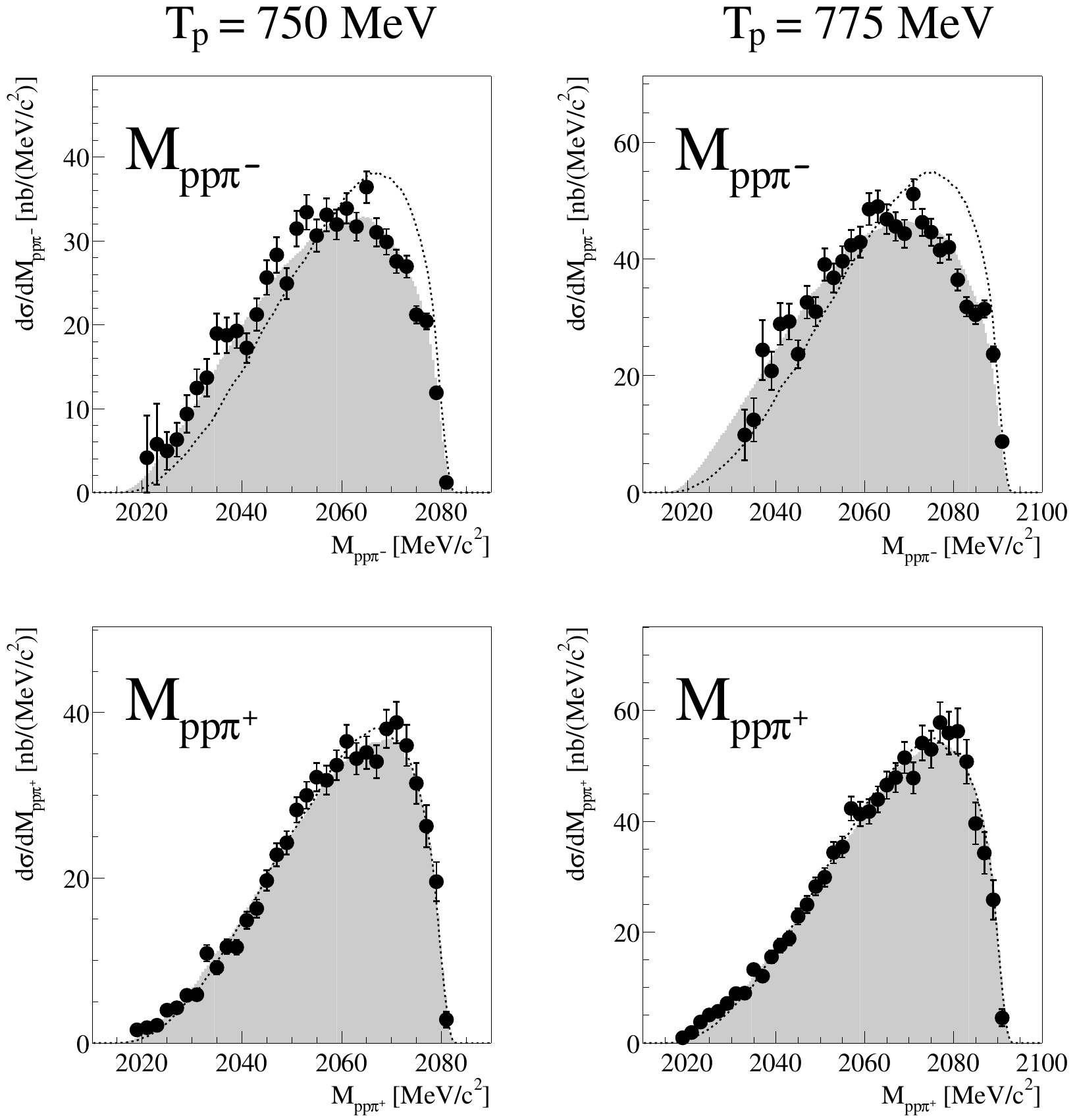}

\caption{Distributions of the $pp\pi^-$ and $pp\pi^+$ invariant masses
  $M_{pp\pi^-}$ (top) and $M_{pp\pi^+}$ (bottom), respectively, obtained from
   measurements of the $pp \to \pi^+\pi^-$ reaction at $T_p$ = 0.75 GeV (left)
   and 0.775 GeV with the WASA/PROMICE detector at CELSIUS. The solid circles
   represent the experimental results. The dotted lines indicate a pure
   phase-space distribution and the shaded areas give a model calculation
   for the excitation of the Roper resonance and its decay into the two-pion
   channel. From Ref.
  \cite{WB1}.  
}
\label{fig-d'}       
\end{figure}

And, indeed, already the first test run for measuring two-pion
production close to threshold revealed a narrow, 4 MeV broad structure in the
$pp\pi^-$ invariant mass spectrum at 2.063 GeV, {\it i.e.} right at the
position, where $d'(2065)$ was expected  \cite{WB}. However, a follow-up
high-statistics run revealed this structure to be partly a statistical
fluctuation and partly an instrumental artifact \cite{WB1}. No statistical
significant ($> 3 \sigma$) narrow structures have been observed, neither in
the $pp\pi^-$ nor in the $pp\pi^+$ spectrum, see Fig.~\ref{fig-d'}.

An upper limit of $\sigma <$ 20 nb has
been derived for the production of any narrow $NN\pi$ dibaryons in the mass
range from 2020 - 2085 MeV. This includes both isoscalar and isotensor
dibaryons. This upper limit is more than an order of magnitude smaller than
estimated \cite{Schepkin3,hcl} for the $d'(2065)$ production in this reaction.
Also a measurement of the $pp \to pp\pi^+\pi^-$ reaction at $T_p$ = 0.793 GeV
at COSY-TOF, which covers $pp\pi$ invariant masses up to 2095 MeV, finds no
evidence for a narrow dibaryon strcuture \cite{AE,AEDiss}.

In conclusion Ref. \cite{WB1} states that
\begin{itemize}
\item either $d'$ does not exist at all, or
\item its production cross section in $pp$ collisions is smaller than
  expected from theoretical estimates \cite{Schepkin3,hcl}, or
\item the mass of the free $d'$ is outside the range investigated here, or
\item it exists only in the nuclear medium. 
\end{itemize}

Effects of the nuclear medium on the effective mass of dibaryons with
particular emphasis on $d'$ have been investigated in Ref. \cite{FBK}. There a
decreasing dibaryon mass with increasing nuclear density is predicted. The
data on the pionic double-charge exchange reaction, indeed, exhibit some
tendency of a 
decreasing peak energy, {\it i.e.} of a decreasing $d'$ mass with an increasing
mass number, {\it i.e.} increasing density of the nucleus
\cite{Draeger}. According to Ref. \cite{FBK} the vacuum dibaryon mass is
expected to be about 15$\%$ larger than that at nuclear density $\rho_0$, {\it
  i.e.} in the interior of $^{208}$Pb. Since the pionic double-charge
exchange reaction occurs predominantly at the nuclear surface, {\it i.e.} at
roughly $\rho_0 / 2$, this would mean that the vacuum mass of $d'$ is expected
to be near 2200 MeV, which is already far above the $NN\pi$ threshold. In
consequence $d'$ could decay easily into the $NN\pi$ system and hence would no
longer appear as a narrow resonance.

\subsection{\it No dibaryon signal in $pp$-induced two-pion production}

\begin{figure} 
\centering

 \includegraphics[width=7cm,clip]{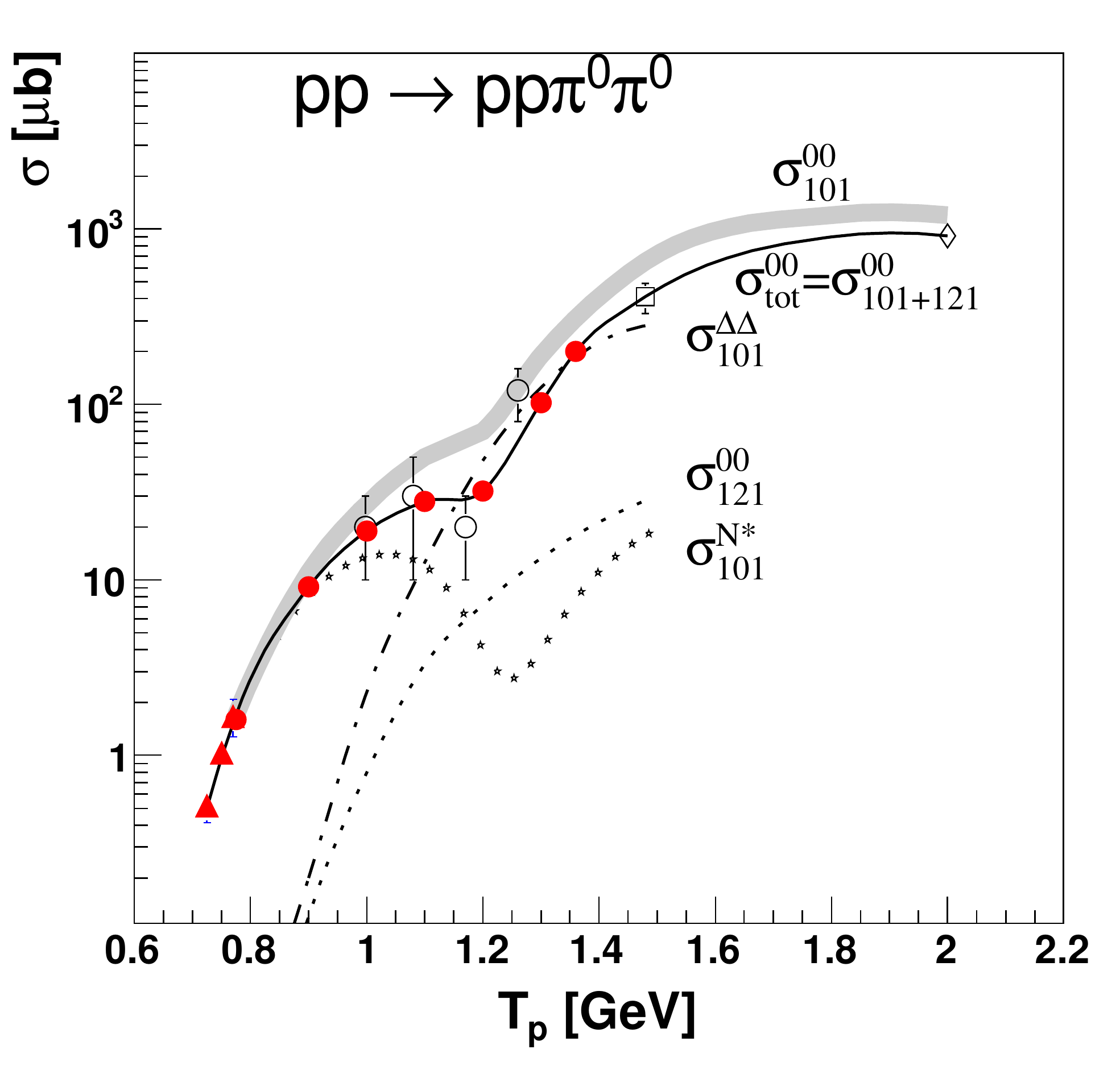}
  \includegraphics[width=7cm,clip]{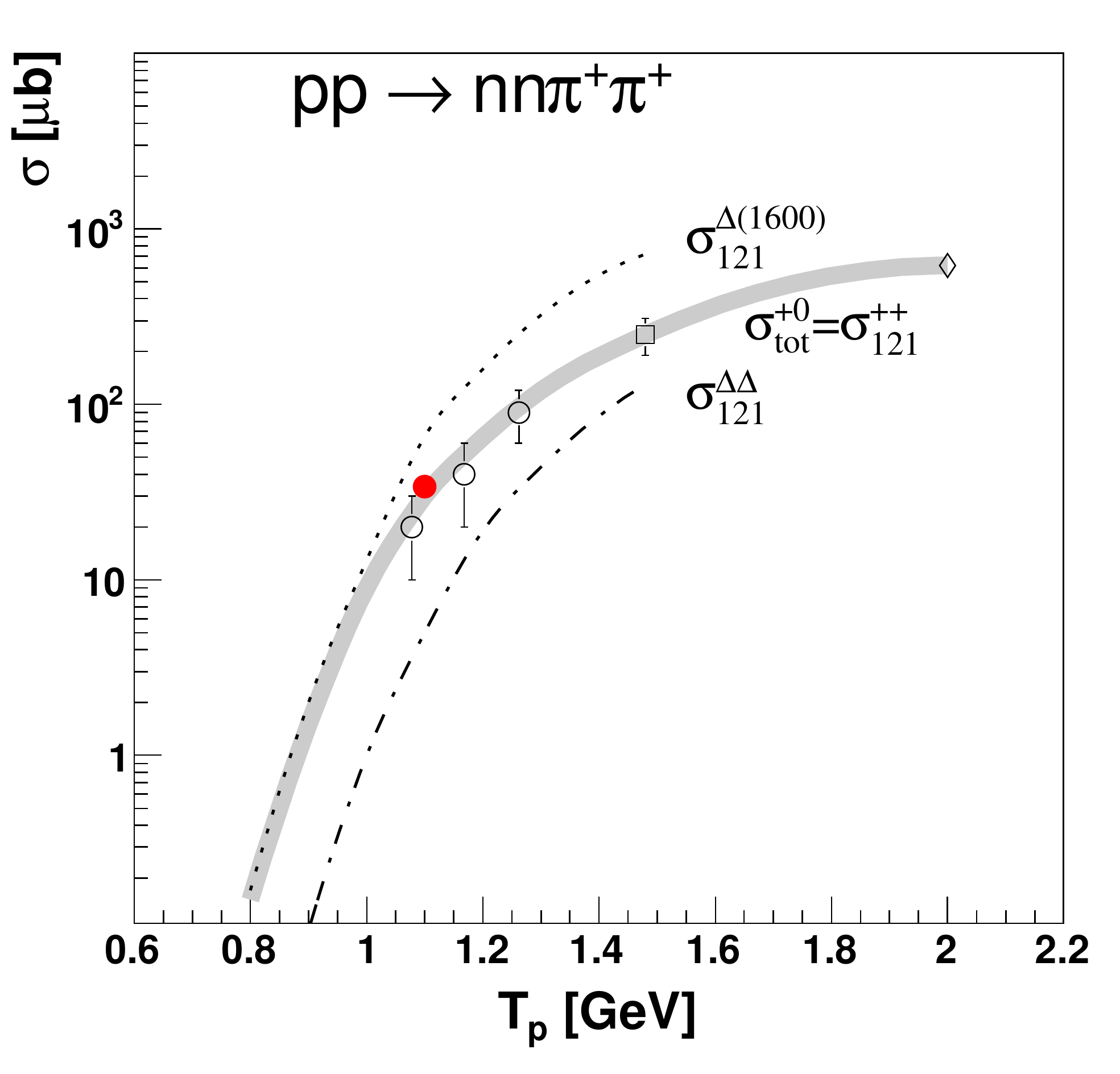}
\includegraphics[width=7cm,clip]{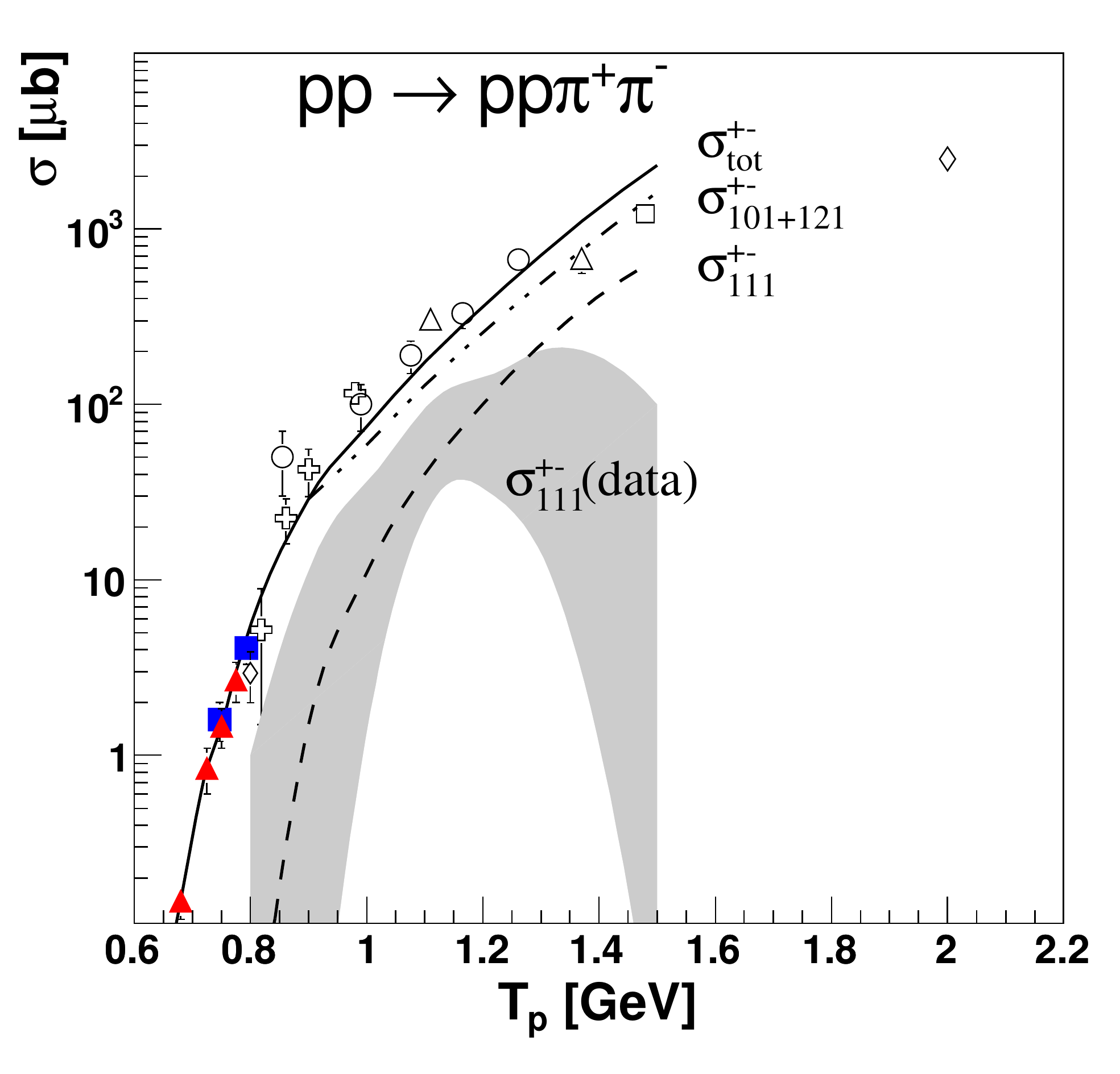}
 \includegraphics[width=7cm,clip]{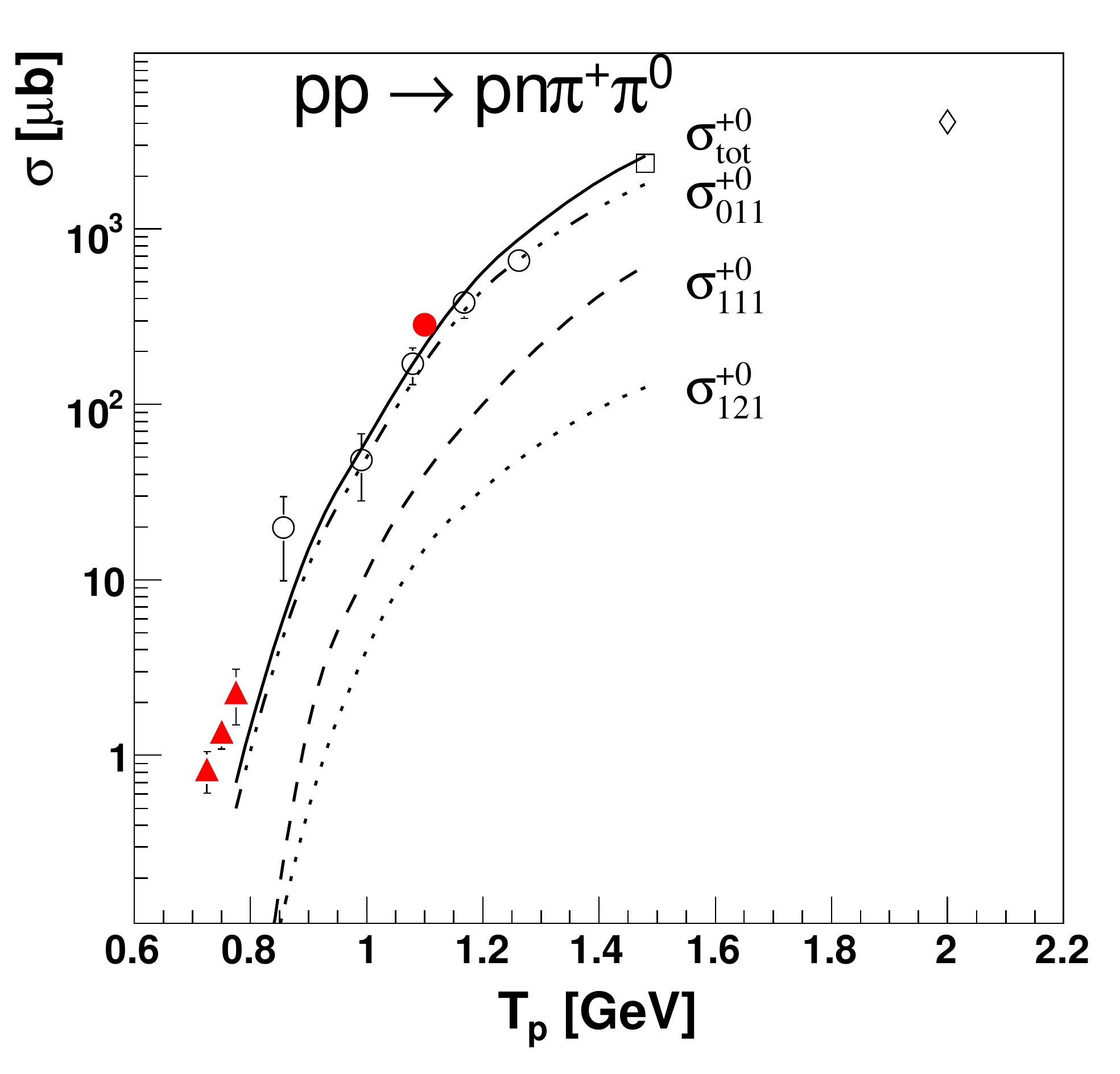}

\caption{Energy dependence of the $pp$-induced total cross sections for
  two-pion production and their isospin decomposition. Open symbols denote
  results from bubble-chamber measurements
  \cite{Dakhno,Brunt,Shimizu,Sarantsev,Eisner,Pickup}, solid symbols represent
  measurements from WASA/PROMICE (triangles) \cite{WB2,JJ,JP}, CELSIUS/WASA
  (circles) 
    \cite{TS,iso,FK,deldel,nnpipi} and COSY-TOF (squares) \cite{AE}. At the
    top the reaction 
    channels 
    $pp \to pp\pi^0\pi^0$ (left) and $pp \to nn\pi^+\pi^+$ (right) are shown
    together with their isospin decomposition. the drawn lines show the
    extracted contributions $\sigma_{121}$, $\sigma_{101}$ and
    $\sigma_{101+121}$ as indicated in the figures. Also shown are the
    decomposition of  $\sigma_{121}$ into contributions from $\Delta\Delta$
    and $\Delta(1600)$ and of  $\sigma_{101}$ into $N^*$ and $\Delta\Delta$
    contributions. At the bottom the reaction channels $pp \to pp\pi^+\pi^-$
    (left) and $pp \to pn\pi^+\pi^0$ (right) are shown . The broken lines show
    the extracted contributions of $\sigma_{121}$, $\sigma_{101}$ and
    $\sigma_{101+121}$. The large shaded area indicates $\sigma_{111}$ as
    derived by direct comparison of the data for the $pp \to pp\pi^+\pi^-$
    reaction with $\sigma_{101+121}$.
 From Ref. \cite{iso}.  
}
\label{fig-pppipi}       
\end{figure}

Before the start of the two-pion production programs at CELSIUS and COSY there
was only very little experimental information on this process. Most of the
previous results originated from low-statistics bubble-chamber
measuerements
\cite{Dakhno,Brunt,Shimizu,Sarantsev,Eisner,Pickup}. Fig.~\ref{fig-pppipi}
shows their 
results for the total cross sections by the open symbols. Due to the very
limited statistics differential distributions were even more scarce and only
available for the $pp\pi^+\pi^-$ channels at energies far above threshold.

In this situation the two-pion production program started at CELSIUS with
exclusive kinematically complete high-statistics measurements with particular
emphasis on the threshold region, the most suitable place for revealing narrow
dibaryon resonances. Starting first with the WASA/PROMICE detector setup,
followed by the completed WASA detector at CELSIUS and finally at COSY
 all $pp$-induced, {\it i.e.} isovector two-pion
production channels from threshold up to $\sqrt s$ = 2.5 GeV have been studied
\cite{WB2,JJ,JP,TS,iso,FK,deldel,nnpipi,tt} complemented by a corresponding
program with polarized beam at COSY-TOF \cite{AE,evd}. The results for the
total cross sections are shown in Fig.~\ref{fig-pppipi} by solid
symbols. Special emphasis 
has been put on the $pp\pi^0\pi^0$ channel, where the bubble-chamber results
suggested a kink in the total cross section at $T_p$ = 1.2 GeV. The WASA
results confirmed this kink in high-statistics measurements. In subsequent
isospin decomposition of the total cross sections according to the formalism
presented in Refs. \cite{Dakhno,Bystricky}  it could be demonstrated
\cite{iso} that
excitation of the Roper resonance by $t$-channel meson exchange and its
subsequent decay into the $N\pi\pi$ channel as well as the mutual excitation
of the colliding nucleons into their 
first excited state, the $\Delta$ resonance with subsequent $N\pi$ decay
($t$-channel $\Delta\Delta$ excitation), are the dominant processes -- as
properly predicted by the Valencia theory group \cite{Luis} and later-on also
by IHEP calculations \cite{Zou}. The $\Delta\Delta$ excitation by conventional
$t$-channel meson exchange is depicted schematically in Fig.~\ref{fig-graphs},
top.  

In this isospin decomposition the total cross sections are
decomposed into contributions from reduced matrix elements
$M_{I_{NN}^fI_{\pi\pi}I_{NN}^i}$, where $I_{NN}^i$ and $I_{NN}^f$ denote the
isospin of the nucleon pair in initial and final state, respectively, and
where $I_{\pi\pi}$ is the isospin of the produced pion pair.     
    For a specific process these matrix elements depend on the isospin
    coupling coefficients. For the $\Delta\Delta$ process, {\it e.g.} the
    matrix elements 
    are proportional to the respective 9j-symbol for isospin recoupling: 
\begin{eqnarray}
M_{I_{NN}^fI_{\pi\pi}I_{NN}^i}^{\Delta\Delta} \sim \hat{I}_{\Delta_1}\hat{I}_{\Delta_2}
\hat{I}_{NN}^f\hat{I}_{\pi\pi}
\left\{
\begin{array}{ccc}
I_{N_1}&I_{\pi_1}&I_{\Delta_1}\\
I_{N_2}&I_{\pi_2}&I_{\Delta_2}\\
I_{NN}^f&I_{\pi\pi}&I_{\Delta\Delta}\\
\end{array}
\right\},
\end{eqnarray}
      where $N_i$ and $\pi_i$ couple to $\Delta_i$ for $i = 1,2$ and
      $\hat{I_\alpha} = \sqrt{2I_\alpha + 1}$ and $I_{\Delta\Delta} =
      I_{NN}^i$ -- as outlined in Refs. \cite{iso,DLS}.

The resulting
partial cross sections $\sigma_{101}$, $\sigma_{011}$, $\sigma_{111}$ and
$\sigma_{121}$ are shown in Fig.~\ref{fig-pppipi} by the drawn lines and shaded areas,
respectively. The isospin decomposition suggests that also a higher-lying
$\Delta$ excitation, preferably the $\Delta(1600)$ could play some role, in
particular in the $nn\pi^+\pi^+$ channel. The isospin decomposition reveals
the kink in the total cross section of the $pp\pi^0\pi^0$ channel as being due
to destructive interference effects between isoscalar and isotensor $\pi\pi$
contributions as well as between $N^*$ and $\Delta\Delta$ contributions. {\it
  I.e.}, in this isospin decomposition concept the kink does not appear to be a
signature of the contribution of two dibaryon 
resonances ($\Delta N$ configurations in $^3F_3$ and $^1G_4$ $pp$ partial
waves), as assumed recently \cite{Platonova}. However, it will be very
interesting to see, whether the alternative concept \cite{Platonova} of
isovector dibaryon resonance 
contributions in combination with low cut-off parameters as discussed at the
end of section 4.4 for single-pion production will lead also here to a
quantitative description of data on all isovector two-pion production channels.

Still much more information about the reaction process is, of course,
contained in 
the various differential distributions, which all are available in kinematically
complete measurements. By fine adjustments of the resonance parameters in the
Valencia model calculations all differential distributions could be
quantitatively 
described by the conventional process of $t$-channel meson exchange leading to
the excitation of the Roper resonance close to threshold followed at higher
energies by the excitation of the $\Delta\Delta$ system. It has been
demonstrated that the Roper excitation and its decay via the interfering
routes $N^* \to N\sigma \to N\pi\pi$ and $N^* \to \Delta\pi \to N\pi\pi$ lead to
very characteristic patterns in the differential distributions, especially in the
$\pi\pi$ invariant-mass distribution \cite{AE,WB2,JP,TS}. In the $\Delta\Delta$
region it is in addition the $p\pi$ invariant mass distributions, which
exhibit a characteristic pattern \cite{deldel}.

As a result of these
systematic studies it was found that isovector induced two-pion production can
be quantitatively well understood by the
conventional process of $t$-channel meson exchange leading to the excitation
of the Roper resonance close to threshold followed by the excitation of the 
$\Delta\Delta$ system at higher energies. To some extent also the
$\Delta(1600)$ excitation is seen to play some role. But no hint for an
exotic resonance production is observed.

\section{Towards the First Extraordinary Dibaryon}
\label{sec-5}

The situation changed drastically, when $pn$-induced two-pion production was
looked at, though the pathway towards a dibaryon resonance was still far from
being straightforward. 

\subsection{\it Following a trace: the ABC effect}

Since the sixties there existed an unsolved problem concerning the two-pion
production in cases, when the participating nucleons merge into a nuclear
boundstate -- the so-called double-pionic fusion reactions.

In 1960 Abashian, Booth and Crowe \cite{ABC} noticed in the inclusively
measured $pd \to ^3$HeX reaction that there is a substantial enhancement in
the $^3$He missing mass spectrum corresponding to the emission of two
pions. This enhancement cumulated at masses close to the threshold of twice the
pion mass. Subsequent measurements demonstrated that this effect --
corresponding to a low-mass enhancement in the spectrum of the $\pi\pi$
invariant mass -- 
occurs in the double-pionic fusion reactions $pn \to d\pi\pi$, $pd \to
^3$He$\pi\pi$  and $dd \to ^4$He$\pi\pi$, but not in the fusion to $^3$H in
the reaction $pd \to ^3$H$\pi\pi$
\cite{bar,abd,hom,hal,ba,ban,ban1,plo,Colin,wur,cod}. Since in the latter case
the pion pair must 
be in an isovector state, it was concluded that the observed enhancement must
be correlated with the creation of an isoscalar pion pair. 

First attempts for explaining this effect by an unusually strong $\pi\pi$
interaction or a narrow $\sigma$ meson had soon to be abandoned due to conflicts
with experimental results from other two-pion production processes.

\begin{figure} 
\centering

 \includegraphics[width=8cm]{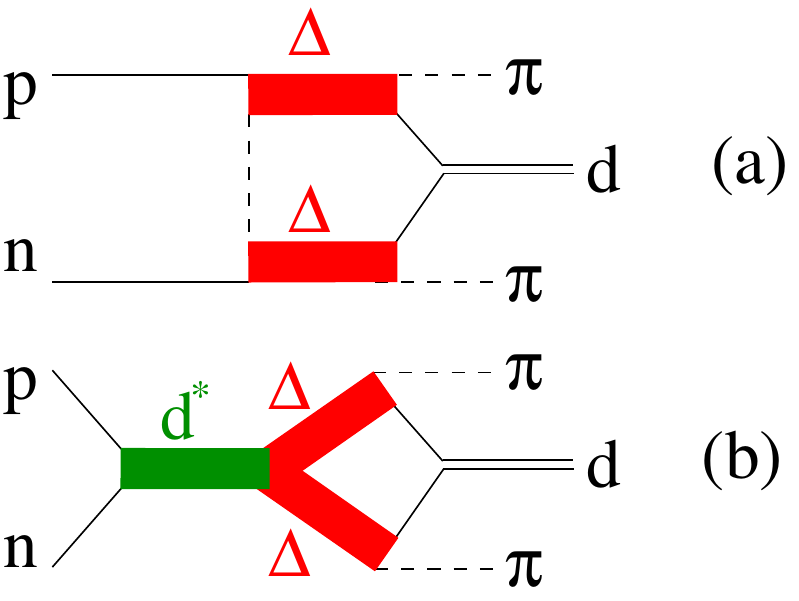}

\caption{Schematic representation of the process $pn \to d\pi\pi$ by
  $\Delta\Delta$ excitation via $t$-channel meson exchange (a) or by
  $s$-channel formation of the $d*(2380)$ resonance with subsequent decay via
  an intermediate $\Delta\Delta$ system (b). 
From \cite{abc}.  
}
\label{fig-graphs}       
\end{figure}

Since this low-mass enhancement was especially pronounced at beam energies,
where the colliding nucleons could undergo a mutual excitation into their first
excited state, the $\Delta(1232)$, it was finally proposed \cite{Risser} that
the conventional $\Delta\Delta$ excitation by $t$-channel meson exchange -- as
depicted schematically in Fig.~\ref{fig-graphs}, top -- might
be the reason for the observed low-mass enhancement. However, according to
this explanation there should also be a high-mass enhancement in the $\pi\pi$
invariant-mass spectrum. And indeed, a number of inclusive measurements 
exhibited such a putative enhancement. But as pointed out in Ref. \cite{Colin}
this high-mass region region corresponded also to the kinematical region of
$\eta$ and three-pion production, which could not be separated in the
inclusive measurements.

So, since finally no conclusive explanation could be found for the observed
low-mass enhancement it was named in the literature "ABC" effect after the
initials of the authors Abashian, Booth and Crowe, who noticed this
enhancement first.

Since previous measurements either were low-statistics bubble-chamber or
inclusive single-arm measurements conducted at a few scattering angles, the
CELSIUS/WASA collaboration decided to reexamine the ABC effect -- with the WASA
detector setup still at CELSIUS -- and to perform exclusive and kinematically
complete measurements of the double-pionic fusion reactions.

In a first step the reactions $pd \to ^3$He$\pi^0\pi^0$ and $pd \to
^3$He$\pi^+\pi^-$ were looked at, for which data already existed from runs
dedicated to the $\eta$ production at threshold.  As a result it was found
\cite{MB} that there is, indeed, a highly visible low-mass enhancement but no
high-mass enhancement as seen in previous inclusive measurements. As already
suspected in Ref. \cite{Colin} this putative high-mass enhancement was in
reality due to $\eta$ and three-pion production. 

Having now an established strong low-mass enhancement, but the $t$-channel
$\Delta\Delta$ 
concept no longer as a valid explanation, it was suggested that the ABC
effect  might be a sign of a quasi-bound $\Delta\Delta$ system
\cite{MB}. However, if this denotes a resonance 
phenomenon in the system of two baryons, then it would need to show up also as
a resonance structure in  the total cross section of the basic double-pionic
fusion, which is the one to the deuteron.

\subsection{\it Exclusive and Kinematically Complete Measurement of the Basic
  Double-Pionic Fusion Process}

For this endeavor the experimental technique had to be
changed, since free neutrons are not available -- neither as beam or as target
particles. Hence the quasi-free process with deuterons being the
source for quasi-free neutrons has been utilized by the WASA collaboration --
a method established before in many other 
experiments, see {\it e.g.} Refs. \cite{EK,HC}. This process has the
additional advantage that the Fermi motion of the neutron within the deuteron
provides a range of collision energies with a single beam energy setting. That
way the energy dependence of a reaction can be conveniently scanned, which is
particularly well suited for the search for narrow resonances. However, a
precondition for a successful use of the quasi-free process is that the
four-momenta of {\bf all} ejectiles are determined experimentally, which
necessitates exclusive and kinematically complete measurements. 

First such measurements were conducted with WASA still at CELSIUS. Though the
statistics was quite limited there due to low beam current, a clear
correlation of the ABC effect with 
a peak-like structure in the total cross section (open triangles in
Fig.~\ref{fig-isofus}, middle panel) could be revealed in the $pn
\to d\pi^0\pi^0$ reaction, which was measured via the quasi-free process $pd
\to d\pi^0\pi^0 + p_{spectator}$ \cite{prl2009}. As it turned out later, this
was the golden channel for the dibaryon issue, since it possesses a very low
background from conventional processes. Experimentally it was only accessible
with instruments like WASA being able to detect both charged and uncharged
particles over essentially the full solid angle. Hence it is not of a
surprise that there are no data for this channel from previous experiments.

\begin{figure} 
\centering
\includegraphics[width=10.7cm,clip]{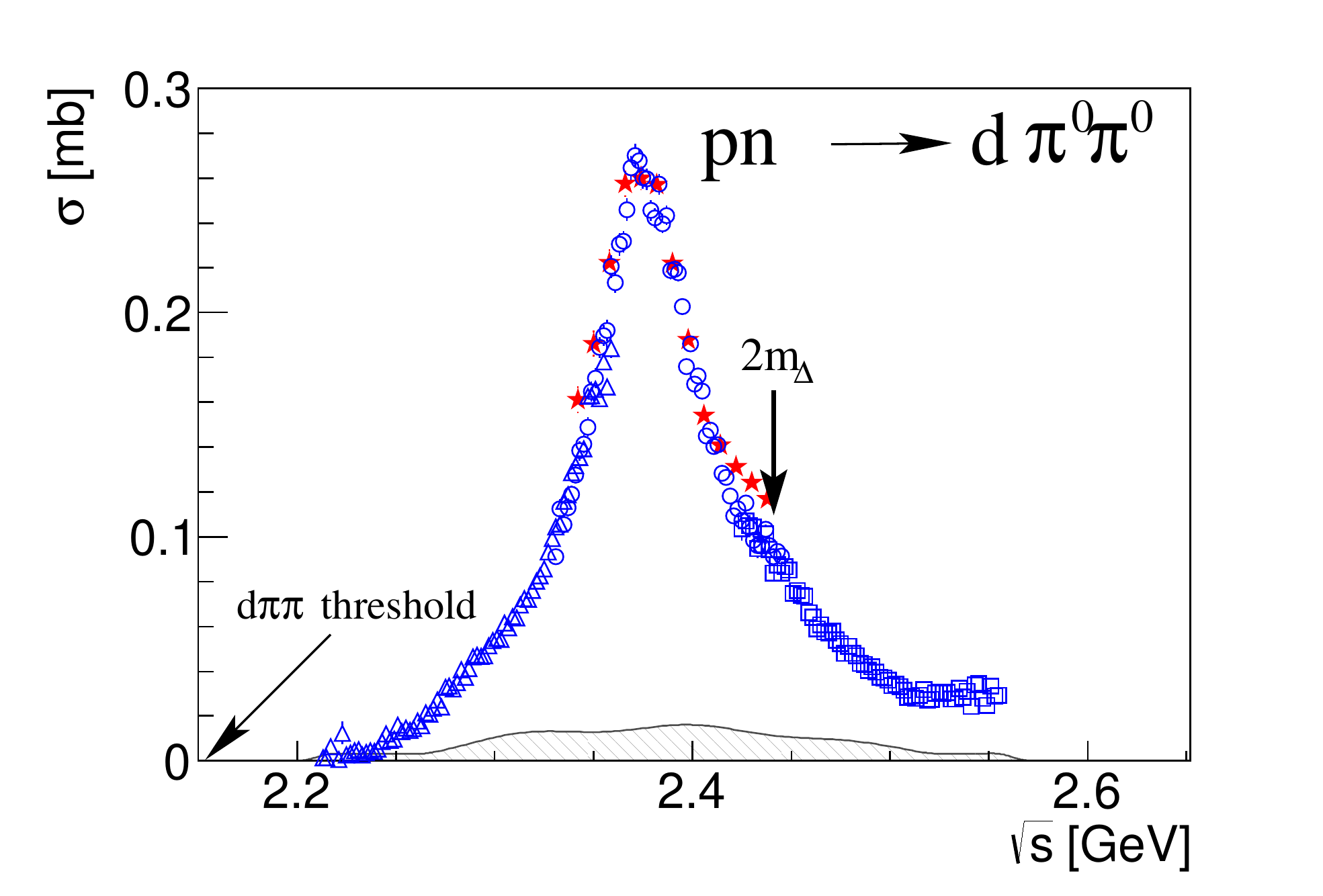}\\
\includegraphics[width=6.27cm,clip]{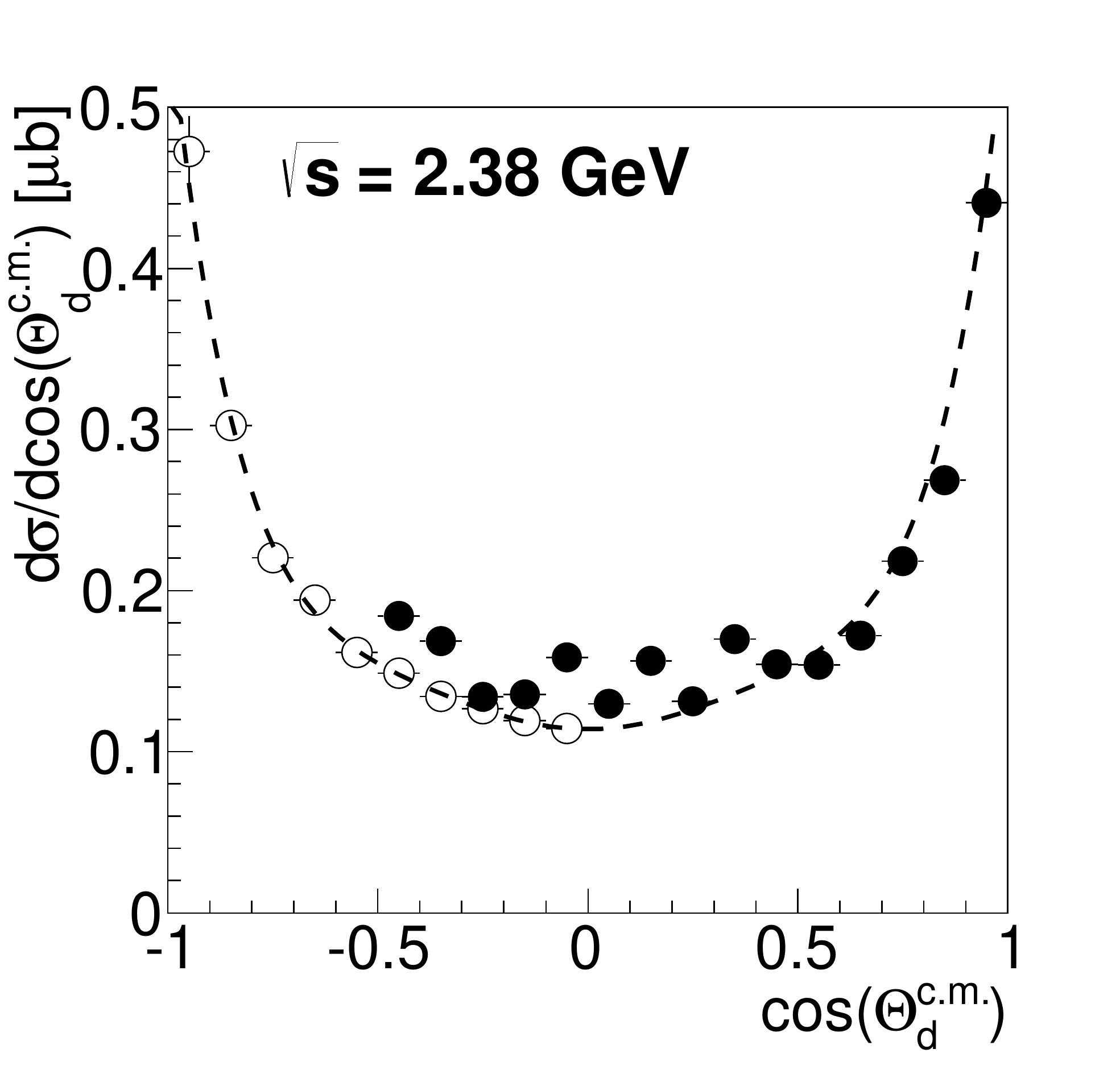}
\includegraphics[width=7cm,clip]{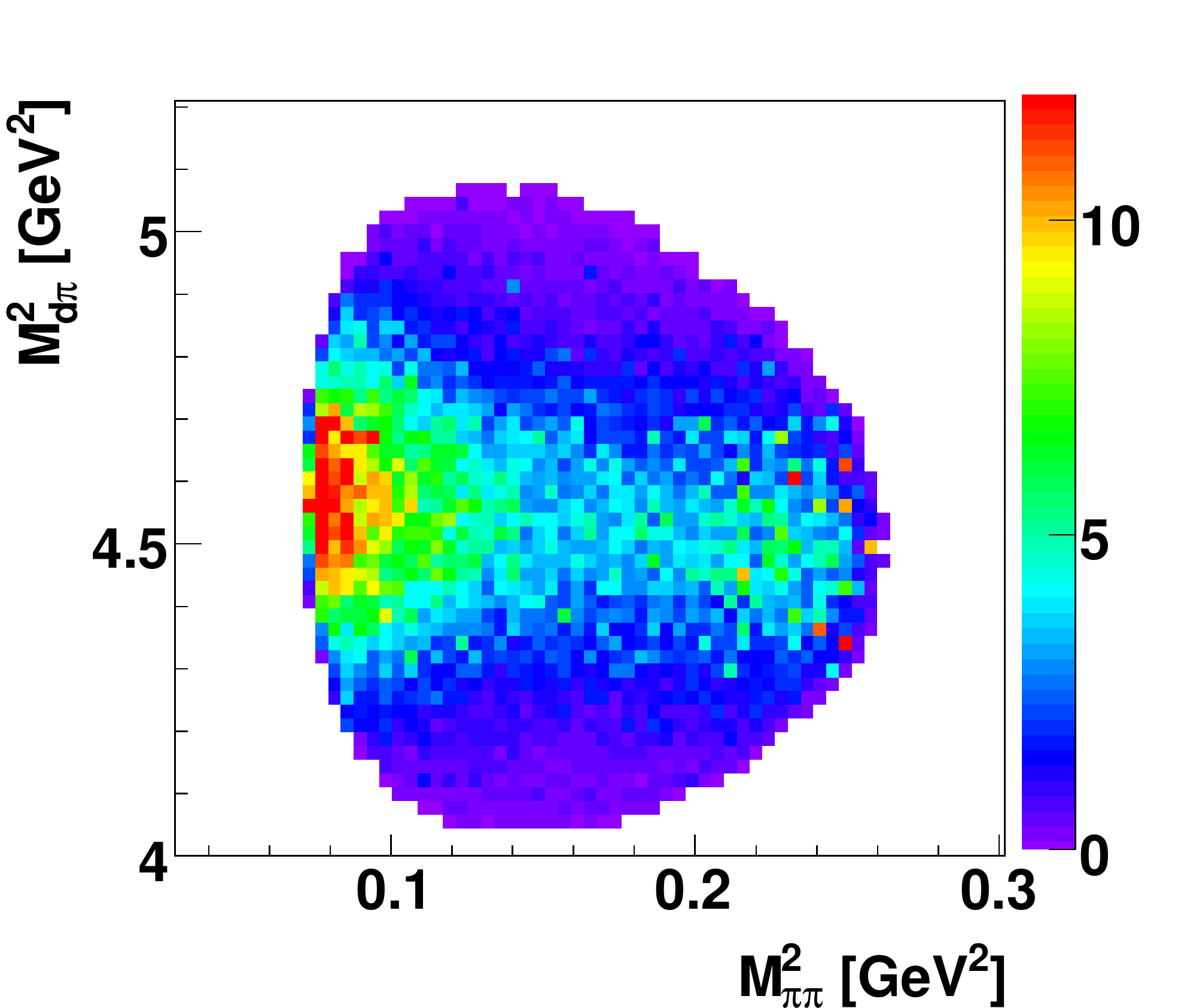}
\includegraphics[width=6.7cm,clip]{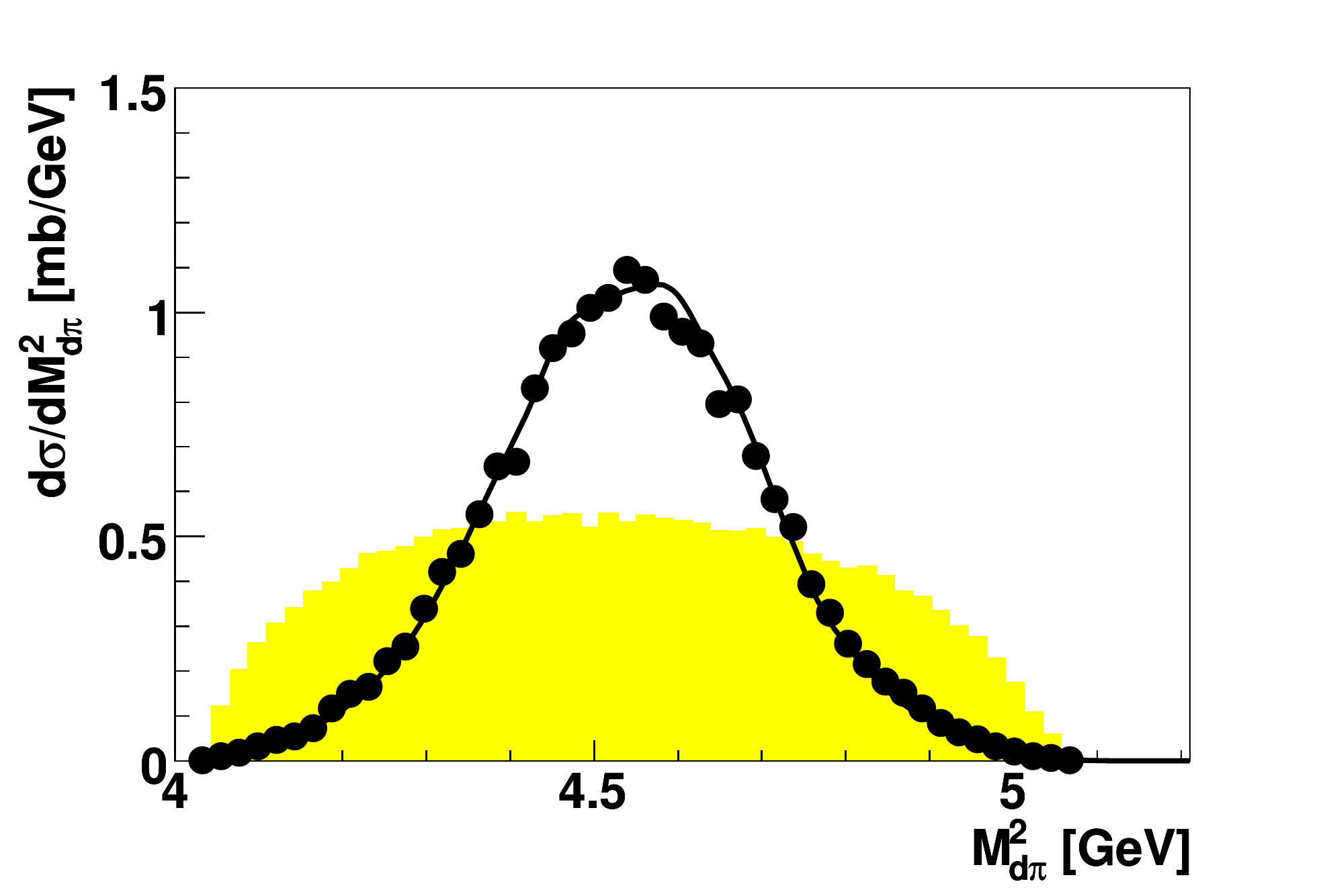}
\includegraphics[width=6.7cm,clip]{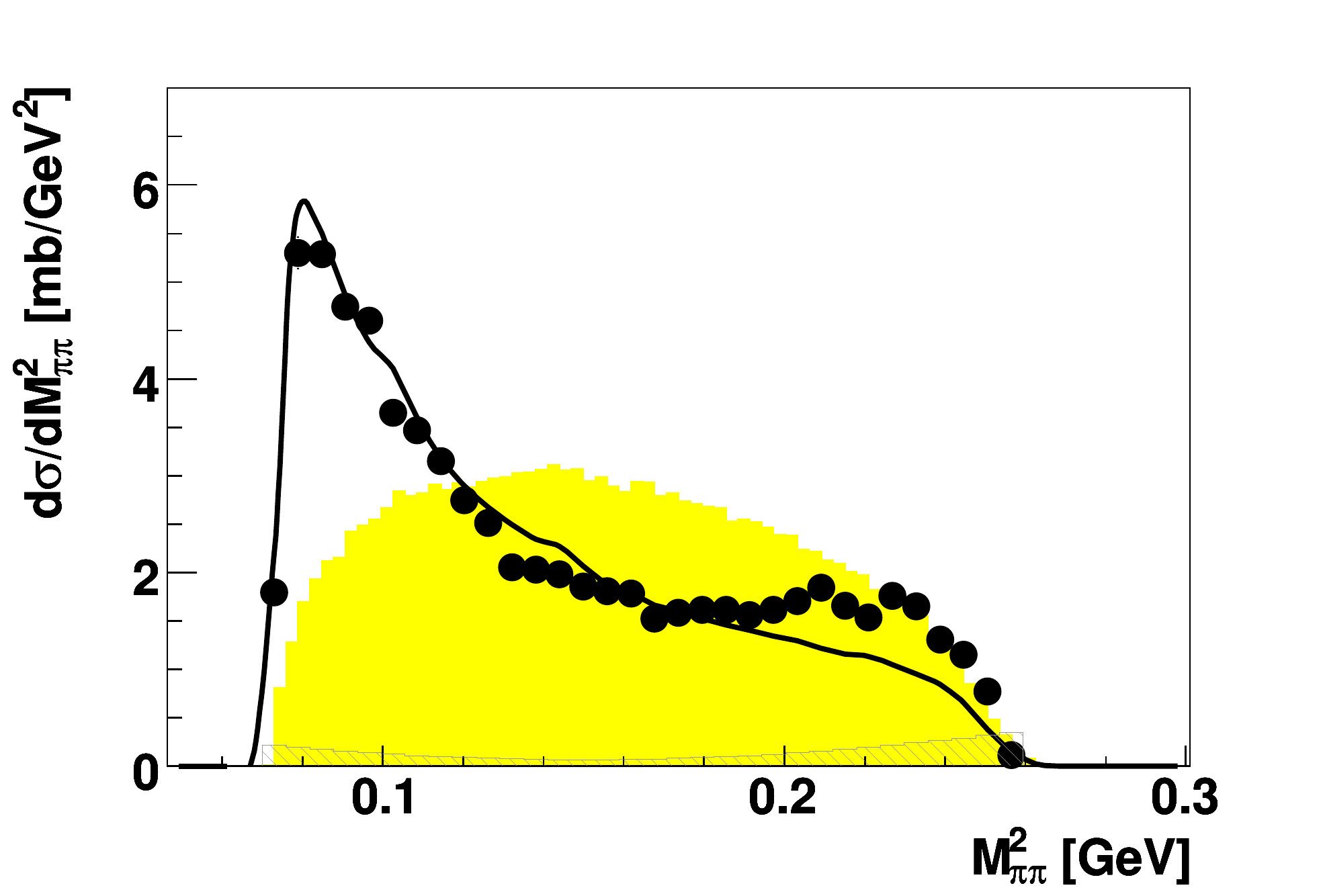}
\caption{Measurements of the golden reaction channel $pn \to d\pi^0\pi^0$
  with WASA at COSY. 
  Top: total cross section exhibiting the pronounced
  resonance effect of $d^*(2380)$. The blue open symbols show the data of
  Ref.~\cite{prl2011} properly normalized in absolute scale to the data  of
  Ref.~\cite{isofus} plotted by red stars. The black shaded area gives an
  estimate of systematic uncertainties.
  Middle: deuteron angular distribution (left) and Dalitz plot (right) at the
  peak energy of $\sqrt s$ = 2.38 GeV. Open and solid dots refer to
  measurements with the spectator proton in the target and in the beam
  (reversed kinematics), respectively. The dashed curve gives a Legendre fit
  with $L_{max}$ = 6 corresponding to J~=~3.
  Bottom: Projections of the Dalitz plot yielding the distributions of the
  squares of the $d\pi^0$ (left) and $\pi^0\pi^0$ (right) invariant
  masses. The large low-mass enhancement in the latter denote the so-called
  ABC effect. The solid lines represent a theoretical description of the
  process $pn \to d^*(2380) \to \Delta^+\Delta^0 \to d\pi^0\pi^0$. From
  \cite{prl2011,Catania2014,poldpi0pi0}.
}
\label{fig-dpi0pi0}       
\end{figure}

\begin{figure} 
\centering
\includegraphics[width=9cm,clip]{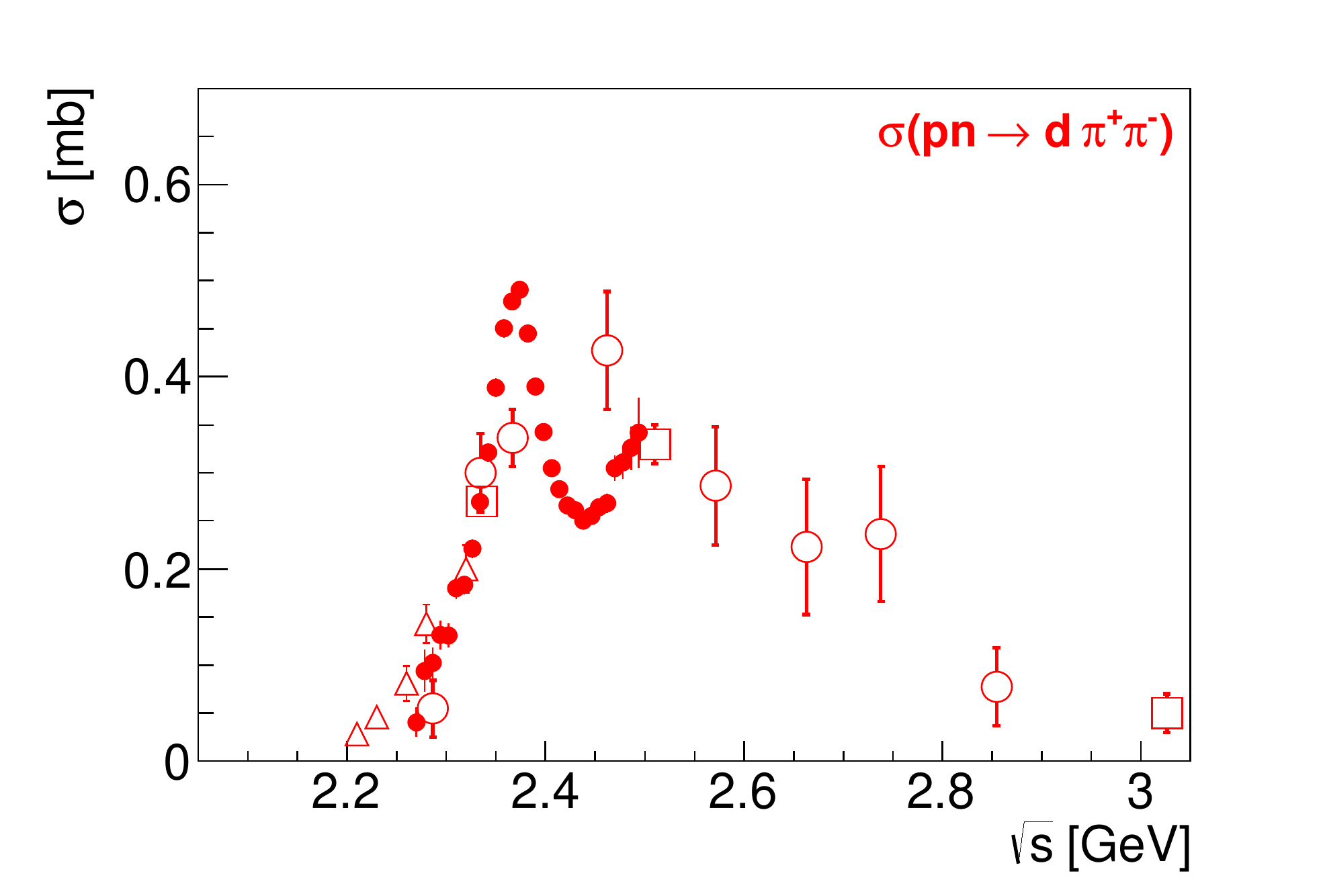}\\
\includegraphics[width=9cm,clip]{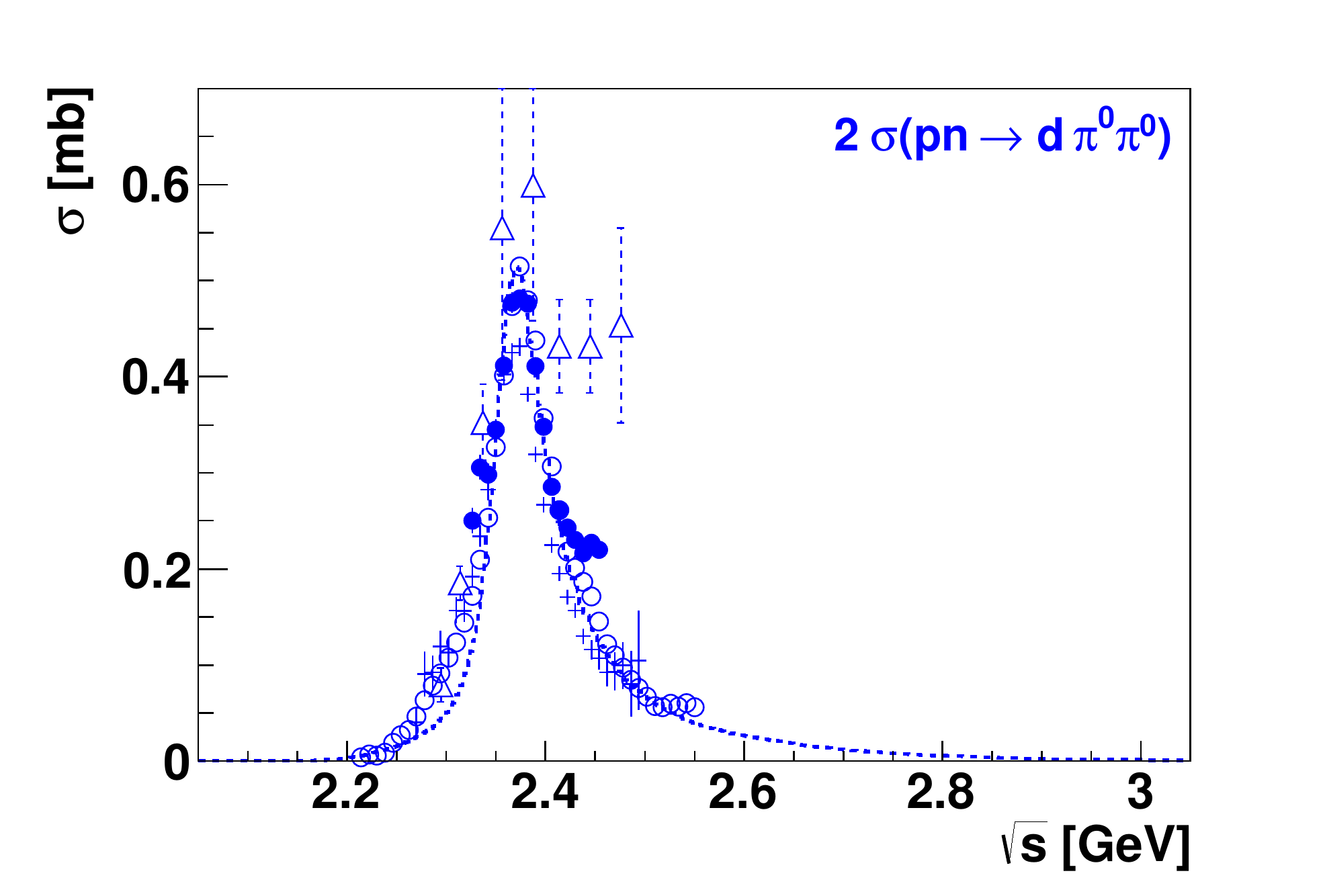}\\
\includegraphics[width=9cm,clip]{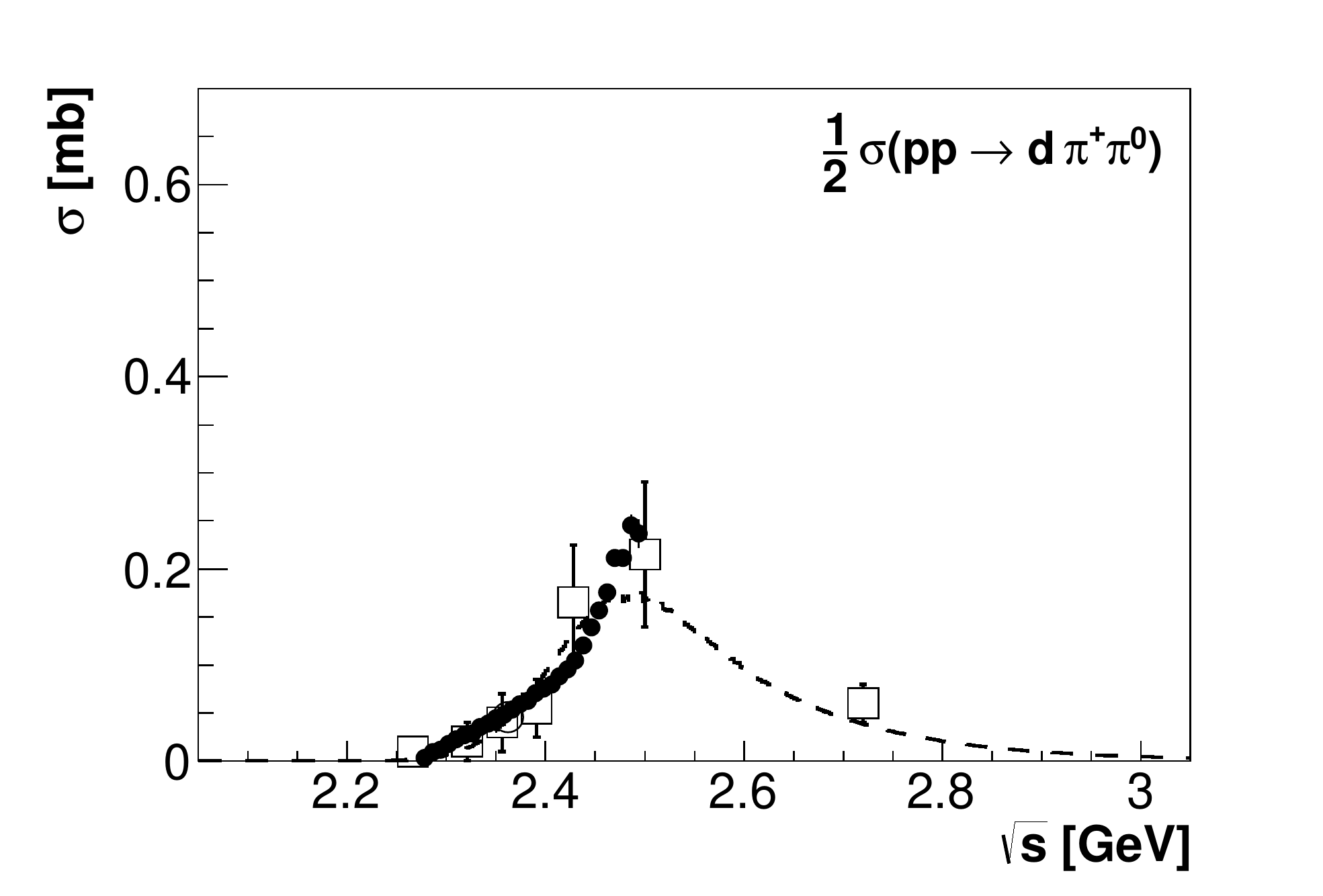}
\caption{
Total cross sections of the double-pionic fusion reactions to deuterium and
their isospin decomposition. Top: the isospin-mixed reaction $pn \to
d\pi^+\pi^-$. Solid dots represent WASA-at-COSY measurements, open symbols
previous bubble-chamber measurements at DESY (circles) \cite{desy}, Dubna
(squares) \cite{dubna} and Gatchina (triangles) \cite{Dakhno}. Middle: the
isoscalar 
reaction $pn \to d\pi^0\pi^0$ plotted as the isoscalar part of the $pn \to
d\pi^+\pi^-$ reaction. The CELSIUS/WASA results \cite{prl2009} are shown by open
triangles. The other symbols refer to WASA-at-COSY measurements
\cite{prl2011,isofus}. The dotted line gives a Lorentzian with m = 2380 MeV
and $\Gamma$ = 70 MeV. Bottom: the isovector reaction $pp \to d\pi^+\pi^0$
plotted as the isovector part of the $pn \to d\pi^+\pi^-$ reaction. Solid dots
represent WASA-at-COSY results, open symbols previous results from
\cite{Shimizu,FK} as well as those given in Ref.~\cite{Bystricky}. The dashed
line shows a $t$-channel $\Delta\Delta$ calculation fitted in height to the
data \cite{FK}. From \cite{isofus}.
}
\label{fig-isofus}       
\end{figure}

Follow-up measurements of this reaction -- meanwhile with WASA at COSY providing two
orders of magnitude higher statistics -- revealed a pronounced Lorentzian
structure in the total cross section corresponding to a resonance at $\sqrt s$
= 2.37 GeV with a width of only 70 MeV \cite{prl2011}, see
Fig.~\ref{fig-dpi0pi0}.  
The resonance structure is about 90 MeV below the mass of two $\Delta$ states
and its width is more than three times narrower than that of a conventional
$t$-channel $\Delta\Delta$ excitation.

Fig.~\ref{fig-dpi0pi0} shows the measurement of the total cross section as
well as a selection of differential distributions at $\sqrt s$ = 2.38 GeV, the
peak energy region. The shown data for the deuteron angular distribution
(Fig.~\ref{fig-dpi0pi0}, on the left of the middle panel) are
the result of two runs, one run with the spectator proton in the target, {\it
  i.e.}, with the reaction $pd \to d\pi^0\pi^0 + p_{spectator}$ (open
circles), and another run with the spectator proton in the beam (reversed
kinematics), {\it i.e.}, with the reaction $dp \to p_{spectator} +
d\pi^0\pi^0$ (solid 
circles). That way the acceptance over the full angular range could be
optimized. The data are in accordance with the Barshay-Temmer theorem
\cite{BTT}, according to which the angular distribution of a purely isoscalar
reaction has to be symmetric about 90$^\circ$ in the center-of-mass system. 
The dashed line shows a fit with an expansion into Legendre Polynomials of
order = 0, 2, 4 and 6 corresponding to a total spin of the resonance of $J$ =
3. Together with the fact that the $pn \to d\pi^0\pi^0$ reaction is purely
isoscalar, we get the quantum numbers $I(J^P) = 0(3^+)$ for this resonance
structure. Due to its isoscalar character it is formally compatible with an
excitation of the deuteron, hence it has been named $d^*$.

In further WASA measurements it has been demonstrated that below and above
this resonance structure the angular distributions
get flatter \cite{poldpi0pi0}. The fact that the angular distribution
flattens out towards lower energies is not unexpected, since towards the
reaction threshold we expect contributions only from lowest partial
waves. However, the observation that the angular distribution gets again
flatter at higher energies is not trivial. It is in support of the fact that
the high spin of $J$ = 3 of the resonance requires an unusually large
anisotropy of the angular distribution, which is larger than that of the
conventional $t$-channel $\Delta\Delta$ process, which dominates at higher
energies.  

The Dalitz plot of this reaction is displayed in Fig.~\ref{fig-dpi0pi0}, on
the right of the middle panel. It exhibits two distinctive features. First,
there is a clear horizontal band, which is in accord with a $\Delta\Delta$
excitation in the intermediate state as it is also corroborated by the Dalitz
plot projection on the vertical axis. This projection gives the distribution
of the square of $d\pi^0$ invariant mass displayed on the left of the bottom
panel. 
The second exceptional feature is the huge enhancement at low $\pi^0\pi^0$
invariant masses as borne out in the Dalitz projection on the horizontal axis.
This provides the distribution of the square of the $\pi^0\pi^0$ invariant
masses as displayed on the right of the bottom panel. This enhancement is the
so-called ABC effect mentioned above. And as the WASA measurements demonstrate
it is strictly correlated with the appearance of the $d^*$ resonance structure
in the total cross section. It may be described phenomenologically by a vertex
function in the decay of this resonance into the intermediate $\Delta\Delta$
state \cite{prl2011}. For alternative descriptions see
Refs. \cite{KukulinABC,abc}. 

The purely isoscalar character of the observed resonance structure has been
confirmed by the subsequent measurement of all 
three fusion reactions $pn \to d\pi^0\pi^0$, $pn \to d\pi^+\pi^-$ and $pp \to
d\pi^+\pi^0$ simultaneously and their isospin decomposition \cite{isofus} -- see
Fig.~\ref{fig-isofus}. Whereas the first reaction is purely isoscalar, the
third one is purely isovector. The second reaction is isospin mixed containing
both isoscalar and isovector components. Its isospin decomposition yields
\cite{Bystricky} 
\begin{equation}
\sigma (pn \to d\pi^+\pi^-) = 2\sigma(pn \to d\pi^0\pi^0) + \frac1 2 \sigma(pp
\to d\pi^+\pi^0).
\end{equation}

Fig.~\ref{fig-isofus} shows the total cross sections of these three reactions
multiplied by factors according to eq.~(4). For the $pn \to d\pi^+\pi^-$ and $pp
\to d\pi^+\pi^0$ reactions there are also some results from previous
measurements \cite{desy,dubna,Dakhno,Shimizu,FK,Bystricky}, they 
are depicted by open symbols in Fig.~\ref{fig-isofus}, top and bottom. As
mentioned already above 
there have been no previous measurements of the $pn \to d\pi^0\pi^0$ reaction,
where the background from the conventional $t$-channel processes is smallest.

As Fig.~\ref{fig-isofus} impressively demonstrates, there is no sign of the
narrow resonance structure around 2.38 GeV to see in the data for the purely
isovector $pp \to d\pi^+\pi^0$ reaction. The broad structure seen in this
reaction with a width of about 2$\Gamma_\Delta$ and peaking around twice the
$\Delta$ mass can be well described by the conventional $t$-channel
$\Delta\Delta$ process \cite{FK} -- see dashed line in the bottom panel of
Fig.~\ref{fig-isofus}.  

By use of isospin decomposition and recoupling of this $\Delta\Delta$ process
-- as described, {\it e.g.}, in Ref. \cite{DLS} -- we may calculate the
contribution of this process to the other double-pionic fusion channels. In
the $pn \to d\pi^+\pi^-$ reaction this process 
should contribute with  9/10 of the strength observed in the  $pp \to
d\pi^+\pi^0$ reaction, {\it i.e.}, this process is expected to 
provide about 0.32 mb at the peak around 2.5 GeV in the total cross
section. We see that this process accounts reasonably well for the measured
data beyond $\sqrt s$ = 2.4 GeV.
 
For the $pn \to d\pi^0\pi^0$ reaction isospin decomposition of the
conventional $\Delta\Delta$ process implies that only 1/5 of the strength
observed in the $pp \to d\pi^+\pi^0$ reaction is to be present in this
channel, which gives a peak contribution of 0.06 mb around 2.5 GeV. This value
is small compared to the value of 0.28~mb observed at the peak of the $d^*$
resonance structure and explains, why this channel is the golden channel with
regard to a small background from conventional processes. We note in passing
that the contribution from the $t$-channel Roper excitation is still much
smaller \cite{LuisPhD}.

\subsection{\it Search for the $d^*$ Resonance Structure in Isoscalar Non-Fusion
  Two-Pion Production }

\begin{figure} 
\centering
\includegraphics[width=8.2cm,clip]{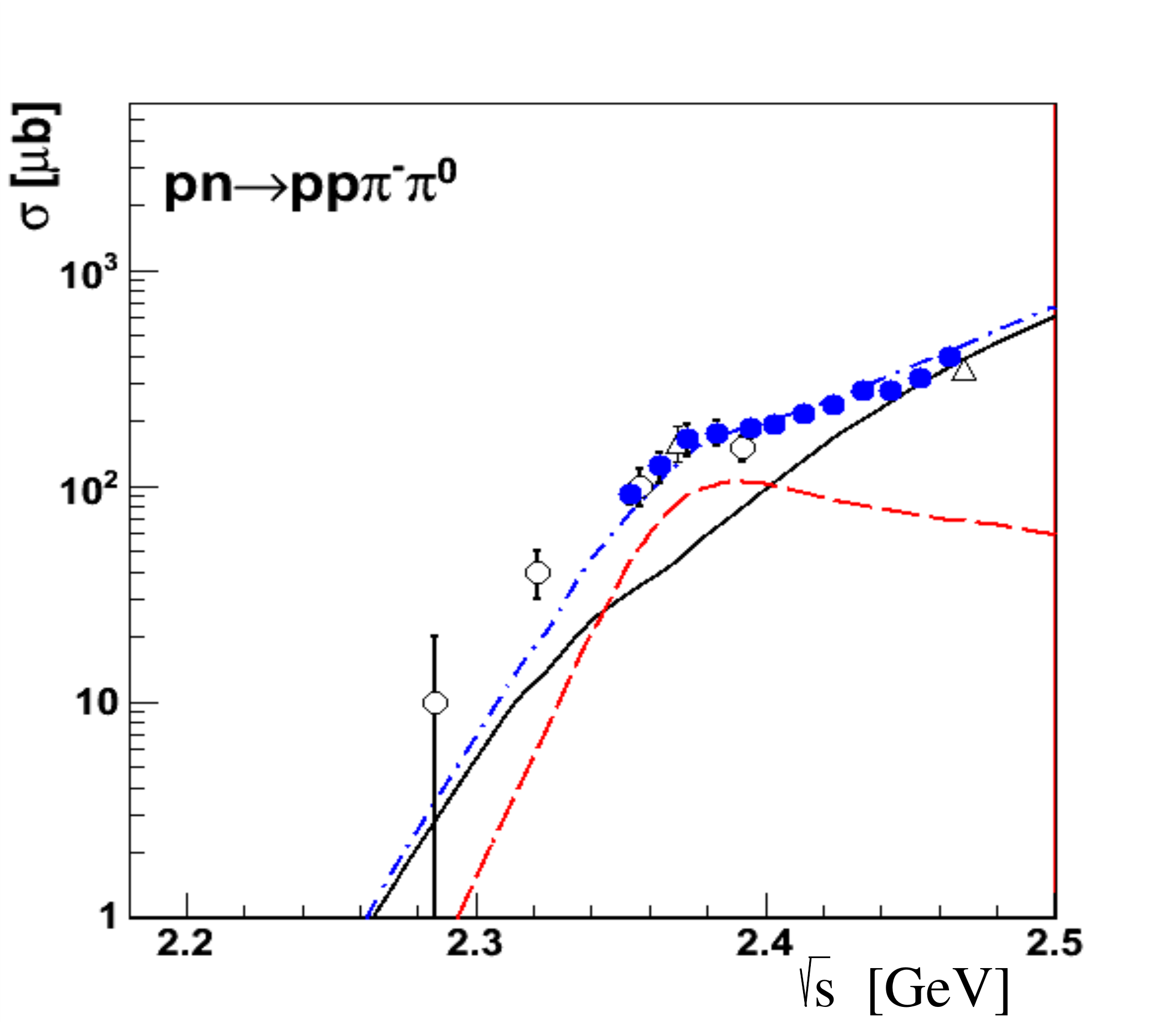}\\
\includegraphics[width=8.2cm,clip]{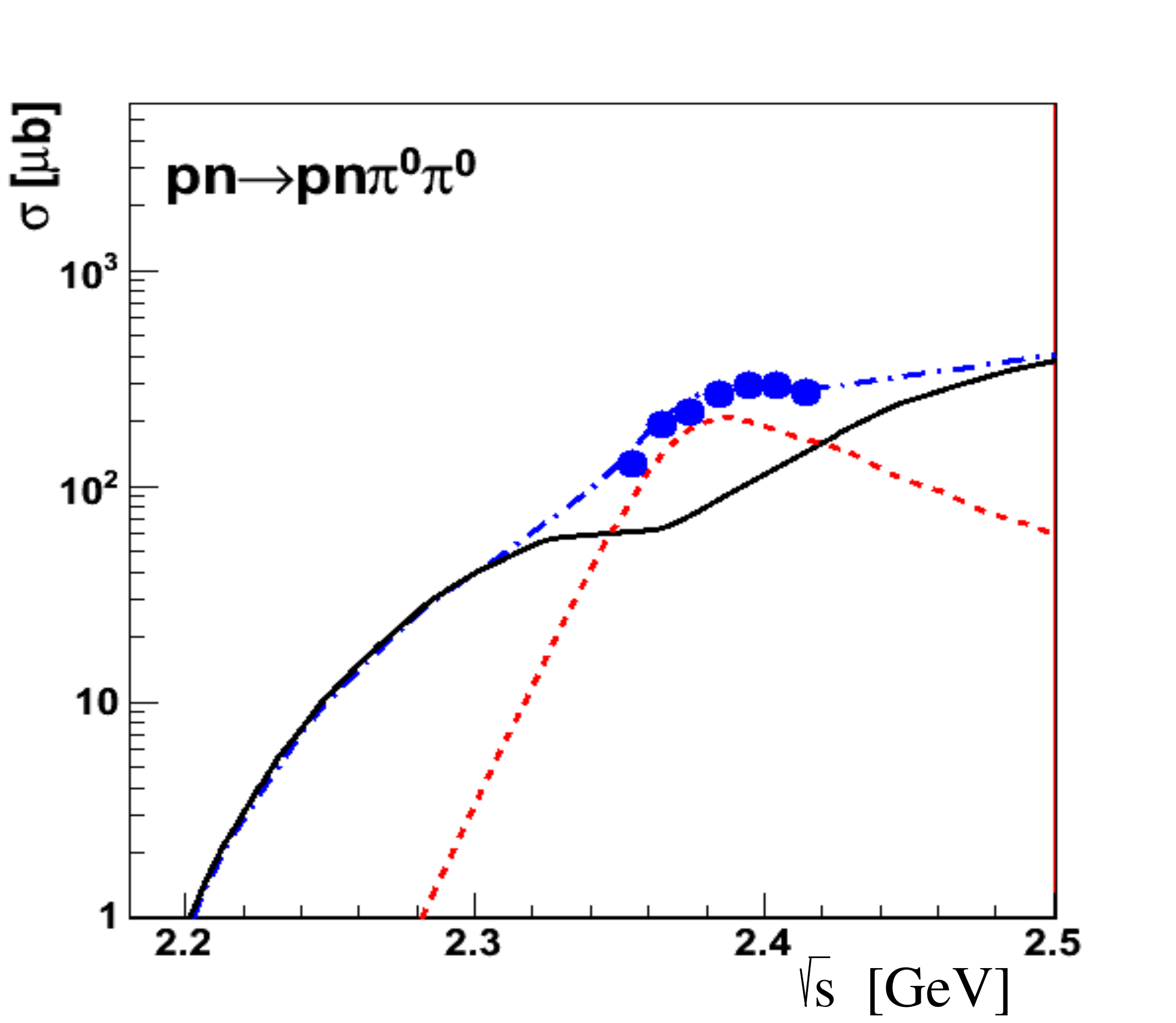}\\
\includegraphics[width=8.2cm,clip]{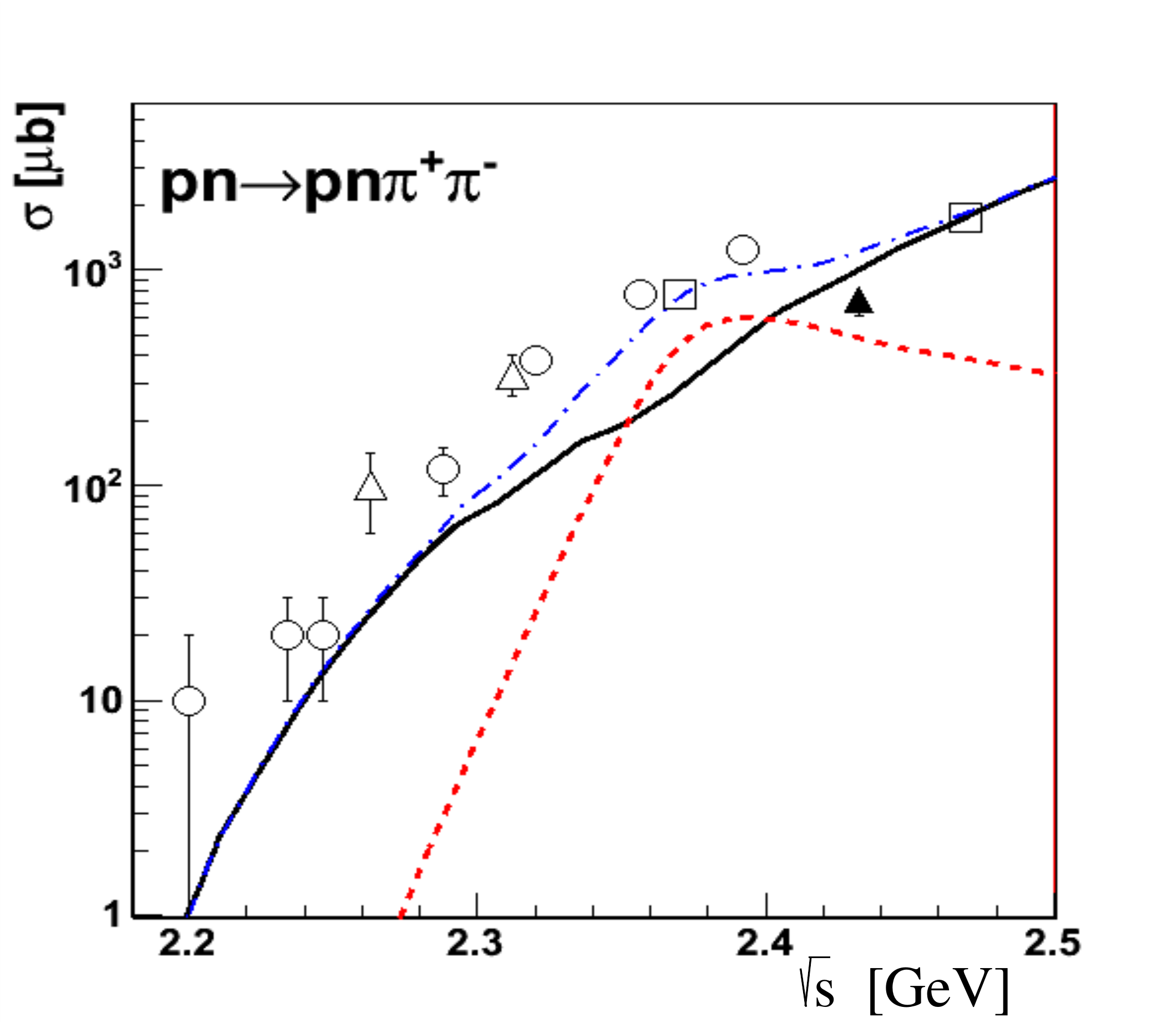}
\caption{
Total cross sections of the two-pion production reactions, where the two
nucleons do not fuse and where there is an isoscalar reaction component,
namely the reactions $pn \to pp\pi^0\pi^-$ (top), $pn \to pn\pi^0\pi^0$
(middle) and $pn \to pn\pi^+\pi^-$ (bottom). Solid dots represent WASA-at-COSY
measurements, open symbols data from previous measurements
\cite{Brunt,kek,dubna}. The solid triangle denotes a recent results from
HADES \cite{Hades}. Solid, dotted and dash-dotted lines show the
contributions from conventional $t$-channel meson-exchange, the $d^*(2380)$
resonance and their coherent sum, respectively.  
}
\label{fig-NNpipiisoscalar}       
\end{figure}

Recently also the non-fusion two-pion production channels $pn \to
pp\pi^0\pi^-$ \cite{pp0-}, $pn \to pn\pi^0\pi^0$ \cite{np00}, $pn
\to pn \pi^+\pi^-$ \cite{STORI2014}, which are partially isoscalar, have been
investigated. The WASA results (solid circles) together with previous data
(open symbols) as well as a recent data point (solid triangle) from HADES
\cite{Hades} are shown for the total cross sections in
Fig.~\ref{fig-NNpipiisoscalar}. Since the four-body phase space grows much
more rapidly with increasing energy than the three-body phase space does, also
the contribution of the conventional $t$-channel processes grows rapidly with
energy. In addition, since these reactions are only partially isoscalar, the
amount of the conventional processes relative to the expected isoscalar $d^*$
contributions is much bigger. Hence the $d^*$ signal is likely to appear only
as a kind of shoulder within the steeply increasing slope of total cross
sections.

The size
of the $d^*$ contribution can be easily estimated by use of the
isospin-decomposition of two-pion production \cite{Dakhno,Bystricky}, where
the $d^*$ contribution enters only in the reduced matrix elements $M_{000}$
and $M_{110}$ introduced in section 8.4. Having the $d^*$ decaying via an
intermediate $\Delta\Delta$ system we get in addition that
$M_{000}^{\Delta\Delta} = \sqrt 2 M_{110}^{\Delta\Delta}$ using isospin
recoupling by means of $9j$ symbols as described in Ref. \cite{DLS} and as
given in eq. (3).
A more detailed treatment takes into account also the different
phase-space situation, when the deuteron is replaced by the unbound $pn$
system \cite{CW,AO}.

For the $pn \to pn\pi^+\pi^-$ reaction the data base is still very limited,
since it has not been investigated by WASA. Hence, so far 
only bubble-chamber data are available for this channel -- with the exception
of new data point from a recent HADES measurement at 2.43 GeV
\cite{Hades}. Further measurements in the $d^*$ energy region are expected to
be performed by HADES. 

As as result all two-pion
production channels are consistent with the hypothesis of an $I(J^P) = 0(3^+)$
dibaryon resonance at 2.37 GeV with a width of 70 MeV. Though these reactions
are only partially isoscalar, the $d^*$ contributions, which are shown by the
dashed lines in Fig.~\ref{fig-NNpipiisoscalar}, constitute still the by far
dominating process in the $d^*$ energy region. There the conventional
$t$-channel processes (solid lines) underpredict the data by factors two to
four.

The observed $d^*$ decay
branchings into the diverse two-pion channels are consistent with
expectations from isospin decomposition \cite{BR} as well as explicit
theoretical calculations \cite{CW,AO} -- as will be discussed in detail in
section 10.4.

\section{$d^*(2380)$ -- a Genuine Dibaryon Resonance}
\label{sec-6}

In order to prove that the resonance structure observed in two-pion production
indeed constitutes a true resonance, {\it i.e.} a $s$-channel resonance, it
has to be sensed also in the entrance channel, {\it i.e.} in $np$
scattering. There it has to be shown
that it produces a pole in the partial waves corresponding to $I(J^P) =
0(3^+)$, {\it i.e.} in the coupled partial-wave system  $^3D_3$ - $^3G_3$. 

From the knowledge of the resonance contribution to the two-pion
channels the expected resonance contribution to  elastic $np$ scattering can 
be estimated \cite{BR} to be in the order of about 170 $\mu$b, which has to
be compared to a total $np$ cross section of nearly 40 mb. The only easily
accessible 
observable, which has the potential to sense such a small contribution, is the
analysing power, since it is composed of only interference terms in the
partial waves and hence sensitive to small contributions in partial
waves. Due to the spin of the resonance its angular contribution in
the analysing power has to be according to the angular dependence of the
associated Legendre polynomial $P^1_3$. Hence the resonance contribution is
expected to be largest at 90$^\circ$, {\it i.e.}, at the angle, where the
differential cross section is smallest. For the sensitivity of other
observables to the resonance contribution see Ref.~\cite{RWnew}

Following these considerations the analyzing power was measured over the
energy region of interest and over practically the full angular range with
WASA at 
COSY. The experiment was carried out again in the quasi-free mode, but in
inverse kinematics by use of a polarized deuteron beam hitting the hydrogen
pellet target.

\begin{figure} 
\centering
\includegraphics[width=10.5cm,clip]{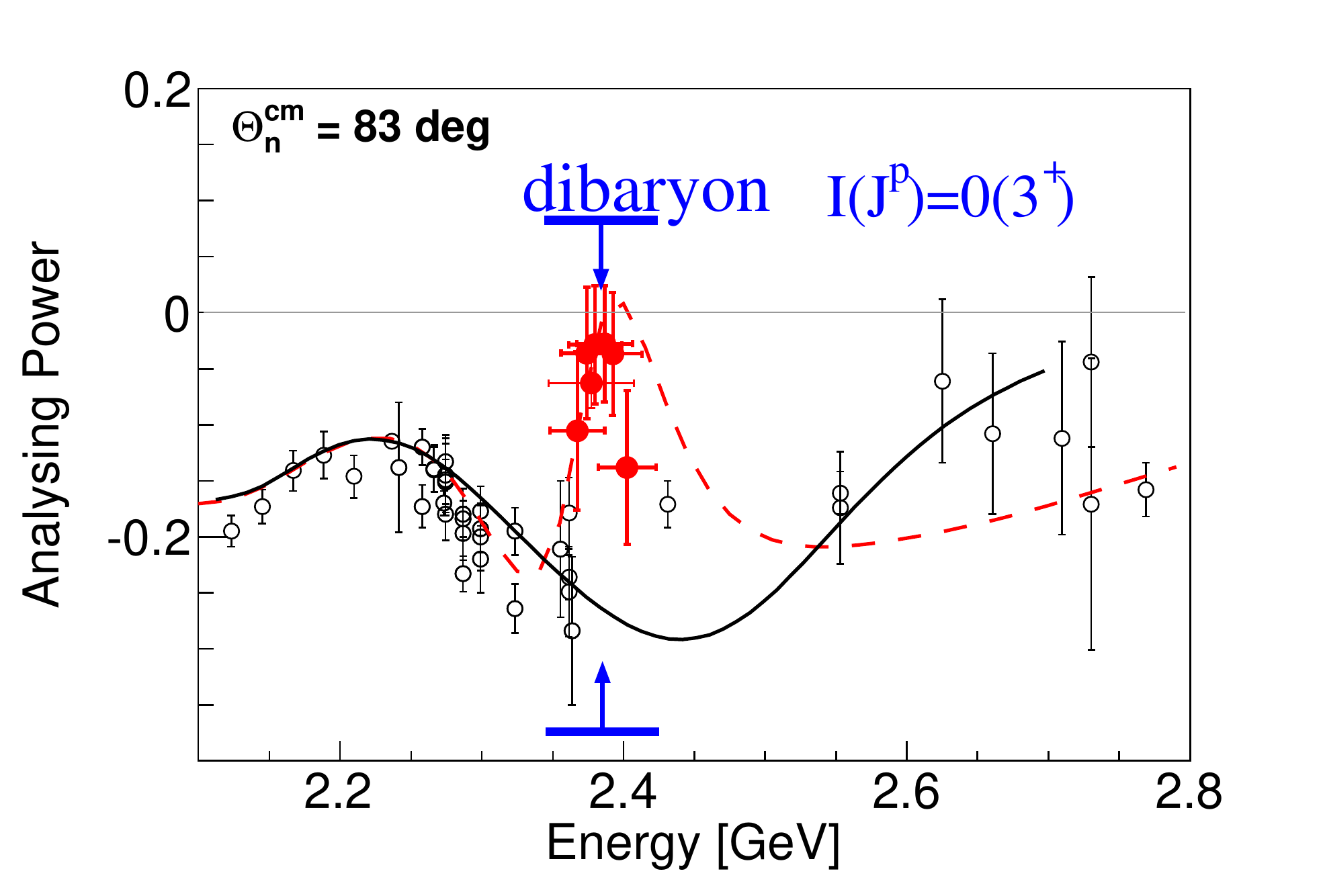}\\
\includegraphics[width=10.5cm,clip]{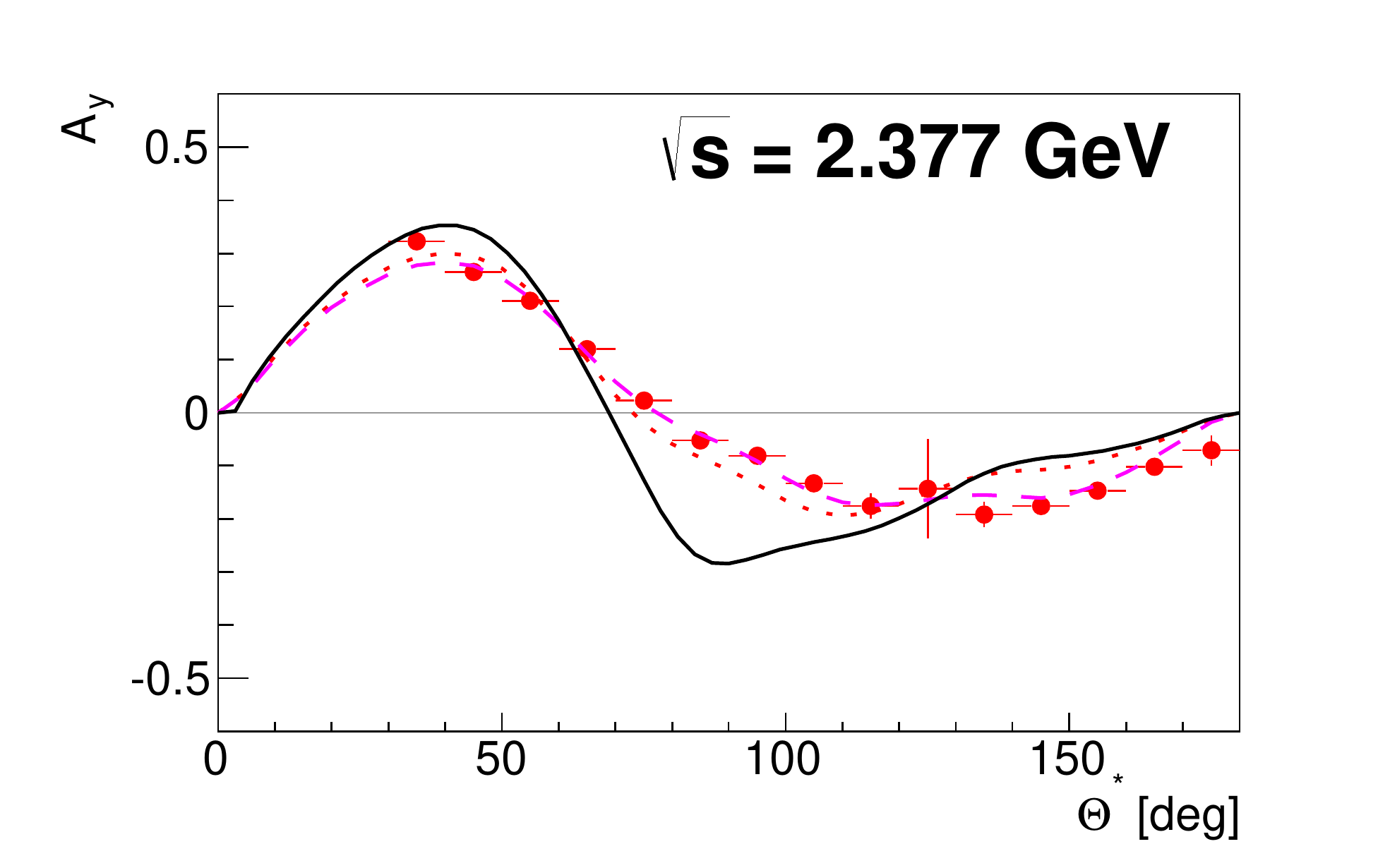}\\
\includegraphics[width=10.5cm,clip]{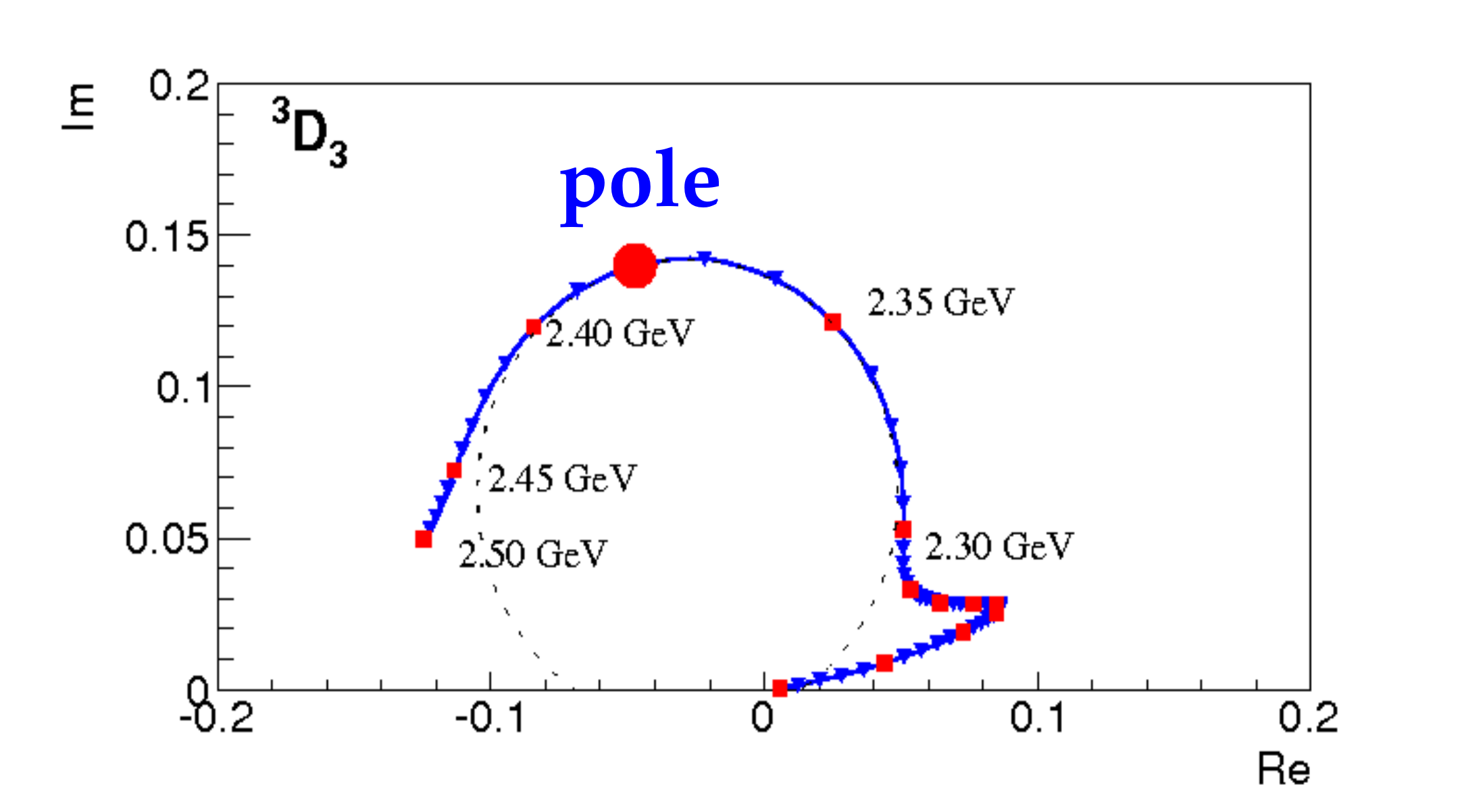}
\caption{Analysing power data for elastic $pn$ scattering in the
  $d^*(2380)$ energy region and their partial-wave analysis
  \cite{prl2014,npfull,RWnew,RW,SM16}.  Top: energy
  dependence of the analysing power in the vicinity of $\Theta_n^{cm} =
  90^\circ$, where the effect of the $d^*(2380)$ resonance is expected to be
  largest. The solid circles denote WASA results, the open symbols previous
  data \cite{Ball,Lesquen,Makdisi,Newsom,Arnold,Ball1,McNaughton,Glass}. The solid line gives the previous SAID partial-wave solution,
  the dashed line the new SAID solution after including the WASA data. Middle: 
  Angular distribution of the analysing power $A_y$ at the resonance
  energy. The curves have the same meaning as in the top figure. Bottom: Argand
  diagram of the new SAID solution for the $^3D_3$ partial wave with a pole at
  2380 MeV. The thick solid circle denotes the pole position.
}
\label{fig-pn}       
\end{figure}

Fig.~\ref{fig-pn} shows the results of the measurements and their subsequent
partial-wave analysis by the SAID group \cite{prl2014,npfull,RWnew,RW,SM16}. The top panel displays the energy
dependence of the analysing power in the region of  90$^\circ$, where the
effect of the resonance contribution is expected to be largest. The new WASA
data (solid circles) exhibit a resonance-like structure right at the position,
where expected and where no data were available from previous measurements
(open symbols)
\cite{Ball,Lesquen,Makdisi,Newsom,Arnold,Ball1,McNaughton,Glass}. The solid
lines shows the previous SAID partial-wave solution SP07 \cite{SAID}, which
completely misses the new data. The dashed line exhibits the new SAID solution
SM16 \cite{SM16} upon inclusion of the new WASA data. 

The middle panel shows the angular
distribution of the analysing power  right at resonance. Clearly, the old SP07
solution misses the data over most of the angular range, but most dramatically
around 90$^\circ$, where $P_3^1$ is largest. The only agreement between old
and new solutions is at 63$^\circ$ and 116$^\circ$, the zeros of $P_3^1$.

The
bottom panel displays the Argand diagram of the new solution for the $^3D_3$
partial wave. It exhibits a pronounced looping of this partial wave in
agreement with a resonant behavior. The new partial-wave analysis SM16 reveals,
indeed, a pole in the coupled $^3D_3 - ^3G_3$ partial waves at the position 
$(2380\pm10)-i(40\pm5)$ MeV \cite{prl2014,npfull,RWnew,RW,SM16} -- fully
consistent with the findings in the two-pion production reactions.

This result establishes the resonance structure observed in two-pion
production as a true $s$-channel resonance in the proton-neutron system. Since
it is of isoscalar character, $d^*(2380)$ has been chosen as its denotation in
analogy to the notation for isoscalar excitations of the nucleon. Also, the
notation $d^*$ was used already by Goldman {\it et.al.} \cite{Goldman} when
predicting the so-called "inevitable dibaryon" with identical quantum numbers.

When investigating the genuineness of a resonance the important question arises,
whether there are nearby thresholds, which possibly could simulate a resonance
effect in form of cusp due to channel-coupling. Indeed, at $\sqrt s$ = 2.38 GeV there
are two thresholds, the $NN\eta$ and the $NN^*(1440)$ production
thresholds. Both processes are both of isoscalar and isovector character and
hence any hypothetical cusp effect would necessarily be observed also in
$pp$-induced two-pion production -- which, however is not the case. This
disproves also the argumentation of Bugg in Ref. \cite{Bugg}, who suggests the
$d^*(2380)$ resonance structure to be the result of the $N^*(1440)$
excitation. As discussed above already in sec. 8.4 this excitation with a
width of about 300 MeV is well understood and contributes only as a marginal,
very broad background process in the golden channel $pn \to d\pi^0\pi^0$.

\subsection{\it Former Occasions for Its Discovery}
\label{sec-7}

Having established now $d^*(2380)$ as a genuine dibaryon resonance we may ask
the question, whether already previous 
experiments could have had the chance to discover it. As already mentioned
above, the golden channel $pn \to d\pi^0\pi^0$ was not accessible
instrumentally before. But what about other channels? 

In the $pn \to
d\pi^+\pi^-$ reaction the resonance effect is no longer as pronounced, but
still could have been observed in principle by previous bubble chamber
experiments 
at DESY \cite{desy} and JINR \cite{dubna} in case of much improved statistics
as well as better neutron beam energy resolution and a careful energy scan,
respectively. 

Previous $np$ scattering experiments at LAMPF and Saclay
were limited by the maximum beam energy available at these
accelerator facilities, which were below the energy region of interest
here. In this respect it was particularly unfortunate that the maximum
available beam energy at Saclay marginally missed the $d^*(2380)$ resonance
region. 

The EDDA experiment at COSY would have had a great chance of its
discovery, if its $pp$ scattering program would have been complimented by a
corresponding $np$ scattering program -- as intended initially. 

Actually groups at Tokyo had been quite close already as early as
1977 considering measurements of the proton polarization in deuteron
photodisintegration $\gamma d \to \vec{p}n$
\cite{Kamae1,Kamae2,IkedaKamae,IkedaKamae1}. They 
observed a strong increase of the polarization at energies beyond $\sqrt s$ =
2.2 GeV, reaching a maximum around 2.38 GeV and decreasing thereafter. From the
fact that the effect appeared to be largest at 90$^\circ$ they concluded that
possibly an $I(J^P) = 0(3^+)$ deeply bound $\Delta\Delta$ system could be the
reason -- making also a 
possible connection to ABC effect and two-pion production channels. Only,
their data favored a broad resonance of about 160 MeV width. Follow-up
measurements and their analysis showed that the data and their analysis are
best described by the assumption of several broad resonances of different
spin-parity with widths larger than 200 MeV \cite{IkedaKamae1}. From
the large widths it looks that these findings correspond to $t$-channel
$\Delta\Delta$ excitations rather than the narrow $d^*(2380)$ dibaryon
resonance. However, we note that recent JLAB measurements of the proton
polarization at 90$^\circ$ \cite{Jlab} confirm a maximum polarization
of -1 at $\sqrt s \approx$ 2.38 GeV with a sharp decrease at higher
energies. This 
indicates that on top of polarization effects due to conventional $t$-channel
$\Delta\Delta$ excitations there might be, indeed, a visible effect from
$d^*(2380)$. If true, it would open the door to
measurements of the electromagnetic excitation of $d^*(2380)$ in such
reactions providing access to the formfactor and thus to the size
of this object.

\subsection{\it $d^*(2380)$ in Nuclear Matter}
\label{sec-8}

If dibaryons exist and if they even survive in a nuclear surrounding, then they
should have an impact on the nuclear equation of state
\cite{faessler,krivo,AT,Kagiyama}, which is of relevance not only for
heavy-ion collisions but also for very compact stellar objects like neutron
stars.  The effect of $\Delta$ and other baryon excitations on the equation of
states has been considered also in Ref. \cite{APR}.

As already expected from the ABC effect observed previously in the
double-pionic fusion reactions to the He isotopes, the exclusive and
kinematically complete measurements of these fusion reactions conducted with
WASA at COSY could demonstrate \cite{3he,4he} that indeed $d^*(2380)$ is
present also in these reactions albeit with a much  increased width due to the Fermi
motion of the nucleons bound in these nuclei.

To investigate this question for still heavier nuclei by exclusive
double-pionic fusion reactions becomes very difficult from the experimental
point of view, since the kinetic energies of the fusion products get such low
that they get below the detection thresholds of
traditional experimental setups and would afford new, very sophisticated
instruments. A way out would be measurements with inverse kinematics, {\it
  e.g.}, using a $^{14}$N beam to conduct the reaction $^{14}$N$d \to
^{16}$O$\pi^0\pi^0$. 

Support for the existence of $d^*(2380)$ in nuclear matter may come from
heavy-ion collisions in connection with the so-called DLS puzzle. In
relativistic heavy-ion collisions an unusual enhancement has been observed in the
spectrum of emitted $e^+e^-$-pairs with invariant masses between $\pi$ and
$\omega$ mass.  
First interpreted as a possible
signal from the quark-gluon plasma it was soon realized that this enhancement
persists also at low collision energies. Dedicated measurements of dilepton
production in nucleon-nucleon collisions by DLS \cite{dls} and HADES
\cite{hades} could trace back this phenomenon to a not-understood enhancement
in $pn$ collisions 
at energies below $\sqrt s \approx$ 3 GeV. 

In a recent publication \cite{DLS}
it could be demonstrated that two-pion production  via $t$-channel
$\Delta\Delta$ excitation as well as via $d^*(2380)$ formation -- where the produced
$\rho$-channel pion pair transforms into a lepton pair -- can account
quantitatively for this enhancement 
and thus solve the DLS-puzzle. Since this enhancement is observed also in
heavy-ion collisions, this means in turn that obviously also $d^*(2380)$ has
been produced in those collisions.

\subsection{\it $d^*(2380)$ in Theory}
\label{sec-9}

As already discussed in section 3 the very first
theoretical paper on dibaryons in connection with quarks  -- or more precise
with SU(6) symmetry breaking -- the one by Dyson and Xuong in 1964 \cite{Dyson},
predicted properly the now discovered state $d^*(2380)$ at a mass
remarkably close to the experimentally observed one. Subsequent, partially very
complex theoretical investigations, turned out to be much less successful. In
particular they predicted often a multitude of states in addition, which were
never observed. 

In connection with the observation of large
polarization effects in deuteron photodisintegration (see Sect.~\ref{sec-7})
also Kamae and Fujita \cite{Kamae} predicted 1977 the mass of the now
observed state properly based on a non-relativistic
one-boson exchange potential model for two $\Delta$ isobars. 

Goldman {\it et al.} \cite{Goldman} pointed out 1989 that due to the unique
symmetries of 
such a state any model based on confinement and effective one-gluon exchange
must predict the existence of this state -- calling it the
"inevitable" dibaryon. However, initially they predicted an enormous binding
energy as large as 350 MeV for such a state. Only recently -- after the
experimental observation -- the calculations of this group
approached the experimental value in the framework of the so-called quark
delocalization and color screening model (QDCSM) \cite{ping,Nanjing}. 

Also the Nijmegen group \cite{Mulders,Aerts,Mulders1,Mulders2} as well as
Mulders and Thomas \cite{MuldersThomas} and also Saito
\cite{Saito} predicted this state correctly at the proper mass in their
bag-model calculations. Only, they simultaneously predicted within the same
framework numerous other unflavored dibaryon states, which were never observed. 

A correct real prediction, {\it i.e.} a prediction before the experimental
discovery of $d^*(2380)$, is the one by the IHEP theory group led by
Z. Y. Zhang, who studied this state in the chiral $SU(3)$ quark model within
the framework of the resonating group method (RGM) \cite{Yuan}.
 
Meanwhile, also a QCD sum rule study finds this state at the right mass
\cite{chen}. On the other hand, a study purely based on quarks and gluons does
not obtain this state, at least not in the energy region of the experimental
observation \cite{Lee}. The reason might be that no coupling to hadrons, where
$d^*(2380)$ finally has to decay into, has been taken into account. A first
result from lattice QCD has been reported at the MPMBI 2015 workshop at Sendai
by K. Sasaki from the HALQCD collaboration \cite{HALQCDd*}. For the
$\Delta\Delta$ system with $J$ = 3 a strongly attractive interaction has been
obtained, which leads to a bound state -- albeit these calculations with a
pion mass of 1015 MeV  are still far away from being very realistic.  

A relativistic
Faddeev-type calculation based on hadronic interactions have been reported
by Gal and Garcilazo recently to see this state, too, correctly at the
experimental mass \cite{Gal,GG1,GG2}. Already before that Garcilazo {\it et al.}
had predicted this state based on two-body interactions derived from the
chiral quark cluster model, though with too small a binding energy of only
29.9 MeV \cite{GarcilazoValcarce,ValcarceGarcilazo} and 7.8 MeV \cite{Mota1},
respectively. 
However, those calculations were performed without coupling to the $NN$
channel, which would lead to larger binding energies. It should be noted that
those calculations give the same sequence of dibaryon states as predicted
previously  by Dyson and Xuong \cite{Dyson}.

More demanding than the mass value appears to be the reproduction of the decay
width of $d^*(2380)$, since it requires also a theoretical treatment of the
dynamics of the decay process. So far there are only three theoretical
predictions for the decay width based either on Faddeev calculations
\cite{Gal,GG1,GG2} or on quark-model calculations within the QDCSM framework
\cite{ping,Nanjing} and the RGM-based chiral $SU(3)$ model
\cite{dong,huang,dong1}, respectively.

In the SAID partial-wave analysis \cite{SM16} the generated pole has an
imaginary part of 40 MeV, which in a Breit-Wigner interpretation corresponds
to a width of 80 MeV. Actually the analysing power data for the energy
dependence around 90$^\circ$ shown in the top panel of Fig.~\ref{fig-pn} would
prefer a somewhat narrower width, but the data base at the high-energy end of
the resonance region is too sparse for a more stringent fixation of the width
from $pn$ scattering. Much better is the data base in two-pion production,
in particular in the golden channel $pn \to d\pi^0\pi^0$, where the Lorentzian
shape of the energy dependence of the total cross section is well
documented by  high-statistics data, see top panel in
Fig~\ref{fig-dpi0pi0}. These data prefer a width of 70 MeV. Though the
background due to conventional processes, essentially $t$-channel
$\Delta\Delta$ excitation, is small, it introduces a small uncertainty in the
exact determination of the width here, too.

From the calculation of the $t$-channel $\Delta\Delta$ process we know that
the plain $\Delta\Delta$ excitation leads to a resonance structure with a
width of about twice the $\Delta$ width \footnote{We note that in a recent
  article \cite{Niskanen1} it is argued that the width of an unbound
  $\Delta\Delta$ configuration is equal to the width of just a single
  $\Delta$.}, {\it i.e.} 230 MeV -- see, {\it e.g.} 
Fig.~\ref{fig-isofus}, bottom. If we consider now the asymptotic configuration
of $d^*(2380)$ as a $\Delta\Delta$ system bound by 80 MeV, then we may easily
estimate its decay width by the known momentum dependence of the $\Delta$
decay width, which is proportional to the third power of the effective 
decay momentum, see, {\it e.g.}, Ref. \cite{Risser}. With having a $\Delta$
mass reduced by 40 MeV due to binding this leads to a reduced width of about
160 MeV for a $\Delta\Delta$ system bound by 80 MeV \cite{BBC}. This is also
the result of Ref.~\cite{ping} for the $d^*(2380)$ decay width. In a more
detailed treatment taking into account correlations the QDCSM
calculations arrive at a width of 110 MeV \cite{Nanjing}. Also, if we
calculate the $d^*$ width by use of eq. (12) in Ref. \cite{abc}, then we
obtain a width of 107 MeV at the $d^*$ mass starting from a width of 230 MeV
for the unbound $\Delta\Delta$ system.  

The Faddeev calculations come closer to the experimental value by obtaining a
width of 94 MeV for a resonance mass of 2383 MeV \cite{Gal,GG1,GG2}
\footnote{The solution with a factor 2/3 applied assumes that there is no
  decay into $pp\pi^0\pi^-$ and $nn\pi^+\pi^0$ channels as well as a reduced
  decay into the $np\pi^+\pi^-$ channel, which is not in accordance with
  experimental results, see discussion of branchings in the next section}. To
this the decay width into the $pn$ system has to be added, so that a total
decay width of slightly above 100 MeV is obtained.

Common to these theoretical investigations is that they overestimate the
width of $d^*(2380)$. This may possibly be related to exotic contributions to
this state - like, {\it e.g.}, hidden color aspects as discussed in
Ref. \cite{BBC}. In fact the calculations of the IHEP group, which include
hidden color configurations arrive at a width of 72 MeV
\cite{dong,huang,dong1} in full agreement with the experimental result. In
these calculations the $d^*(2380)$ wave function contains about 67$\%$ hidden
color components, which can not decay easily and hence reduce the decay width
to the experimental value. Also the Nanjing group finds a reduction of 25 MeV
in the decay width, if coupling to hidden color channels is taken into account
\cite{Nanjing}. Hidden-color configurations for a hexaquark state with the
quantum numbers of $d^*(2380)$ have been investigated recently also in
Ref.~\cite{An}. 

These studies suggest that the observed unusually small decay width of
$d^*(2380)$ signals an exotic character of this dibaryon resonance and points
to a compact hexaquark nature of this object as discussed in detail in
Refs. \cite{huang,zhang,lue}. These calculations obtain a
root-mean-square radius of 0.8 fm for the size of this object. Also the Nanjing
calculations give similar results for the size of $d^*(2380)$. If true then
this resonance indeed constitutes a very compact hexaquark system.
Reservations about the compact hexaquark interpretation of $d^*(2380)$ have
been voiced recently by Gal \cite{Galpriv}.

\subsection{\it The Branchings of the $d^*(2380)$ Decays}

The decays of $d^*(2380)$ into $pn$ and $NN\pi\pi$ channels derive from the
analysis of $pn$ elastic scattering and the measurements of the $d^*(2380)$
signal in the various two-pion production reaction channels. The branching
ratios into these channels have been evaluated from the data in
Ref.~\cite{BR} and are presented in Table~\ref{tab:branching}.  

A hypothetical decay $d^*(2380) \to NN\pi$ has been investigated by
analyzing WASA data for the single-pion production reactions $pp \to pp\pi^0$
and $np \to pp\pi^-$, in order to extract the isoscalar single-pion production
cross section in the energy region of the $d^*(2380)$ resonance. The resulting
data on the isoscalar single-pion production give no evidence for a sizeable
contribution from the $d^*(2380)$ decay into the isoscalar $NN\pi$ channel
\cite{NNpi}.

\begin{table}
\begin{center}
\begin{minipage}[t]{16.5 cm}
\caption{Branching ratios in percent of the $d^*(2380)$ decay into $pn$, $NN\pi$ and
  $NN\pi\pi$ channels. The experimental results
  \cite{BR,NNpi} are compared to results from a theoretical calculation
  \cite{dong1} starting from the theoretical $d^*$ wave function and including
  isospin breaking effects. They are also compared to values expected from
  pure isospin recoupling of the various $NN\pi\pi$ channels according to
  eqs. (3) and (5), respectively. In the latter case the branching into the
  $d\pi^0\pi^0$ channel is normalized to the data.
}  
\label{tab:branching}
\end{minipage}
\begin{tabular}{llll}
\\ 
\hline\\

decay channel&experiment&theory \cite{dong1}
&$NN\pi\pi$ isospin recoupling\\ 

\hline\\

$d\pi^0\pi^0$&~~$14\pm1$&~~~~~~~12.8&~~~~~~~~~~13\\
$d\pi^+\pi^-$&~~$23\pm2$&~~~~~~~23.4&~~~~~~~~~~26\\
$np\pi^0\pi^0$&~~$12\pm2$&~~~~~~~13.3&~~~~~~~~~~13\\
$np\pi^+\pi^-$&~~$30\pm5$&~~~~~~~28.6&~~~~~~~~~~32.5\\
$pp\pi^0\pi^-$&~~~$6\pm1$&~~~~~~~~4.9&~~~~~~~~~~~6.5\\
$nn\pi^+\pi^0$&~~~$6\pm1$&~~~~~~~~4.9&~~~~~~~~~~~6.5\\
$(NN\pi)_{I=0}$&~~~~~$<9$&~~~~~~~~0&~~~~~~~~~~~--\\
$np$&~~$12\pm3$&~~~~~~~12.1&~~~~~~~~~~~--\\

\hline\\
$\sum(total)$&~$103\pm7$&~~~~~~100&~~~~~~\\
\hline
 \end{tabular}\\
\end{center}
\end{table}

As already discussed in sections 8.4, 9.2 and 9.3 the cross sections of the
various two-pion 
production channels are closely linked by isospin relations
\cite{Dakhno,Bystricky,DLS}, where the reduced matrix elements
$M_{I_{NN}^fI_{\pi\pi}I_{NN}^i}$ enter. For formation and decay of $d^*(2380)$
only the matrix elements $M_{000}$ and $M_{110}$ are of relevance. 

In the
following we discuss, how these branchings are sensitive to the
intermediate asymptotic hadronic configuration, where $d^*(2380)$ decays
to. Whereas Dyson and Xuong \cite{Dyson} as well as QCD-based quark models
\cite{Mulders,Aerts,Mulders1,Mulders2,MuldersThomas,Saito,Goldman,ping,Nanjing,Yuan,dong,huang,dong1}
predict a $\Delta\Delta$ system as hadronic intermediate configuration, the
work of Kukulin and Platonova \cite{Platonova,KukulinABC} as well as that of Gal
and Garcilazo \cite{GG1,GG2}  prefer a $D_{12}\pi$ intermediate
configuration, where the $\Delta N$ threshold state $D_{12}$ decays
subsequently into $NN$ and $NN\pi$ final states.

\subsubsection{\it compatibility with an asymptotic $\Delta\Delta$
  configuration}

For an intermediate $\Delta\Delta$ system we have $M_{000}^{\Delta\Delta} = -\sqrt 2
M_{110}^{\Delta\Delta}$ from eq. (3). If we put this relation into the
expressions of reduced matrix elements for the various two-pion cross sections,
then we obtain the branchings into the various $NN\pi\pi$ channels as given in
the fourth column of Table~\ref{tab:branching}. As discussed in
Ref.~\cite{isofus} the isospin breaking due to mass differences between neutral
and charged pions as well as between proton and neutron -- which entail
differences in $NN\pi\pi$ reaction thresholds -- causes changes in these values
in the order of 10 - 20$\%$. The same is true for the comparison of $d\pi\pi$
and unbound $NN\pi\pi$ channels due to different phase spaces as treated in
more detail in 
Refs.~\cite{CW,AO}. With this in mind we find excellent agreement between the
experimentally determined branchings and the ones expected for the case that
$d^*(2380)$ decays into an intermediate $\Delta\Delta$ system. The branching
into the $pn$ system is then obtained by using unitarity as detailed in
Ref.~\cite{BR}. As also discussed in that work there is no known process,
which could feed the isoscalar $NN\pi$ channel starting from an intermediate
isoscalar $\Delta\Delta$ configuration. 

In Ref. \cite{dong1} the branching has been calculated starting directly from
the theoretical $d^*$ wave function obtained in a chiral $SU(3)$ quark model
calculation. Asymptotically this wave function represents a bound
$\Delta\Delta$ configuration containing hidden color. This calculation, which
also includes isospin breaking effects due to different masses of charged and
neutral pions and baryons, is shown in the third column of
Table~\ref{tab:branching}. It is in excellent agreement with the experimental
results given in the second column of Table~\ref{tab:branching}.

\subsubsection{\it compatibility with a decay route via $D_{12}$?}

Kukulin and Platonova \cite{Platonova,KukulinABC} propose as an alternative to
an asymptotic $\Delta\Delta$ configuration the decay of $d^*(2380)$ via the
$\Delta N$ threshold state $D_{12}$ by emission of a $p$-wave pion. They
argue that due to the small size of the compact six-quark object $d^*(2380)$
it is much more natural for it to decay into another small six-quark object,
$D_{12}$, by pion emission rather than into a much more extended $\Delta\Delta$
system. 

Gal and Garcilazo \cite{GG1,GG2} have taken up this idea, since it provides the
possibility to 
reduce the four-body $NN\pi\pi$ system effectively to a three-body $\Delta
N\pi$ system with an intermediate $D_{12} \pi$ configuration, which they then
can handle in the Faddeev formalism.

Indeed, a $D_{12}\pi$ configuration with the pion in relative $p$-wave is
kinematically hardly to distinguish from a $\Delta\Delta$ configuration. The
Dalitz plot would look the same. So the only way to check this alternative
hypothesis appears to be the branchings of the $d^*(2380)$ decay, which
contain also the dynamics and structure informations.

Assuming again isospin invariance we may obtain the branching for the
$D_{12}\pi$ scenario simply by considering the decays of $D_{12}$ into the
various $NN\pi$ channels. By isospin recoupling we obtain then \cite{Galpriv}

\begin{eqnarray}
M_{I_{NN}^fI_{\pi\pi}I_{NN}^i}^{D_{12}\pi_2} \sim (-)^{I_{NN}^f}
  \hat{I}_{NN}^f\hat{I}_{\Delta}
\left\{
\begin{array}{ccc}
I_{N_1}&I_{N_2}&I_{NN}^f\\
I_{D_{12}}&I_{\pi_1}&I_\Delta\\
\end{array}
\right\},
\end{eqnarray}

      where the $NN$ system in the final state couples with $\pi_1$ to $D_{12}$
      and further-on with $\pi_2$ to $d^*(2380)$.

Of relevance for the branchings into the various $NN\pi\pi$ channels are again
only the matrix elements $M_{000}$ and $M_{110}$, which according to eq. (5)
are related by $M_{000} = -\sqrt 2 M_{110}$. Since the cross sections depend
just on these matrix elements squared, we obtain identical
branchings into the various $NN\pi\pi$ channels for both scenarios.

Besides of the decays of $D_{12}$ into the isovector $NN\pi$ channels,
$D_{12}$ decays also into the isovector $NN$ system. Hence $d^*(2380)$ decays
also into the isoscalar $NN\pi$ channel in this scenario via the
route $d^*(2380) \to [D_{12}\pi]_{I=0} \to [NN\pi]_{I=0}$. The SAID
partial-wave analyses \cite{said4,SAID,SM16} of elastic $pp$ scattering give a
modulus of the residue at the $D_{12}$ pole of G = 10 - 11 MeV, which results
in a branching of the $D_{12} \to NN$ decay of 16 - 18$\%$.
The isoscalar single-pion production has been analyzed at WASA with
the aim to search experimentally for a possible decay branch $d^*(2380) \to
[NN\pi]_{I=0}$. No indication in the data for such a decay  has
been found, only an upper limit of 9 $\%$ (90 $\%$ C. L.) could be derived
\cite{NNpi}. 

Note that different from the case with a $\Delta\Delta$ intermediate state
there is no decay route into the $np$ system in
this scenario with an intermediate $D_{12}\pi$ configuration. Hence the decay
route into the $np$ channel, which has been measured to have a branching of
$12\pm 3\%$, has to be added by hand in this scenario.

In the model of Gal and Garcilazo \cite{GG1,GG2} only decay modes
with $I_{NN}$ = 0 are allowed in leading approximation, {\it i.e.} there are
no $I_{NN}$ = 1 contributions of the decays into the $pp\pi^0\pi^-$,
$nn\pi^+\pi^0$ and $pn \to pn\pi^+\pi^-$ channels. In a forthcoming paper
\cite{Galpriv} various possibilities are discussed, how agreement can be
achieved between experimental branching ratios and the ones obtained in a
$D_{12} \pi$ scenario. 

Summarizing, in both scenarios we find identical decay branchings into the
various $NN\pi\pi$ channels. In the $D_{12}\pi$ scenario we expect a
substantial decay into the isoscalar $NN\pi$ system with no decay into the
$np$ channel, whereas in the $\Delta\Delta$ scenario the situation is just
opposite. However, their $D_{12}$ and therefore also $d^*(2380)$ are not
compact six-quark objects.

\subsection{\it Size and Structure of $d^*(2380)$}

One of the most important questions with regard to $d^*(2380)$ is, whether it
constitutes a large-sized molecule-like object or a compact hexaquark
system. So far we do not have any conclusive answer to this from experiment. A
possibility to approach this 
question experimentally will be discussed in the next section. 

Also for the recently observed tetraquark and pentaquark systems there is as
of yet no experimental knowledge about the size of these objects. Here the
situation is even more confusing, since these objects have a very small decay
width, though they have masses very close to adjacent thresholds and hence
appear to be very loosely bound. In the theoretical discussion of these
objects Piccinini {\it et al.} proposed \cite{Erice,Piccinini} a rough radius
estimate for a system of two quark clusters based on the Heisenberg
uncertainty relation and depending just on the binding energy $\epsilon$ of
the object in question: 
\begin{eqnarray}
r \approx \frac {\hbar c} {\sqrt {2m\epsilon}}
\end{eqnarray}
In case of the much-debated tetraquark system $X(3872)$, which is bound by
only few keV, this leads to a radius of the order of 10 fm, which means a much
extended molecular system. If we apply this 
formula to $d^*(2380)$ being bound by 80 MeV relative to the nominal
$\Delta\Delta$ threshold, then we obtain a radius of about 0.5 fm. Such a
small radius means that the two quark clusters must strongly overlap and hence
are expected to form a compact hexaquark system. This is also the conclusion
obtained in detailed quark-model calculations, which give a radius of 0.8 fm
 \cite{huang,zhang,lue}.  

\subsubsection{\it electromagnetic excitation of $d^*(2380)$}

An electromagnetic excitation of the deuteron groundstate to the $d^*(2380)$
resonance is highly informative, since its transition formfactor gives
access to size and structure of this resonance.

Judging just from the electromagnetic coupling constant, we expect
electromagnetic decays to be suppressed by two order of magnitudes --
as is borne out, {\it e.g.} in the decay of the $\Delta$ resonance. A
technically feasible excitation of $d^*(2380)$ would start by photo- or
electro-excitation from the deuteron groundstate. A real or virtual photon
would need then to transfer two units of angular momentum, {\it i.e.} be of E2
or C2 multipolarity, which lowers the transition probability further. In
addition, the overlap in the wavefunctions of $d$ and $d^*(2380)$ enters
profoundly. We are aware of two 
theoretical calculations dealing with such a scenario \cite{wong,qing}, where
cross sections in the range pb/sr - nb/sr are predicted for the forward
angle region. This order of magnitude may be expected also by the simple
consideration that the $d^*$ photo-production is facilitated by a double
spinflip in the six-quark configuration of the deuteron. Since the latter is
in the order of $10^{-3}$ -- see section 2.1.1., the order of cross section,
given by this number multiplied with the electromagnetic coupling constant,
should be $10^{-5}$ that of the corresponding hadronic cross section
\cite{MBpriv}. {\it 
  I.e.},  since the peak cross section of the $pn \to d \pi^0\pi^0$ reaction
is about 200 $\mu$b, the cross section of the $\gamma d \to d\pi^0\pi^0$
reaction 
should then be in the order of a few nanobarn. 

The same order of magnitude is expected for the $d^*$ photoproduction in the
$d\pi^+\pi^-$ channel. But the $\gamma d \to d\pi^+\pi^-$ cross section due to
conventional processes is two orders of magnitude larger. Hence it appears
very difficult to sense the resonance excitation under such conditions. 

In this respect the reaction $\gamma d \to d\pi^0\pi^0$ appears to be
attractive, since there the conventional processes are expected to be
particularly small \cite{Arenhoevel,Fix}. In the energy region of $d^*(2380)$
the conventional cross section is predicted to be below 20 nb. In fact,
first preliminary results from MAMI for this reaction \cite{Guenther} are in very good
agreement with these predictions for incident photon energies $E_\gamma >$ 600
MeV, where the calculations predict cross sections of 50 - 70 nb. Below that
energy the calculations predict the cross section to be strongly decreasing, whereas
the preliminary data show still appreciable strength, which possibly could be in
accordance with a $d^*$ excitation \cite{MBpriv}. Meanwhile also a measurement
of this reaction at ELPH has been reported \cite{ELPH}, which sees a similar
energy dependence of the cross section, though its absolute value is larger by
nearly a factor of two. Which of the two measurements is correct, has to be
still evaluated.

Another way to sense such small contributions are polarization measurements. The
situation looks similar to 
the one in elastic $np$ scattering. As we have shown above, the $d^*(2380)$
resonance contribution is about 0.17 mb, which is more than two orders below
the total elastic cross section. However, with help of the analyzing power,
which consists only of interference terms in partial waves, it was possible to
filter out reliably the resonance contribution. 

The analogous case in electro- or photo-excitation of $d^*(2380)$ constitute 
measurements of the polarization of the outgoing proton in the reactions
$\gamma d \to n\vec{p}$ and $\gamma^* d \to n\vec{p}$, respectively, where
$\gamma^*$ stands for a virtual photon created in inelastic electron
scattering on the deuteron. As in the analyzing power of $np$ scattering the
angular dependence of the resonance effect in the polarization of the outgoing
proton should be proportional to the associated Legendre polynomial
$P_3^1(cos~\Theta)$ \cite{npfull}. Therefore, the maximal resonance effect is
expected to be at a scattering angle of $\Theta$ = 90$^\circ$.

In fact, such an effect has already been looked for previously by Kamae
{\it et al.} in corresponding data from the Tokyo electron synchrotron
\cite{Kamae1,Kamae2,IkedaKamae,IkedaKamae1}. In order to describe the observed
large polarizations in 
the region of $d^*(2380)$, they fitted a number of resonances to the data,
among others also a $J^P = 3^+$ state. However, presumably due to the limited
data base they only obtained very large widths for these resonances in the
order of 200 - 300 MeV -- as one would expect from conventional $\Delta\Delta$
excitations. 

Recently, new polarization measurements from JLAB appeared \cite{Jlab}. Their
lowest energy point is just in the $d^*(2380)$ region and is compatible
with a maximal polarization of $P = -1$. It confirms thus the old Tokyo results
in the sense that in this region there is a build-up of a very large
polarization, which rapidly decreases both towards lower and higher energies,
see Fig.~1 in Ref.~\cite{Jlab}. Of course, a dedicated measurement over the
region of interest is needed, in order to see, whether a narrow structure with
the width of $d^*(2380)$ can be observed in this observable.

\subsubsection{\it the ABC effect and the structure of $d^*(2380)$}

 Coming back to the trace followed for the detection of d*(2380):  the ABC
 effect. As demonstrated by the experiments, the ABC effect is strictly
 correlated with the occurrence of this resonance. But what is its explanation?
 In Ref.~\cite{prl2011} {\it ad hoc} a vertex function for the decay vertex
 $d^*(2380) \to \Delta\Delta$ was introduced for its description. This vertex
 function parameterized in form of a monopole form-factor gives a good account
 for the data and with it for the ABC effect -- see Fig.~\ref{fig-dpi0pi0} --,
 if a cutoff parameter $\Lambda$  as small as $\Lambda$ = 0.16 GeV/c is
 used. On a first glance this appears to be a surprisingly small value being
 three to four times smaller than usually used in the description of hadronic
 processes. However, the vertex functions used there are conventionally
 employed for $t$-channel meson exchanges. In our case of an
 $s$-channel resonance decay the assumed vertex function is actually identical
 to that used for the $\Delta \to N\pi$ decay with identical cutoff parameter
 \cite{Risser,abc}. In fact, such vertex functions are commonly used for the
 description of the decay of baryon resonances \cite{teis} with cutoff
 parameters being similarly small as the value obtained for the $d^*(2380)$
 case. 

However, the unique
 situation in case of double-pionic fusion --  where the momentum difference
 between nucleons cancels when neglecting the Fermi motion --  is that the
 $\pi\pi$-invariant mass spectrum happens to map just the momentum dependence
 of the $d^*(2380) \to \Delta\Delta$ decay, which is given by that of the
 vertex function. Since this special kinematic condition holds no longer in
 case of non-fusion, this scenario also naturally explains the non-appearance
 of the ABC effect in non-fusion reactions. For a detailed recent discussion
 of this issue see Ref.~\cite{abc}, where also alternative concepts and 
 other small contributions to the ABC effect are discussed. 

Kukulin and Platonova \cite{Platonova,KukulinABC}, {\it e.g.}, proposed an
alternative scenario for the $d^*(2380)$ decay and the ABC effect. Instead of
the decay route $d^*(2380) \to \Delta\Delta$ they propose the dominating decay
to be $d^*(2380) \to D_{12}\pi$  with a relative $p$-wave between $D_{12}$ and
$\pi$ -- as discussed in some detail already above in section 10.4.2. In addition
to this route Kukulin and Platonova propose a 5$\%$ decay route $d^*(2380) \to
d\sigma$ with a very light and narrow $\sigma$ meson in relative $d$-wave,
in order to account for the ABC effect. Though there have been reports on the
observation of narrow and/or light $\sigma$ mesons
\cite{Abraamyan,Abraamyan1,MBgg}, a convincing clear-cut experimental
confirmation is still missing. For the CELSIUS/WASA workshop contribution
\cite{MBgg}, {\it e.g.}, it finally could not be excluded that the observed
narrow structure possibly resulted from detector artifacts. 
Aside from the question about a light and narrow $\sigma$ meson
 a problem arises for isoscalar non-fusion two-pion production. In such
cases no ABC effect is observed. However, there is no obvious reason, why
there the postulated $d^*(2380) \to d\sigma$ decay branch should be
suppressed. Kukulin and Platonova argue \cite{Platonova} that due to the much
increased phase space for the unbound $NN$ system the centrifugal barrier for
the $d$-wave emission of the $\sigma$ meson plays here a much bigger role and
might suppress its emission more than in the $d\pi\pi$ case. Unfortunately, a
convincing quantitative calculation has not yet been presented by these authors.

Also other small $D$-wave admixtures as they may originate from a $D$-wave
admixture in the intermediate $\Delta\Delta$ system~\cite{Yuan,zhang} or by a
hypothetical decay into a $N^*(1440)N$ system, as considered to some extent by
Bugg \cite{Bugg}, may lead to an ABC-effect like behavior as discussed in
detail in Ref.~\cite{abc}.

If decays such as $d^* \to D_{12}\pi$ and $d^* \to N^*(1440)N$ exist, then
there must be also  a $d^*$ decay branch into the isoscalar $NN\pi$ channel
because of the decays $D_{12} \to NN$ and $N^*(1440) \to N\pi$. The analysis of
WASA data on $np \to pp\pi^-$ and $pp \to pp\pi^0$ reactions with the goal to
extract the isoscalar single-pion production cross section in the $d^*(2380)$
energy region serves the aim to provide an answer for this issue. As mentioned
above the preliminary results do not give any indication for such a decay
\cite{NNpi}. 

In conclusion, the ABC effect turns out to be yet another manifestation that
the dominant asymptotic configuration of $d^*(2380)$ is a $\Delta\Delta$
system in relative $S$ wave.

\section{Are There More Extraordinary Dibaryons?}

\subsection{\it Search for the Isospin $I$ = 3 State $D_{30}$}

Having found one extraordinary dibaryon, of course, raises the question, whether
there are more exceptional dibaryon states. In particular, a state with
quantum numbers mirrored to those of $d^*(2380)$, {\it i.e.}, with
$I(J^P)=3(0^+)$, appears to be particularly appealing. In its highest charge
state it would consist of just six up quarks. It has been predicted to have a
mass similar to that of 
$d^*(2380)$ by a number of theoretical calculations
\cite{Dyson,GG2,Nanjing,Li}, which have been able to predict $d^*(2380)$ at
the proper mass. The calculations of Mota {\it et al.} \cite{Mota1} predict
this state to be right at the $\Delta\Delta$ threshold, {\it i.e.}, much less
bound than $d^*(2380)$. 

Since such a dibaryon is $NN$-decoupled due to its large 
isospin, it can only be produced in combination with other, associatedly
produced particles. For the latter pions are ideally suited due to their large
isospin. Hence Dyson and Xuong \cite{Dyson}, who were the first to propose
such a state (denoted by $D_{30}$, where the first index refers to the isospin
and the second one to the spin of this state), suggested to look for this
state in the reaction $pp \to D_{30}^{++++}\pi^-\pi^- \to
\Delta^{++}\Delta^{++} \to pp\pi^+\pi^+\pi^-\pi^-$. Due to $I$ = 3 the
$\Delta^{++}\Delta^{++}$ configuration is the most preferred $\Delta\Delta$
combination, where $D_{30}^{++++}$ decays into. For an early attempt to search
for an $I$ = 3 dibaryon state in proton-nucleus collisions see Ref.\cite
{IgorI=3}. 

Because of the COSY shutdown a dedicated search run could no longer be
performed by the WASA collaboration. However, using runs from the WASA
data base primarily performed for $\omega$ and $\eta$' production at $\sqrt
s$ = 2.72 and 2.88 GeV, data sets for exclusive and kinematically complete
measurements of four-pion production could be extracted
\cite{I3dibaryon}. These constitute the first data in the low-energy regime of
this reaction. The obtained $pp\pi\pi$-invariant mass spectra do not show any
significant narrow structures and are essentially compatible with a $t$-channel
$N^*(1440)N^*(1440)$ process. By use of  detailed peak analyses upper limits for
the existence of such an $I$ = 3 dibaryon in the mass range 2280 – 2500 MeV
could be deduced. Of course, above the $\Delta\Delta$ threshold of nominal
$2m_\Delta$ = 2464 MeV such a dibaryon state, if existent, would undergo
fall-apart decay with an accordingly huge width, which makes any experimental
detection practically impossible.

Though no solid evidence for the existence of an $I$ = 3 dibaryon state below
the $\Delta\Delta$ threshold was found, the results are still significant. The
currently achievable bounds for $I$ = 3 dibaryon production are primarily not
due to accuracy and statistics of the measurements, but due to the insufficient
quantitative knowledge of the conventional mechanisms in four-pion
production. So the current limits may be improved, as soon there is progress
in the theoretical description of the conventional processes. 

Nevertheless,
these currently obtained upper limits are more than three orders of magnitude
smaller than the formation cross section found for $d^*(2380)$. More
importantly --  with the possible exception of a dibaryon mass of about 2380 MeV
combined with a width of about 100 MeV -- the upper limits are significantly
smaller than the 
cross section for conventional $\Delta^{++}\Delta^{++}$ production in the
$pp\pi^+\pi^+\pi^-\pi^-$ channel – again in sharp
contrast to the corresponding situation for $d^*(2380)$ formation, where this is
an order of magnitude larger than in conventional $\Delta\Delta$ formation. If
the interaction between the two $\Delta^{++}$ particles produced side-by-side
in the decay of the intermediate $N^*N^*$ system would be attractive, then the
probability to form a dibaryon should be quite large – as it is obviously the
case for $d^*(2380)$ formation in the presence of an isoscalar
$\Delta^+\Delta^0$ system. However, the WASA results suggest that the
probability for dibaryon formation in the presence of a
$\Delta^{++}\Delta^{++}$ system in the intermediate state is small (with the
possible exception of the case mentioned above). This is in
support of the findings of Goldman {\it et al.} \cite{Goldman}, who  predicted
an attractive interaction between the $\Delta\Delta$ pair in case of
$d^*(2380)$, but repulsion in case of $D_{30}$ and hence no dibaryon formation. 

\subsection{\it Search for the Isospin $I$ = 2 State $D_{21}$}

According to Dyson and Xuong \cite{Dyson} and supported by recent Faddeev
calculations \cite{GG2} there should exist a $NN$-decoupled dibaryon state
with $I(J^P) = 2(1^+)$ corresponding asymptotically to a $\Delta N$ system and
decaying into $NN\pi$. These calculations predict such a state, called
$D_{21}$ by Dyson and Xuong, with a mass slightly below the $\Delta N$
threshold and with a width close to that of the $\Delta$ -- featuring thus a
state like $D_{12}$ discussed in chapter 4.4, but with mirrored quantum
numbers.

It appears natural to search for such a state in the $pp \to pp\pi^+\pi^-$
reaction, where such a state would be produced associatedly with an additional
pion via the route $pp \to D_{21}^{+++}\pi^- \to pp\pi^+\pi^-$. Unfortunately
there are no high-quality data for this reaction in the energy region of
interest. From WASA and COSY-TOF there exist exclusive and kinematically
complete measurements from threshold up to $T_p$ = 0.8 GeV ( $\sqrt s$ =) 2.24
GeV. Beyond that there are only poor statistics bubble-chamber
measurements. The nominal $\Delta N\pi$ threshold is at $T_p$ = 0.96 GeV
($\sqrt s$ = 2.31 GeV), {\it i.e.}, for a $D_{21}$ search data above this
energy are needed. 

In principle there are high-statistics experimental data available from WASA at
$T_p$ = 1.2 GeV ($\sqrt s$~=~2.4 GeV) and in particular at $T_p$ = 1.4 GeV
($\sqrt s$ = 2.48 GeV), which originally were intended for $\eta$ production
and decay studies. An appropriate analysis of these measurements would allow
the search for $D_{21}$ in 
$pp\pi^+$ and $pp\pi^-$ subsystems up to masses of 2.34 GeV. Due to isospin
coupling $D_{21}$ should show up in the $pp\pi^+$ subsystem six times stronger
than in the $pp\pi^-$ subsystem, where primarily the reflection of this
resonance should be seen.
  
Though the search procedure would be similar to that used for the $D_{30}$ issue
discussed above in the previous section, the search for $D_{21}$ would be under
much better conditions. Experimentally there are ten thousands of
kinematically overdetermined events expected to be obtainable from the WASA
data base.
Also with respect to conventional reaction contributions the situation is
much superior in this case, since the proton-proton induced two-pion
production by $t$-channel meson exchange is understood here on a quantitative
level – also with the additional help of isospin relations and isospin
decomposition \cite{iso}. As shown in Ref.~\cite{iso} 
there is a problem with understanding the total cross section beyond $T_p$ = 0.9
GeV ($\sqrt s$ = 2.29 GeV), which happens to coincide with the $\Delta N\pi$
threshold. Beyond this energy the measured total cross sections are too high
by up to 30$\%$ compared to the prediction by isospin decomposition. However,
all those data originate still from old bubble-chamber measurements.

A hint for the existence of $D_{21}$ comes also from the double-pionic charge
exchange reaction on nuclei considered in chapter 5. There the forward-angle
cross section of so-called non-analog transitions exhibits an unexpected
resonance-like behavior in the region of the $\Delta$ resonance. For its
explanation the so-called DINT (Delta-nucleon INTeraction) mechanism was
invented \cite{Johnson,Johnson1,JohnsonKisslinger,Wirzba}, which in essence
can be thought of a $\Delta N$ system with $I(J^P) = 2(1^+)$ in the
intermediate state \cite{JohnsonKisslinger} -- see chapter 5.

\subsection{\it Renaissance of the $\Delta N$ Region: more Resonance Structures}

In chapter 4.4 we discussed the situation in the region of the $\Delta N$
threshold, where clear resonance structures with the width of the $\Delta$
resonance were observed in the past. Especially pronounced appear these
structures in the $pp \to d\pi^+$ reaction. From a combined SAID partial-wave
analysis of this reaction with $pp$ elastic scattering a looping in the
Argand diagram has been observed for the $^1D_2$, $^3P_2$-$^3F_2$, $^3F_3$ and
$^3F_4$-$^3H_4$ $pp$-partial waves corresponding to isovector states with $J^P
= 2^+, 2^-, 3^-$ and $4^-$. The first one represents the by far most
pronounced resonance structure. It corresponds asymptotically to a slightly
bound $\Delta N$ system in relative $S$ wave, whereas in the other cases this
system has to be in relative $P$ wave.

\begin{figure} 
\centering
\includegraphics[width=10.5cm,clip]{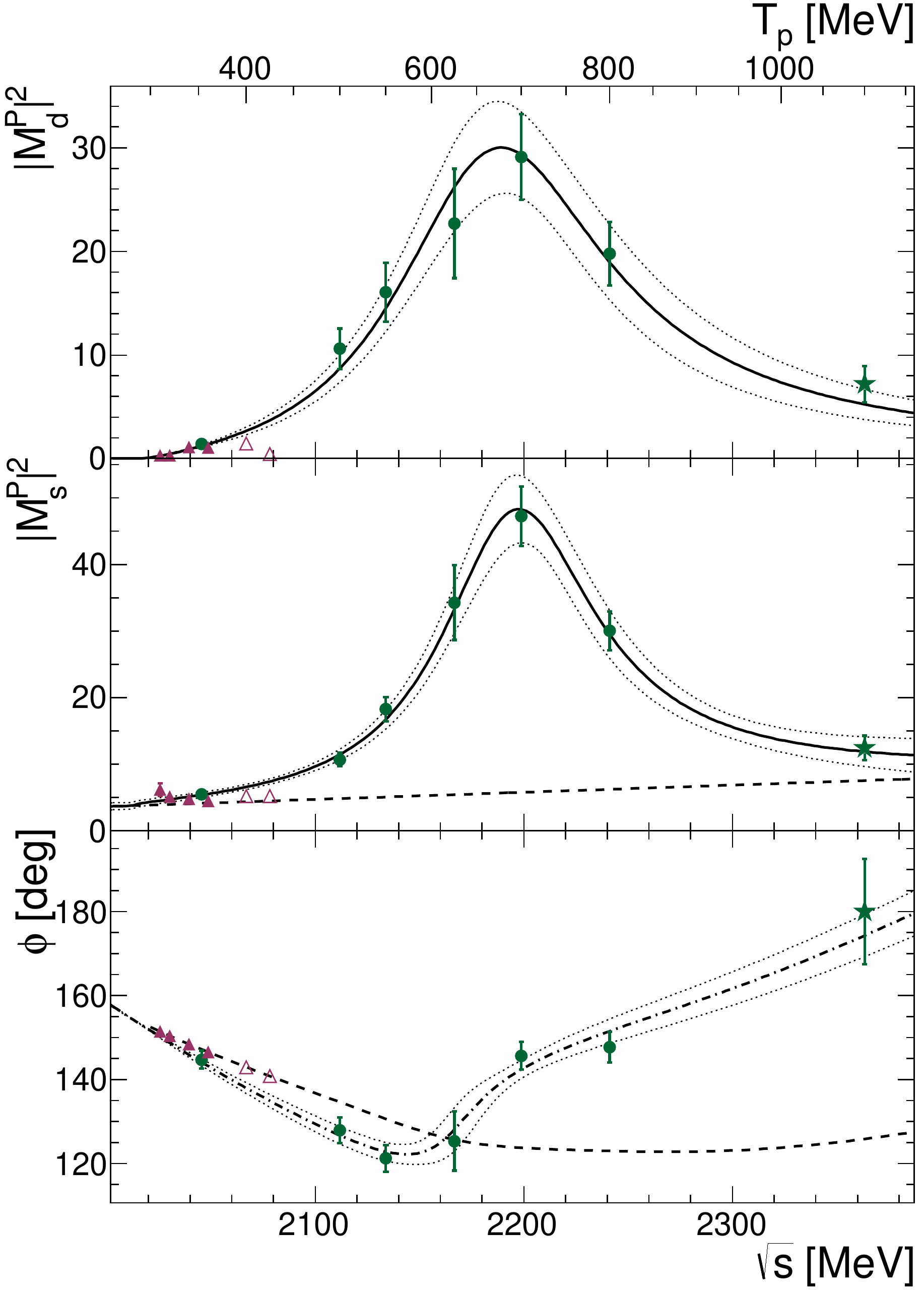}\\
\caption{Energy dependence of the transition amplitudes squared for the
  transitions $^3P_2 \to ^1S_0d$ (top) and $^3P_0 \to ^1S_0s$ (middle) as well
  as their relative phase (bottom). Filled circles and star denote ANKE
  measurements \cite{ANKE}, whereas triangles give PROMICE/WASA results
  \cite{Jozef}. The solid lines represent Breit-Wigner resonance fits with an
  empirical ansatz (dash-dotted line) for the relative phase, the
  corridors indicated by the dashed lines give the 68$\%$ confidence
  interval. The dotted line in the middle panel shows the assumed background
  and the dotted line in the bottom panel the relative phase of the $^3P_0$
  and $^3P_2$ amplitudes in elastic $pp$ scattering. From \cite{ANKE}.    
}
\label{fig-anke}       
\end{figure}

In a very recent series of measurements carried out at the ANKE detector at
COSY the $\vec p p \to pp\pi^0$ reaction has been investigated under the
condition that the emitted proton pair is in relative $S$-wave, {\it i.e.} in
the $^1S_0$ state \cite{ANKE}. That way the ANKE measurements are
complementary to the reaction $pp \to d\pi^+$, where the nucleons within the
deuteron are in relative $^3S_1$ state -- neglecting the small $D$-wave
admixture in the deuteron.  

In a very careful phenomenological analysis of their data as well as of 
corresponding data from PROMICE/WASA \cite{Jozef} over the energy range from
close-to-threshold up to $T_p$ = 1.1 GeV ($\sqrt s$~=~2.36 GeV) the ANKE
collaboration finds the transitions $^3P_0 \to ^1S_0s$ and $^3P_2 \to ^1S_0d$
to exhibit pronounced Lorentzian-shaped resonance structures. Fitted by an
Breit-Wigner ansatz these structures correspond to $I(J^P) = 1(0^-)$ and
$1(2^-)$ resonances with masses $2201\pm5$ MeV and $2197\pm8$ MeV,
respectively, and with widths of $91\pm12$ MeV and $130\pm21$ MeV,
respectively. These values for mass and width suggest the resonance structures
to represent $\Delta N$ configurations, where due to spin and parity $\Delta$
and $N$ have to be in relative $P$ wave. 

The  $I(J^P) = 1(2^-)$ resonance structure has been observed already before
in SAID partial-wave analyses. The ANKE results for mass and width of this
resonance structure are in reasonable agreement with the pole position of
$(2167 - i 86)$ MeV obtained in the SAID analysis of $pp$ scattering
\cite{FA91}. According 
to Ref. \cite{ANKE} the pole position obtained in the more recent
(unpublished) SAID solution SP96 for the $pp \to d\pi^+$ reaction is $(2192 -
i~64)$ MeV, which is even in perfect agreement with the ANKE result.  

Recently Kukulin and Platonova have demonstrated that the spin correlation
parameters are particularly sensitive to this resonance
\cite{Platonova1}. That way they are able to describe all measured cross
section and polarisation data of the $pp \to d\pi^+$ reaction for the first
time in a comprehensive and quantitative manner.

The finding of a  $I(J^P) = 1(0^-)$ state is new, the corresponding transition
$^3P_0 \to ^1S_0s$  is forbidden in the $pp \to d\pi^+$ reaction.  Whereas the
width of the  $I(J^P) = 1(0^-)$ state appears to be slightly less than the
width of the $\Delta$ isobar, the width of the   $I(J^P) = 1(2^-)$ tends to be
slightly larger.

The masses of both resonance structures coincide within their uncertainties
and are about 30 MeV above the nominal $\Delta N$ mass. In ref.~\cite{ANKE} it
is argued that the orbital $P$-wave motion should lead to an increase of the
effective mass by roughly 60 MeV. Relative to that the observed masses appear
to signal a binding of about 30 MeV indicating thus an attractive $P$-wave
$\Delta N$ interaction.

At the CLAS spectrometer at JLAB the $\gamma d \to d\pi^+\pi^-$ reaction has
been investigated recently. Preliminary results have been reported at the APS
Spring meeting 2014 and at BARYON 2016. No sign of $d^*(2380)$ has been
observed as actually expected -- see chapter 10.5.1 --, since this reaction
channel is overwhelmed by conventional processes. However, the Dalitz plot of
the $d\pi^+$-invariant mass-squared versus the $d\pi^-$-invariant mass-squared
exhibits a pronounced broad resonance line symmetric in both invariant mass
systems slightly below the nominal $\Delta N$ threshold. Spin and parity of
this isovector structure have not yet been reported.

\subsection{\it Upcoming Results for the $\Omega^- p$ system ($S$ = -3)} 

Since in heavy ion collisions hyperons are produced copiously, this situation
can be used to investigate the hyperon-hyperon interaction by measuring
hyperon-hyperon correlations. Such correlations of Hanbury-Brown-Twiss
\cite{HBT} type are usually used to fix the size of the source, which emits
the particular particles. But if the source size is known, {\it e.g.} from the
study of other channels, such correlations can be used to deduce scattering
length and effective range of particle-particle interactions
\cite{MoritaOmegaN}. After the 
deduction of the $\Lambda\Lambda$ scattering length in search of the $H$
dibaryon-- see section 7 -- the STAR collaboration has now also measured the
$\Omega^- p$ correlation. The result is expected to be released in nearest
future. 

From the theoretical side a $\Omega^- p$ bound state has been predicted by
several groups. Goldman {\it et al.} found by use of two different quark
models that there should be two bound states with $J$ = 1 and 2 being stable
against strong decay \cite{GoldmanS=3}. Oka found out by use of a constituent
quark model that there should be a quasi-bound state with $I(J^P)$ = $\frac 1 2
(2^+)$ \cite{Oka1a}. Also Li and Shen observe this state to be bound within
their 
chiral quark model approach \cite{LiShen}. And very recently lattice QCD
calculations by the HALQCD collaboration report such a state being bound by 19
MeV -- though the pion mass of 875 MeV used in these calculations is still far
off the physical value \cite{HALQCDOmegaN}.

\subsection{\it Are there Dibaryons in the Heavy-Quark Sector?}

The most convincing evidence for the existence of tetraquark and pentaquark
systems comes from experiments performed in sectors, where charmed and bottom
quarks are involved. This suggests that the attraction between quark clusters
is largest, if each of the two clusters contain a heavy quark (or anti-quark).
This has been worked out recently also for the dibaryon scenario by Karliner
{\it et al.} \cite{karliner1,karliner2,karliner3,karliner4} predicting, {\it
  e.g.}, a doubly heavy $\Sigma_b^+\Sigma_b^-$ dibaryon. 

Pion assisted dibaryon candidates with charm have been discussed recently by
Gal {\it et al.} \cite{Gal,Galcharm} within the chiral constituent quark model
\cite{Galconstituent}. In this work the same formalism as applied
earlier to the $\Lambda N \pi$ system \cite{GarcilazoGal} with strangeness $S$ =
-1 was applied to 
the $\Lambda_c N\pi$ system with charm $C$ = 1. That way $\Lambda(1116)$ and
$\Sigma(1385)$ have been replaced by the charmed analogs $\Lambda_c(2286)$ and
$\Sigma_c(2520)$. Faddeev equations with relativistic kinematics were solved
to look for bound states and resonances with $I(J^P) = \frac 3 2
(2^+)$. Depending on the tested model parameters a bound state or resonance
was generated, which may be viewed as $\Sigma_c(2520) N$ bound state. It could
likely be the lowest lying charmed dibaryon -- below the mass of about 3500 MeV
predicted recently in Ref. \cite{Bayar} for a $DNN$ bound state. 

The $DNN$ bound state
would be analogous to the $\bar K NN$ system discussed in section 6. by
substituting the $\bar K$ by a $D$ meson. Due to the attractive interaction in
the isoscalar part of the $DN$ interaction a $DNN$ system with $I(J^P) =
\frac 1 2 (0^-)$ is expected to be bound by about 250 MeV relative to the
$DNN$ threshold with a narrow width of only 20 - 40 MeV. Hence it should be
observable much easier than the $\bar K NN$ state and may be interpreted
as a quasibound $\Lambda_c(2595)N$ system \cite{Bayar}.

There are not yet any experimental results for this issue. Possible searches
on this issue should be feasible at J-PARC, RHIC, LHC. Also, if PANDA at FAIR
should be realized in some foreseeable future, it would be a top installation for
dibaryon searches in the heavy-quark sector.

\section{Conclusions and Outlook}
\label{sec-10}

The history of dibaryons has been marked by a period of enthusiasm, the
dibaryon rush era, followed by a period of great frustration, where the
buzzword "dibaryon" caused immediate aversions in audiences and
committees. The reason was that  in the nineties an overwhelming number of
dibaryon claims turned out not to survive a careful experimental examination.  

The only remaining experimentally established resonance-like structures
from this period have been the narrow structure at the $\Sigma N$ threshold
and the broad structures near the $\Delta N$ threshold with widths close to
that of the $\Delta$ resonance. The narrow structure at the $\Sigma N$
threshold has
been reexamined by exclusive and kinematically complete measurements of the
$pp \to \Lambda p K^+$ reaction. It agrees with the cusp effect due to the
coupling of $\Lambda p$ and $\Sigma N$ channels. The broad $\Delta N$ resonance
structures have been further established by observation in various
channels. Whether they constitute genuine $s$-channel resonances is still not
definitely settled. Anyway they resemble loosely bound $\Delta N$
systems -- presumably without significant exotic quark configurations. 

So we are left so far with a single firmly established non-trivial dibaryon
state, the one with $I(J^P)=0(3^+)$ at 2380 MeV. Its dynamic decay
properties point to an asymptotic $\Delta\Delta$ configuration, bound by 80 -
90 MeV. Its width of only 70 MeV -- being more than three times smaller than
expected for a conventional $\Delta\Delta$ system excited by $t$-channel meson
exchange -- points to an exotic origin like hidden-color effects.

Though the observation of such a state came for many as a surprise, it was
predicted properly already as early as 1964 by Dyson and Xuong.
After having established experimentally the first non-trivial dibaryon state
the agreement with the early prediction by Dyson and Xuong appears to be very
remarkable. All four states predicted to be coupled to the
$NN$ system have now been observed -- with the first three being the "trivial"
states deuteron, virtual $^1S_0$ state and $^1D_2$ $NN$ scattering state at the
$ \Delta N$ threshold. The remaining two states of the predicted sextet are
$NN$-decoupled having I = 2 and 3, respectively, but may be searched for in
$NN$-initiated multi-pion production. 
Though the 
hadron physics program has finished at COSY, there is still a wealth of WASA
data available, which are suited to search also for these states.   

It comes a bit as a surprise that not a single dibaryonic state has yet been
established in the strange sector -- despite of many dedicated
searches. This includes also the $H$ dibaryon, which once initiated the
dibaryon rush. Though many sophisticated and state-of-the-art experimental
searches have been conducted in recent times, no sign of its existence has
been found -- be it as bound or resonant object. From 
hyperon-nucleon studies we know that the hyperon-nucleon interaction is
substantially weaker than the nucleon-nucleon interaction. From this we
understand that there is no bound strange deuteron. Obviously it is also too
weak to allow for resonances. Whether the current hints for a loosely bound
molecular-like $ppK^-$ system will materialize has to be seen by future
experiments. 
From RHIC experiments we see that the
hyperon-hyperon interaction tends to be still weaker. However, there is hope
that these kind of heavy-ion experiments may give us an answer, whether in
the $S$ = -3 system the $\Omega^- p$ configuration has a bound state.

The fact that tetra- and penta-quark systems have been observed not in the
strange, but in heavy-quark sector, gives confidence that the baryon-baryon
interaction there is more attractive than in the strange sector. This is also
borne out by a number of model calculations, which support the hope that in
this sector also dibaryonic hexaquark systems will be found in near future.  

\section{Acknowledgments}
\label{sec-11}

The work on this issue would not have been possible without intense
communications and 
discussions lasting partly over decades, for which I am particularly grateful
to Dick Arndt, W. Briscoe, Stanley J. Brodsky, D. Bugg, Y. B. Dong,
E. Friedman, A. Gal,
T. Goldman, W. Gibbs, Ch. Hanhart, M. B. Johnson, V. Kukulin, E. Oset,
M. Platonova, P. N. Shen, M. Schepkin, K. K. Seth, I. Strakovsky, 
A. Valcarce, F. Wang, C. Wilkin, R. Workman and Z. Y. Zhang. I am particularly
indebted to my T\"ubingen 
colleagues M. Bashkanov, A. Buchmann, A. Faessler, Th. Gutsche, H. M\"uther, T.
Skorodko and G. J. Wagner -- and all my former students -- for their enduring
help in all kinds of problems.  
The finally successful dibaryon search would not have been possible without
the intense support by the WASA collaboration and the staffs of CELSIUS and
COSY. 

I am grateful to all of those, 
who were very sceptical about the
dibaryon issue and thus forced us to go through all the cumbersome tests of
the dibaryon hypothesis.

This work has been supported by DFG(CL 214/3-1 and 3-2) and in the years
before by BMBF and Forschungszentrum J\"ulich (COSY-FFE).  

\end{document}